\documentclass[a4paper,12pt]{article}
\usepackage{fullpage}
\usepackage{verbatim}
\usepackage{bm,latexsym,amsmath,amssymb,amsfonts,mathrsfs,amsthm,mathtools}
\usepackage{bbm} 
\usepackage{graphicx}
\usepackage[nosort]{cite}
\usepackage{color}
\input{colordvi.tex}
\usepackage{latexsym}
\usepackage{epsf}
\usepackage{slashed} 
\usepackage{array} 
\usepackage{booktabs}

\usepackage{feynmp-auto}

\renewcommand{\d}{\textrm{d}}
\newcommand{\e}{\textrm{e}}

\newcommand{\ba}{\begin{equation}\begin{aligned}}
\newcommand{\ea}{\end{aligned}\end{equation}}

\renewcommand{\d}{\textrm{d}}

\def\cald         {{\cal D}}

\def\cali         {{\cal I}}

\def\call         {{\cal L}}

\def\calo         {{\cal O}}

\newcommand{\de}{\partial}

\newcommand{\tr}{\textrm{tr\hskip0.1em}}
\newcommand{\Id}{\mathbbm{1}}

\allowdisplaybreaks

\allowdisplaybreaks[3]

\usepackage[hang]{footmisc}
\usepackage[compress,square,numbers]{natbib}
\usepackage[colorlinks,linktocpage,linkcolor=blue,citecolor=blue,urlcolor=blue]{hyperref}

\renewcommand{\d}{\mathrm{d}}

\newcommand{\nl}{\notag \\ &\quad\,}
\newcommand{\nll}{\notag \\ &}

\begin{document}

\numberwithin{equation}{section}

\thispagestyle{empty}

\begin{flushright}
\small KOBE-COSMO-18-01\\
\small MAD-TH-17-07
\normalsize
\end{flushright}
\vspace*{1cm}

\begin{center}

{\LARGE \bf A Tower Weak Gravity Conjecture}

\vspace{0.5cm}

{\LARGE \bf from Infrared Consistency}

\vspace{1cm}
{\large Stefano Andriolo${}^{1}$, Daniel Junghans${}^{2}$, Toshifumi Noumi${}^{3}$ and Gary Shiu${}^{1,4}$}\\

\vspace{1cm}
${}^1$ Department of Physics and Jockey Club Institute for Advanced Study\\
Hong Kong University of Science and Technology, Hong Kong\\

\vspace{0.5cm}
${}^2$ Institut f{\"{u}}r Theoretische Physik, Ruprecht-Karls-Universit{\"{a}}t Heidelberg,\\
Philosophenweg 19, 69120 Heidelberg, Germany\\

\vspace{0.5cm}
${}^3$ Department of Physics, Kobe University\\
Kobe 657-8501, Japan\\

\vspace{0.5cm}
${}^4$ Department of Physics, University of Wisconsin-Madison\\
Madison, WI 53706, USA\\

\vspace{1.5cm}

\begin{abstract}
We analyze infrared consistency conditions of 3D and 4D effective field theories with massive scalars or fermions charged under multiple $U(1)$ gauge fields. At low energies, one can integrate out the massive particles and thus obtain a one-loop effective action for the gauge fields. In the regime where charge-independent contributions to higher-derivative terms in the action are sufficiently small, it is then possible to derive constraints on the charge-to-mass ratios of the massive particles from requiring that photons propagate causally and have an analytic S-matrix. We thus find that the theories need to contain bifundamentals and satisfy a version of the weak gravity conjecture known as the convex-hull condition. Demanding self-consistency of the constraints under Kaluza-Klein compactification, we furthermore show that, for scalars, they imply a stronger version of the weak gravity conjecture in which the charge-to-mass ratios of an infinite tower of particles are bounded from below. We find that the tower must again include bifundamentals but does not necessarily have to occupy a charge (sub-)lattice.
\end{abstract}

\end{center}

\newpage

\setcounter{tocdepth}{2}
\tableofcontents

\section{Introduction}

Many aspects of quantum gravity can conveniently be analyzed within the framework of effective field theory (EFT). At low energies, the dynamics are expected to be governed by an effective action with only a small number of degrees of freedom, while most of the complicated details of the underlying microscopic theory do not play a role. A natural question to ask is then whether all EFTs one can write down arise as the low-energy limit of a consistent quantum gravity theory. Based on thought experiments and general expectations about the properties of quantum gravity, it has been argued that this is most likely not the case. Instead, the EFTs which admit a UV completion into a theory of quantum gravity (termed the ``landscape'') are distinguished from those which do not (the ``swampland'') by a number of rules and consistency conditions.

\medskip
A well-known proposal for such a condition is the weak gravity conjecture (WGC) \cite{ArkaniHamed:2006dz}, which asserts that a $U(1)$ gauge theory coupled to gravity needs to contain at least one particle with mass $m$ and charge $q$ whose charge-to-mass ratio satisfies a lower bound
\begin{equation}
z = \frac{gq}{m} \ge \mathcal{O}(1) \label{wgc}
\end{equation}
in Planck units. Here, $g$ is the gauge coupling constant, and the precise numerical value of the bound depends on the considered theory. Depending on the version of the conjecture, the particle may also be required to satisfy additional properties such as being the lightest particle in the theory.

\medskip
The WGC has a natural generalization to $p$-branes charged under $(p+1)$-form fields. In particular, the $0$-form (axion) version of the conjecture has generated a lot of activity in recent years since it may imply strong constraints on large-field inflation \cite{Rudelius:2014wla, delaFuente:2014aca, Montero:2015ofa, Brown:2015iha, Brown:2015lia, Hebecker:2015rya, Bachlechner:2015qja, Junghans:2015hba, Rudelius:2015xta, Heidenreich:2015wga, Kooner:2015rza, Hebecker:2015zss, Palti:2015xra, Baume:2016psm, Hebecker:2016dsw, Hebecker:2017wsu, Hebecker:2017uix, Blumenhagen:2017cxt} (see also \cite{Ibanez:2015fcv,Hebecker:2015zss,Brown:2016nqt}  for an application to cosmological relaxation). Moreover, it has recently been realized that the conjecture is closely related to 
the swampland conjecture \cite{Ooguri:2006in, Klaewer:2016kiy, Palti:2017elp}, to cosmic censorship \cite{Crisford:2017gsb,Cottrell:2016bty} and to instabilities of non-supersymmetric AdS vacua \cite{Ooguri:2016pdq, Danielsson:2016mtx, Freivogel:2016qwc, Ooguri:2017njy, Danielsson:2017max}. In particular, the last relation implies an intriguing constraint that the WGC imposes on neutrino physics \cite{Ooguri:2016pdq}.
Possible {\it correlated} consequences in
particle physics and cosmology
were explored in \cite{Ibanez:2017kvh, Ibanez:2017oqr, Hamada:2017yji}. Various other extensions/applications of the WGC and related quantum gravity conjectures have furthermore been discussed in the recent works \cite{Montero:2017yja, Montero:2017mdq, Hebecker:2017lxm, Lust:2017wrl, Ibanez:2017vfl, rudelius2017}.
See \cite{Brennan:2017rbf} for a recent review.

\medskip
Another way to generalize the WGC is to consider theories in which the gauge group contains multiple $U(1)$ factors. In \cite{Cheung:2014vva}, it was argued based on black hole arguments that such theories are only consistent with quantum gravity if they satisfy a convex-hull condition, i.e., they require a set of particles such that the convex hull of the charge-to-mass vectors contains a ball of radius $\mathcal{O}(1)$. An even stronger version of the WGC---the so-called lattice WGC---can be motivated if one additionally demands that the WGC is self-consistent under Kaluza-Klein compactification \cite{Heidenreich:2015nta}.
Requiring consistency under dimensional reduction suggests that a bound on the charge-to-mass vectors has to be satisfied by 
the whole charge lattice. This stronger version of the WGC was subsequently shown to not always hold \cite{Montero:2016tif,Heidenreich:2016aqi}, though there are examples in string theory where the particles satisfying the WGC occupy a proper sub-lattice \cite{Montero:2016tif,Heidenreich:2016aqi}.\footnote{There appear to be string theory examples where BPS states do not span a (sub-)lattice. We thank Eran Palti for private communication on this point. See \cite{palti2018} for work relating this to the swampland and weak gravity conjectures.}

\medskip
In view of the potentially far-reaching implications of the WGC, it is of obvious importance to understand which of the many versions of it, if any, holds in quantum gravity. While the conjecture was originally motivated by general black hole arguments and circumstantial evidence in string theory, there have been efforts in the recent literature to make this more precise and bring us closer to proving the conjecture. Indeed, this has been achieved with some success, at least in specific setups, using AdS/CFT \cite{Nakayama:2015hga, Harlow:2015lma, Benjamin:2016fhe, Montero:2016tif} or from entropy considerations \cite{Cottrell:2016bty}.\footnote{Subsequent works \cite{Fisher:2017dbc,Cheung:2018cwt} using arguments along the lines of \cite{Cottrell:2016bty} have appeared. However, despite what their titles suggest, \cite{Fisher:2017dbc,Cheung:2018cwt} do not present proofs of the WGC. The entropy corrections formulae used in \cite{Fisher:2017dbc} cannot be applied in the regime of macroscopic black holes, nor away from extremality, which is where conflicts with the WGC were argued to arise. Ref.~\cite{Cheung:2018cwt} made an interesting connection between the WGC and the positivity of entropy corrections. It is not known, however, if the latter follows from some fundamental consistency conditions. Logarithmic corrections to extremal black hole entropy are not universally positive. See, e.g., \cite{Sen:2011ba, Sen:2014aja}.}

\medskip
Another possibility is to derive WGC-like bounds from infrared consistency conditions of the EFTs. In \cite{Adams:2006sv}, it was shown that the requirements of causal photon propagation and an analytic S-matrix imply that the coefficients of certain higher-derivative terms in the effective action must be positive or zero. Since these coefficients receive loop corrections from charged particles, one can reformulate the positivity constraints in terms of a bound of the form \eqref{wgc}. This idea was used in \cite{Cheung:2014ega} to show that the WGC must indeed hold in simple EFTs with a single $U(1)$ gauge field, provided that a certain parameter containing the charge-independent contributions to the higher-derivative terms is sufficiently small. The value of this parameter depends on the UV completion of the EFT and can thus be interpreted as encoding the microscopic properties of the quantum gravity theory.

\medskip
In the present paper, we apply the ideas of \cite{Adams:2006sv, Cheung:2014ega} to 3D and 4D EFTs with multiple 
$U(1)$ gauge fields. We find that causality and analyticity constraints then again lead to lower bounds for the charge-to-mass ratios of particles in the regime where charge-independent contributions to the higher-derivative terms in the effective action are small. We are thus able to recover the convex-hull condition without making reference to arguments involving black holes. However, compared to the single-$U(1)$ case analyzed in \cite{Cheung:2014ega}, we also find qualitatively new effects. Specifically, one of the constraints we find yields an \emph{upper} bound instead of a lower bound on the charge-to-mass ratios, unless the theory contains particles charged under multiple $U(1)$'s. This constraint is due to the requirement that photons travel subluminally in backgrounds generated by different gauge fields and does therefore not appear in theories with just a single $U(1)$ factor. We interpret it as evidence that the WGC for theories with multiple $U(1$)'s should be stronger than the convex-hull condition (which can also be satisfied in theories with a diagonal charge matrix).

\medskip
In order to substantiate this claim, we then analyze the self-consistency of the causality and analyticity constraints under the compactification of a class of 4D EFTs on a circle. We find that, due to the Kaluza-Klein gauge field, the constraints become stronger in the compactified theories. In particular, for scalar theories, they cannot be satisfied anymore by proposing a finite number of particles with bounded charge-to-mass ratios. In order that both ordinary and Kaluza-Klein photons travel subluminally, the theories instead need to contain an infinite tower of particles satisfying such a bound. Interestingly, we find that the tower must include bifundamentals but does not necessarily have to occupy the full charge lattice. Our result thus suggests a very specific version of the WGC, which, to our knowledge, is compatible with all known examples in string theory.

\medskip
Let us take stock of our findings. The requirement for an ultraviolet-completable theory to be well-behaved upon compactification has been used as a guiding principle for distinguishing the landscape from the swampland \cite{Brown:2015iha,Heidenreich:2015nta,  Montero:2017yja}. 
Rather than assuming consistency of some conjectured principles 
upon Kaluza-Klein reduction, the present work examines potential inconsistencies directly in the lower-dimensional theory.
In a sense, our result is a more direct test of consistency of the theory under dimensional reduction, as causality and unitarity are well-tested principles of Nature. While our analysis applies to theories with multiple $U(1)$'s, we consider for simplicity phases of the theories where all the $U(1)$ gauge fields remain massless. It would be interesting to extend our study to cases where some of the $U(1)$'s gain a mass in the infrared \cite{Shiu:2015uva, Shiu:2015xda, Saraswat:2016eaz}. We leave this investigation to a future work.

\medskip
This paper is organized as follows. In Sec.~\ref{sec:3d}, we derive causality and analyticity constraints for 3D EFTs with multiple $U(1)$ gauge fields and use them to obtain bounds on the charge-to-mass ratios of charged particles. In Sec.~\ref{sec:4d}, we repeat our analysis for 4D EFTs. In Sec.~\ref{sec:comp}, we study the Kaluza-Klein compactification of 4D EFTs on a circle and argue that causality and analyticity then imply a strong form of the WGC in which the charge-to-mass ratios of an infinite tower of particles are bounded from below. We conclude in Sec.~\ref{concl} with a discussion of our results. The details of several longer computations can be found in Apps.~\ref{app:formulae}--\ref{app:reduction}.

\section{Infrared Consistency in $D=3$}
\label{sec:3d}

In the proceeding two sections, we derive a class of bounds on charge-to-mass ratios of matter in theories with multiple $U(1)$ gauge symmetries, by generalizing the analysis in~\cite{Cheung:2014ega}. We focus on 3D in this section and then extend our arguments to 4D in the next section.

\medskip
As we explain in Sec.~\ref{setup3d}, our starting point is the low-energy EFT of multiple photons, whose EFT parameters depend on the charge-to-mass ratios of matter fields that have been integrated out. The positivity bounds on these EFT parameters, derived in Secs.~\ref{causality3d}--\ref{analyticity3d} from causality and analyticity, are then translated into bounds on the charge-to-mass ratios in Sec.~\ref{ex:3d}. We demonstrate that, in addition to an ordinary WGC-type lower bound, this includes a new upper bound on the charge-to-mass ratios unless the theory contains particles charged under multiple $U(1)$'s. As we discuss in Sec.~\ref{sec:comp}, our new bound turns out to be crucial to motivating the tower WGC.

\subsection{Setup}
\label{setup3d}

Let us suppose that the dynamics below a cutoff scale $\Lambda$ is captured by massive charged particles coupled to gravity and $N$ $U(1)$ gauge fields. For concreteness, we consider a Wilsonian effective action of the form\footnote{Throughout the paper, we use the mostly-plus convention for the metric.}
\begin{align}
\label{startingEFT_fermion}
\Gamma & = \int \d^3 x \sqrt{-g} 
\bigg[
\frac{M_3}{2} R - \frac{1}{4} \sum_iF_i^2
\bigg]
+\text{C.S.}
+\text{H.O.}
+ \left\{\begin{matrix*}[l]
  \Gamma_\text{scalar} \\
  \Gamma_\text{fermion} \end{matrix*}\right.
\,,
\end{align}
where we consider either scalar or fermionic matter fields with
\begin{align}
\Gamma_\text{scalar} &= \int \d^3 x \sqrt{-g}\, \sum_a \left(-|D_{\mu} \phi_a|^2
- m_a^2 |\phi_a|^2\right), \\
\Gamma_\text{fermion} &= 
\int \d^3 x \sqrt{-g}\, \sum_a\bar\psi_a (-\slashed{D} - m_a) \psi _a\,. \label{3dfermion-action}
\end{align}
Here, $M_3$ is the 3D Planck mass. In what follows, we use $i,j,...$ to label the $N$ photons and $a,b,...$ for the massive charged particles. The covariant derivative is defined by
\begin{equation}
D_{\mu} = \nabla_\mu  + i \sum_iq_{ai} g_i A_{i\mu}
\,,
\end{equation}
where $q_{ai}$ is the $i$-th $U(1)$ charge of the particle $a$ and $g_i$ is the gauge coupling of the $i$-th $U(1)$.
``C.S." denotes parity-violating Chern-Simons terms which can in general appear in the effective action (notice that parity is already violated by the presence of fermion masses in \eqref{3dfermion-action}). ``H.O." stands for higher-dimensional operators, which depend on the UV completion beyond the cutoff $\Lambda$ and are given by combinations of Riemann tensors and gauge field strengths. In 3D, the Riemann tensor is completely determined by the Ricci tensor, and terms involving the latter can be eliminated by a field redefinition at the four-derivative level (see App.~\ref{simplify_action}).\footnote{This is, however, different in 4D, see Sec.~\ref{setup4d}.} The general form of the higher-dimensional operators is therefore
\begin{align}
\label{HOops}
\text{H.O.}
= \sum_{i,j,k,l} c_{ijkl} (F_i \cdot F_j) (F_k \cdot F_l)
\end{align}
up to terms with more than four derivatives.

\medskip
The charge-to-mass ratio of a scalar or a fermion is defined by
\begin{align}
\label{3dz}
z_{ai}\equiv \frac{q_{ai}g_i\sqrt{M_3}}{|m_a|}\,,
\end{align}
and this is what we would like to constrain in the following by requiring that the EFT is consistent in the IR.

\medskip
For this purpose, we integrate out the massive charged particles to obtain a 1-loop effective action of gravity and $N$ photons. Since the calculation is quite long, we only present the results here and refer the interested reader to App.~\ref{app:heatkernel} for the details. As we derive there, integrating out the particles yields higher-derivative corrections to the effective action which are given by products of Riemann tensors and gauge field strengths. At the four-derivative level, they are schematically of the form $R^2$, $R F^2$ and $F^4$. However, as stated above, the curvature dependence of these terms can be eliminated by a field redefinition such that, subsequently, all corrections are of the form $F^4$ (see App.~\ref{simplify_action}). Up to terms with more than four derivatives, we thus find
\begin{equation}
\label{3dlagrangian}
\Gamma_1 = \int \d^3 x \sqrt{-g}
\left[
\frac{M_3}{2} R
- \frac{1}{4} \sum_{i,j}\delta_{ij} F_i \cdot F_j 
+ \sum_{i,j,k,l} C_{ijkl} (F_i \cdot F_j) (F_k \cdot F_l) \right]
+\widetilde{\text{C.S.}}
\end{equation}
with
\begin{align}
\label{cijkl}
C_{ijkl} &= c_{ijkl} 
 + \sum_a\frac{1}{1920\pi|m_a|M_3^2} \cdot \left\{\begin{matrix*}[l]
   \left[ \frac{7}{8}z_{ai}z_{aj}z_{ak}z_{al} + \frac{3}{2} z_{ai}z_{aj} \delta_{kl}  - z_{ai}z_{ak} \delta_{jl} \right. \\[0.5em] \left. \qquad + \frac{1}{2}\delta_{ij}\delta_{kl} + \delta_{ik}\delta_{jl} \right] & \text{(scalars)} \\[1.2em]
  \left[ z_{ai}z_{aj}z_{ak}z_{al} + z_{ai}z_{aj} \delta_{kl}  - \frac{3}{2} z_{ai}z_{ak} \delta_{jl}  \right. \\[0.5em] \left. \qquad -\frac{1}{2}\delta_{ij}\delta_{kl} + \frac{3}{2} \delta_{ik}\delta_{jl}\right] & \text{(fermions)\,.} \end{matrix*}\right.
\end{align}
Here, $\widetilde{\text{C.S.}} = \text{C.S.} + \text{C.S.}_{\text{1-loop}} $ is the 1-loop corrected Chern-Simons term, which generates a mass for the corresponding photons if nonzero. Indeed, the Chern-Simons level is shifted by fermion loop effects \cite{Dunne:1998qy}. In this paper, we would like to analyze massless $U(1)$'s and therefore focus on the case where the total Chern-Simons term vanishes,
$\widetilde{\text{C.S.}}=0$.\footnote{It was argued in \cite{Montero:2017yja} that EFTs consistent with quantum gravity must contain phases in which a different type of Chern-Simons term is nonzero, which involves a coupling of a $U(1)$ gauge field to other form fields. For simplicity, we will assume a phase in which such terms are not present.}

\begin{figure}[t]
\centering
\includegraphics[scale=1.1]{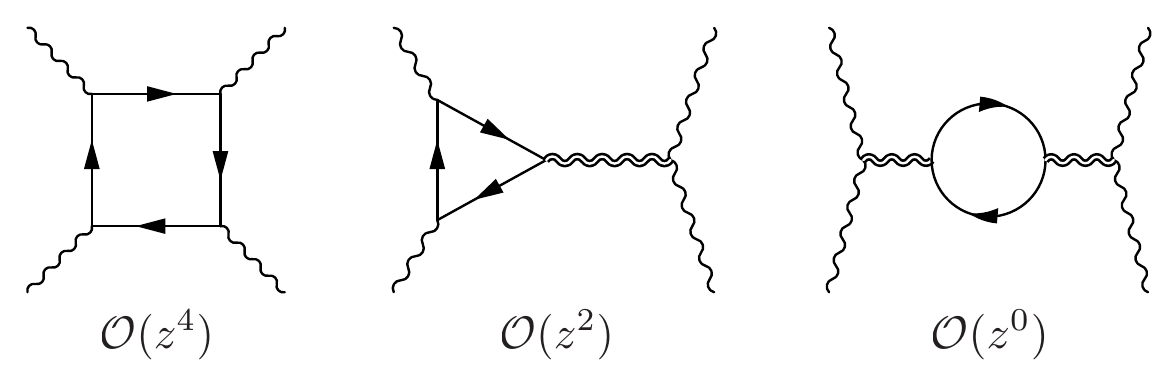}
\caption{\label{fig:diagrams}\emph{Typical diagrams for the effective $F^4$ operator after integrating out scalars/fermions. In the left, the scalar/fermion loop induces $F^4$ through four gauge couplings. In the middle, the loop induces an $RF^2$ term through two gauge couplings and one gravitational coupling, hence it is $\mathcal{O}(z^2)$. After using the tree-level equation of motion, $R\sim F^2$, it is converted to $F^4$. In the right, the loop induces $R^2$, which is converted to $F^4$ with an $\mathcal{O}(z^0)$ coefficient.}}
\end{figure}

\medskip
According to \eqref{cijkl}, the $C_{ijkl}$ coefficients are given schematically by
\begin{align}
C_{ijkl} \sim \calo(z^4) + \calo(z^2) + \calo(z^0)
\,.
\end{align}
As illustrated in Fig.~\ref{fig:diagrams}, diagrams involving loops of the scalars/fermions we integrated out contribute to all three types of terms in $C_{ijkl}$. The $\calo(z^0)$ term furthermore receives contributions from photon loops and the higher-order operators denoted by ``H.O." in \eqref{startingEFT_fermion}. This implies in particular that the magnitude of the $\calo(z^0)$ term depends on the UV completion of the EFT and should consequently be viewed as an (a priori unknown) boundary condition encoding the microscopic properties of the quantum gravity theory \cite{Cheung:2014ega}. 

\medskip
An interesting observation by Cheung and Remmen~\cite{Cheung:2014ega} is that the positivity of the EFT parameters $C_{ijkl}$ implies a WGC-type lower bound on the charge-to-mass ratios if the $\mathcal{O}(z^0)$ terms mentioned above are in a certain range. This was shown in \cite{Cheung:2014ega} for EFTs with a single $U(1)$ gauge field, where the effective action only depends on one parameter $C_{1111}$. We extend their argument to the multiple-$U(1)$ case in the rest of this section. First, in the next two subsections, we show that both causality and analyticity imply a positivity bound on (a particular combination of) the $C_{ijkl}$'s.
In Sec.~\ref{ex:3d}, we then use this bound to constrain the charge-to-mass ratios. There, we show that an ordinary WGC-type lower bound follows in a certain range of the $\mathcal{O}(z^0)$ term, which can be understood as a 3D analogue of the convex-hull condition. Interestingly, we also find that a \emph{new} upper bound on the charge-to-mass ratios shows up unless the theory contains particles charged under multiple $U(1)$'s. This new constraint will be crucial in order to motivate the tower WGC in Sec.~\ref{sec:comp}.

\medskip
Before proceeding with the discussion, let us briefly summarize the parameter range where our argument is applicable. Throughout the discussion, we assume that the gauge and gravitational interactions are in the perturbative regime:
\begin{align}
\frac{|qg|}{\sqrt{|m|}}\ll1\,,
\qquad
\frac{|m|}{M_3}\ll1\,,
\end{align}
where we dropped the $(a,i)$ indices. Moreover, restricting to terms with at most four derivatives in the effective action is tantamount to working in the weak-field limit (for both gravity and photons). This means working in the regime where 
\begin{equation}
\frac{|qgF| }{m^2} \ll 1
\,, \qquad
\frac{|R|}{m^2}  \ll 1 
\,.
\end{equation}
Since the charge-to-mass ratio~\eqref{3dz} takes the form
\begin{align}
z=\frac{qg}{\sqrt{|m|}}\left|\frac{m}{M_3}\right|^{-1/2}\,,
\end{align}
we may cover a parametrically wide range of the charge-to-mass ratios:
\begin{align}
\left|\frac{qg}{\sqrt{|m|}}\right|\ll|z|
\ll \left|\frac{m}{M_3}\right|^{-1/2}\,.
\end{align}
In particular, $z\sim\mathcal{O}(1)$ near the WGC bound is in our regime of validity.

\subsection{Causality Constraints}
\label{causality3d}

We now study the IR consistency of the effective Lagrangian~\eqref{3dlagrangian} with vanishing Chern-Simons term,
\begin{equation}
\label{photon-eft}
\Gamma_1 = \int \d^3 x \sqrt{-g}
\left[
\frac{M_3}{2} R
- \frac{1}{4} \sum_{i,j}\delta_{ij} F_i \cdot F_j 
+ \sum_{i,j,k,l} C_{ijkl} (F_i \cdot F_j) (F_k \cdot F_l) \right]
\,,
\end{equation}
where $C_{ijkl}$ is given by \eqref{cijkl} and depends on the charges and masses of the matter fields that have been integrated out.
In 3D, a massless vector field is dual to a massless scalar field. Instead of \eqref{photon-eft}, we can therefore consider the dual scalar theory
\begin{equation}
\label{finalEFT_dualised}
\Gamma_1  = \int \d^3 x \sqrt{-g} 
\left[
\frac{M_3}{2} R 
- \frac{1}{2} \sum_{i,j}\delta_{ij} \de\phi_i \cdot \de\phi_j
+ 4 \sum_{i,j,k,l}C_{ijkl} (\de\phi_{i} \cdot\de\phi_j)(\de\phi_{k}\cdot \de\phi_l) \right]
\,,
\end{equation}
which is valid up to terms with more than four derivatives and obtained by the usual procedure of integrating out an auxiliary field (see App.~\ref{app:dualization}).

\medskip
Following the ideas in \cite{Adams:2006sv, Cheung:2014ega}, we can now derive a bound on the $C_{ijkl}$'s by requiring that fluctuations of the fields around nontrivial backgrounds are subluminal.\footnote{To be precise, we need to discuss the global causal structure. It turns out, however, that the subluminality argument we make here practically reproduces the same condition. See~\cite{Adams:2006sv} for more details.} For this purpose, we expand $g_{\mu\nu}$ and $\phi_i$ around their background values, denoted with a bar:
\begin{align}
g_{\mu\nu} = \bar g_{\mu\nu} + h_{\mu\nu}
\,,
\qquad
\phi_i = \bar\phi_i + \varphi_i 
\,.
\end{align}
Since the graviton is non-dynamical in 3D \cite{Deser:1983tn}, we set $h_{\mu\nu} = 0$.

\medskip
For simplicity, let us assume a constant electromagnetic background field $\overline{\de_\alpha\phi_i} = w_{i \alpha}$. Here and in what follows, we take the local Lorentz frame and use $\alpha,\beta,\ldots$ for local Lorentz indices. The metric is given by $\eta_{\alpha\beta}=(-++)$ in particular.
At quadratic order in the fluctuations $\varphi_i$, the Lagrangian then takes the form\footnote{We define symmetrized and anti-symmetrized quantities as $A_{(ij)}=A_{ij}+A_{ji}$ and $A_{[ij]}=A_{ij}-A_{ji}$, respectively.}
\begin{equation}
\label{finalEFT_fluctuations}
\call  =
-\frac{1}{2} \sum_{i,j}\delta_{ij} \de\varphi_i \cdot \de\varphi_j
+ 4 \sum_{i,j,k,l}C_{(ij)(kl)} (w_i \cdot \de\varphi_j) (w_k \cdot \de\varphi_l) 
\,,
\end{equation}
where we used the leading-order equations of motion $\de^2 \varphi_i = 0$ (amounting to a field redefinition) to simplify the expression. In momentum space, it may be rewritten as
\begin{align}
\mathcal{L}=\frac{1}{2}\sum_{i,j}K_{ij}(w\cdot k)\varphi_{i}(k)\varphi_{j}(-k)\,,
\end{align}
where $K_{ij}$ is the kinetic operator
\begin{align}
K_{ij}(w\cdot k)=-\delta_{ij}k^2+4D_{ij}(w\cdot k)\,,
\end{align}
and $D_{ij}$ is the correction to the dispersion,
\begin{align}
\label{dispersion3D}
D_{ij}(w\cdot k)=\sum_{k,l}\left( C_{(ik)(jl)}+C_{(jk)(il)}\right)(w_{k}\cdot k)(w_l\cdot k).
\end{align}

\medskip
To discuss subluminality, let us diagonalize the kinetic operator as
\begin{align}
\widetilde{K}_{ij}={\rm diag}\Big(-k^2+4D_1(w\cdot k),-k^2+4D_2(w\cdot k),\ldots , -k^2+ 4D_N(w\cdot k)\Big)\,,
\end{align}
where we denote by $D_i$ the eigenvalues of the matrix $D_{ij}$. We then have $N$ modes with the dispersion relations
\begin{align}
-k^2+ 4D_i(w\cdot k)=0\,. \label{dispersion1}
\end{align}
The subluminality of the fluctuations therefore implies
\begin{align}
D_i(w\cdot k)\geq0 \quad\text{on the shell}. \label{dispersion2}
\end{align}
Here, the dispersion relations \eqref{dispersion1} should be considered order by order in the weak-field expansion. Recall that the corrections $D_i(w\cdot k)$ to the leading-order equation $k^2=0$ originate from four-derivative terms in the effective action \eqref{finalEFT_dualised}. Up to higher-order corrections which we neglect, it is then valid to rephrase \eqref{dispersion2} as
\begin{align}
D_i(w\cdot k)\geq0 \quad\text{for any null vector $k_\alpha$ and any $w_{i\alpha}$}.
\end{align}
In terms of the original matrix $D_{ij}$, we thus have
\begin{align}
\sum_{i,j}D_{ij}(w\cdot k)u_iu_j\geq0
\end{align}
for arbitrary real $u_i$ and $w_{i\alpha}k^\alpha$. Writing $v_i \equiv w_{i\alpha}k^\alpha$ for convenience, we can rewrite this as
\begin{align}
\label{bound_on_D_3d}
\sum_{i,j,k,l}C_{(ij)(kl)}u_iv_ju_kv_l\geq0
\end{align}
for arbitrary real $\vec{u}$ and $\vec{v}$. In the following, without loss of generality, we assume that $\vec{u}$ and $\vec{v}$ are unit vectors.

\medskip
The positivity constraint \eqref{bound_on_D_3d} is one of the main results of this paper. We will see below that it can also be obtained from requiring analyticity of the photon S-matrix and that it can be used to obtain WGC-like bounds on the charge-to-mass ratios of the particles we integrated out before.

\subsection{Analyticity Constraints}
\label{analyticity3d}

\begin{figure}[t]
\centering
\includegraphics[scale=1.2]{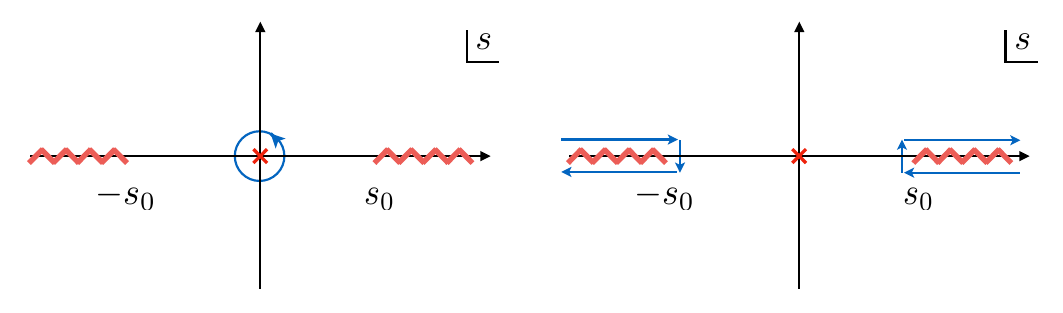}
\caption{\label{fig:contour}\emph{The blue curve in the left figure is the integration contour for Eq.~\eqref{int_IR}, which captures the IR physics. On the other hand, the one in the right figure is for Eq.~\eqref{int_UV}, which carries the UV information. The integrand accommodates a pole at the origin and discontinuities on the real axis associated to on-shell intermediate states  (depicted by red).}}
\end{figure}

We now derive the same positivity constraint by using the optical theorem and analyticity of scattering amplitudes.  The key is that we may relate IR amplitudes to the UV ones by virtue of analyticity. Following~\cite{Adams:2006sv}, let us consider a contour integral,
\begin{align}
\oint \frac{\d s}{2\pi i}\frac{\mathcal{M}(1_i,2_j,3_k,4_l;s)}{s^3}\,,
\end{align}
of the photon forward scattering amplitude $\mathcal{M}(1_i,2_j,3_k,4_l;s)$, where, e.g., $1_i$ means that the first photon is for the $i$-th $U(1)$ and $s$ is the Mandelstam variable satisfying $s=-(k_1+k_2)^2$. The integration contour is defined such that it encircles the origin $s=0$ (see Fig.~\ref{fig:contour}). We then evaluate this integral in two different ways, based on the IR and UV viewpoints. First, our effective Lagrangian~\eqref{finalEFT_dualised} tells us that the photon forward scattering takes the form
\begin{align}
\mathcal{M}(1_i,2_j,3_k,4_l;s)=\left( C_{(ij)(kl)}+C_{(kl)(ij)}+C_{(il)(kj)}+C_{(kj)(il)}\right)s^2+\mathcal{O}(s^3)\,,
\end{align}
where $C_{ijkl}$ is defined in Eq.~\eqref{cijkl}. The integral is then expressed in the IR language as
\begin{align}
\label{int_IR}
\oint \frac{\d s}{2\pi i}\frac{\mathcal{M}(1_i,2_j,3_k,4_l;s)}{s^3}= C_{(ij)(kl)}+C_{(kl)(ij)}+C_{(il)(kj)}+C_{(kj)(il)}\,.
\end{align}

\medskip
It is further convenient to introduce a crossing-symmetric combination of amplitudes,
\begin{align}
\mathcal{M}(s)=\sum_{i,j,k,l}u_iv_ju_kv_l\mathcal{M}(1_i,2_j,3_k,4_l;s)\,,
\end{align}
where $\vec{u}$ and $\vec{v}$ are arbitrary real unit vectors and
\begin{equation}
\oint \frac{\d s}{2\pi i}\frac{\mathcal{M}(s)}{s^3} = 4\sum_{i,j,k,l}u_iv_ju_kv_l C_{(ij)(kl)}\,.
\end{equation}
We may then deform the integration contour as (see Fig.~\ref{fig:contour})
\begin{align}
\label{int_UV}
\oint \frac{\d s}{2\pi i}\frac{\mathcal{M}(s)}{s^3}
=\left(\int_{-\infty}^{-s_0} + \int_{s_0}^\infty\right) \frac{\d s}{2\pi i}\frac{{\rm Disc}[\mathcal{M}(s)]}{s^3}\,,
\end{align}
where we assumed that the scattering amplitude is analytic away from the real axis of $s$ and it enjoys the Froissart bound to drop the boundary contributions. $s_0$ is the square of the lowest energy for which the non-analyticity shows up. The analyticity also implies that the discontinuity function is nothing but the imaginary part of the amplitude:
\begin{align}
{\rm Disc}[\mathcal{M}(s)]
&=\mathcal{M}(s+i\epsilon)-\mathcal{M}(s-i\epsilon) =2i\,{\rm Im}\,\mathcal{M}(s+i\epsilon)\,.
\end{align}
Here, we used the Schwarz reflection principle, which implies that $\mathcal{M}(s-i\epsilon)=\mathcal{M}^*(s+i\epsilon)$ for real $s$.
We therefore have
\begin{align}
\label{eq:UV_IR}
\oint \frac{\d s}{2\pi i}\frac{\mathcal{M}(s)}{s^3}
=\left(\int_{-\infty}^{-s_0} + \int_{s_0}^\infty\right) \frac{\d s}{\pi}\frac{{\rm Im}\,\mathcal{M}(s)}{s^3}
=\frac{2}{\pi} \int_{s_0}^\infty \d s\frac{{\rm Im}\,\mathcal{M}(s)}{s^3}\,,
\end{align}
where at the second equality we used the crossing-symmetric property of $\mathcal{M}(s)$.
Notice here that the l.h.s.\ is evaluated in the IR, whereas the r.h.s.\ is an integration over the UV region. This is how the UV information is encoded into the IR observables.

\medskip
To show that the r.h.s.\ is positive, we use the optical theorem, which states that
\begin{align}
2{\rm Im}\,\mathcal{M}(1_i,2_j,3_k,4_l;s)
=\sum_{n}\mathcal{M}_{ij\to n}(s)\mathcal{M}_{kl\to n}^*(s)\,,
\end{align}
where the r.h.s.\ is a sum over the partial waves with the intermediate on-shell state $n$. It is easy to see that the imaginary part of the crossing-symmetric amplitude is positive,
\begin{align}
2{\rm Im}\,\mathcal{M}(s)=\Big|\sum_{i,j}\mathcal{M}_{ij\to n}u_iv_j\Big|^2\geq0\,.
\end{align}
All in all, we arrive at the positivity bound
\begin{align}
4\sum_{i,j,k,l}u_iv_ju_kv_l C_{(ij)(kl)}
=\oint \frac{\d s}{2\pi i}\frac{\mathcal{M}(s)}{s^3}
=\frac{2}{\pi} \int_{s_0}^\infty \d s\frac{{\rm Im}\,\mathcal{M}(s)}{s^3}\geq0\,,
\end{align}
which is the same as the one derived in the previous subsection from causality.

\subsection{Bounds on Charge-to-Mass Ratios}
\label{ex:3d}

We now use the positivity condition~\eqref{bound_on_D_3d} on the EFT parameters to derive bounds on the charge-to-mass ratios of the scalars and fermions.
Using \eqref{cijkl}, we find
\begin{align}
\nonumber
0 &\le M_3^2\sum_{i,j,k,l}C_{(ij)(kl)}u_iu_kv_jv_l
\\
&=\sum_{a}\frac{1}{480\pi|m_a|}\bigg[
|\vec{u}\cdot \vec{z}_a|^2|\vec{v}\cdot \vec{z}_a|^2
- \frac{3}{8}|\vec{u}\cdot \vec{z}_a|^2
- \frac{3}{8}|\vec{v}\cdot \vec{z}_a|^2
+ \frac{1}{4}(\vec{u}\cdot\vec{v})(\vec{u}\cdot \vec{z}_a)(\vec{v}\cdot \vec{z}_a)
\bigg] \nl
+\gamma_f(\vec{u},\vec{v}) \label{inequality1}
\end{align}
for fermions and 
\begin{align}
\nonumber
0 &\le M_3^2\sum_{i,j,k,l}C_{(ij)(kl)}u_iu_kv_jv_l
\\
&=\sum_{a}\frac{7}{3840\pi|m_a|}\bigg[
|\vec{u}\cdot \vec{z}_a|^2|\vec{v}\cdot \vec{z}_a|^2
 -\frac{2}{7}|\vec{u}\cdot \vec{z}_a|^2
 -\frac{2}{7}|\vec{v}\cdot \vec{z}_a|^2
+\frac{8}{7}(\vec{u}\cdot\vec{v})(\vec{u}\cdot \vec{z}_a)(\vec{v}\cdot \vec{z}_a)
\bigg] \nl
+\gamma_s(\vec{u},\vec{v}) \label{inequality2}
\end{align}
for scalars.
Here, the functions $\gamma_{f/s}(\vec{u},\vec{v})$ are defined such that they contain all $\mathcal{O}(z^0)$ contributions to the inequalities, i.e., those which are independent of the $U(1)$ charges. They are given by
\begin{align}
\gamma_f(\vec{u},\vec{v})&=
\sum_{a}\frac{1}{480\pi|m_a|}
\left(
\frac{3}{4}
+\frac{1}{4}(\vec{u}\cdot\vec{v})^2
\right)
+M_3^2\sum_{i,j,k,l}c_{(ij)(kl)}u_iu_kv_jv_l\,,\\
\gamma_s(\vec{u},\vec{v})&=
\sum_{a}\frac{7}{3840\pi|m_a|}
\left(
\frac{4}{7}+\frac{8}{7}(\vec{u}\cdot\vec{v})^2
\right)
+M_3^2\sum_{i,j,k,l}c_{(ij)(kl)}u_iu_kv_jv_l\,.
\end{align}
As we mentioned in Sec.~\ref{setup3d}, the coefficients $c_{ijkl}$ in the above expressions depend on the details of the UV completion of the EFT, so that we leave them arbitrary numbers. The values of $\gamma_f(\vec{u},\vec{v})$ and $\gamma_s(\vec{u},\vec{v})$ are therefore in general unknown. Let us stress, however, that $\gamma_f(\vec{u},\vec{v})$ and $\gamma_s(\vec{u},\vec{v})$ depend on the cutoff scale $\Lambda$ of the EFT. For example, we can imagine increasing/decreasing $\Lambda$ such that some particles whose masses were originally above the cutoff scale are now below it or vice versa. In general, the inequalities \eqref{inequality1} and \eqref{inequality2} are trivially satisfied if $\gamma_f(\vec{u},\vec{v})$, $\gamma_s(\vec{u},\vec{v})$ are positive and large enough. However, whenever they are sufficiently small in a given EFT at some energy scale $\Lambda$, this leads to nontrivial bounds on the charge-to-mass ratios.

\medskip
Since $\vec{u}$ and $\vec{v}$ are arbitrary unit vectors, we may obtain the strongest bounds on the charge-to-mass ratios by scanning over all the choices of $\vec{u}$ and $\vec{v}$. However, it turns out to be illustrative and interesting enough for our purpose to focus on two extremal cases: $\vec{u}=\vec{v}$ and $\vec{u}\cdot\vec{v}=0$. Let us begin with the case $\vec{u}=\vec{v}$, under which the positivity conditions \eqref{inequality1} and \eqref{inequality2} are reduced to
\begin{align}
& \sum_{a}\frac{1}{480\pi|m_a|}
|\vec{u}\cdot \vec{z}_a|^2\Big(|\vec{u}\cdot \vec{z}_a|^2- \frac{1}{2}\Big)
+\gamma_f(\vec{u},\vec{u})\geq0 && \text{(fermions)}\,, \label{inequality3a} \\
& \sum_{a}\frac{7}{3840\pi|m_a|}
|\vec{u}\cdot \vec{z}_a|^2\left(|\vec{u}\cdot \vec{z}_a|^2 + \frac{4}{7}\right)
+\gamma_s(\vec{u},\vec{u})\geq0 && \text{(scalars)}\,. \label{inequality3b}
\end{align}
Note that the inequalities have a different $z$-dependence in the fermion and scalar case, respectively. In particular, the scalar contribution to \eqref{inequality3b} is always positive such that the condition is trivially satisfied unless $\gamma_s(\vec{u},\vec{u})$ is negative. In the fermionic case \eqref{inequality3a}, the condition is trivially satisfied if $\gamma_f(\vec{u},\vec{u})$ is positive and large enough but provides nontrivial bounds on charge-to-mass ratios when $\gamma_f(\vec{u},\vec{u})$ is in a certain range. As an illustrative case, let us consider $\gamma_f(\vec{u},\vec{u})=0$: this value requires the existence of a particle satisfying\footnote{Our numerical bound differs from the one obtained in \cite{Cheung:2014ega} for the single-$U(1)$ case due to a different convention for the charge-to-mass ratio, i.e., $z_\text{\cite{Cheung:2014ega}}=\frac{\sqrt{2} qg\sqrt{M_3}}{|m|}$ in units where $M_3=\frac{1}{2}$. Similarly, in 4D, $z_\text{\cite{Cheung:2014ega}}=\frac{\sqrt{2}qgM_4}{|m|}$ in units where $M_4=\frac{1}{\sqrt{2}}$.}
\begin{align}
|\vec{u}\cdot\vec{z}_a|^2\geq \frac{1}{2}\,. \label{3dchc}
\end{align}
Since this condition has to be satisfied for an arbitrary unit vector $\vec{u}$, it means that the charge-to-mass vectors $\vec z_a$ span a convex hull which contains a ball of radius $1/\sqrt{2}$.
Notice that the bound on the charge-to-mass ratio becomes stronger (weaker) if $\gamma_f(\vec{u},\vec{u})$ gives a negative (positive) contribution. This is a natural extension of the original Cheung-Remmen argument~\cite{Cheung:2014ega} to the multiple $U(1)$ case.

\medskip
Our result is the 3D analogue of the convex-hull condition, which was originally motivated in 4D using black-hole arguments \cite{Cheung:2014vva}. Since there are no black holes in 3D (unless one considers AdS boundary conditions), there is no extremality bound to compare with the numerical factor in our bound \eqref{3dchc}. 
Nevertheless, our bound states that the convex hull of the charge-to-mass vectors $\vec z_a$ must contain a ball of radius $\mathcal{O}(1)$. 
3D EFTs in which $\gamma_f(\vec{u},\vec{u})$ or $\gamma_s(\vec{u},\vec{u})$ is small enough at a given cutoff scale $\Lambda$ must therefore satisfy a convex-hull condition even though this is not required by any arguments involving black hole decay.\footnote{For nonzero $\gamma_f(\vec{u},\vec{u})$, $\gamma_s(\vec{u},\vec{u})$ with general $c_{ijkl}$, the numerical bound depends on $\vec u$ and is therefore not necessarily isotropic, i.e., the object contained in the convex hull of the charge-to-mass vectors need not necessarily be a ball.}

\medskip
Perhaps even more interestingly, a new type of bound may be obtained by choosing $\vec{u}\cdot\vec{v}=0$ in \eqref{inequality1} and \eqref{inequality2}. This yields
\begin{align}
& \sum_{a}\frac{1}{480\pi|m_a|}\bigg[
|\vec{u}\cdot \vec{z}_a|^2|\vec{v}\cdot \vec{z}_a|^2
- \frac{3}{8}|\vec{u}\cdot \vec{z}_a|^2
- \frac{3}{8}|\vec{v}\cdot \vec{z}_a|^2\bigg]
+\gamma_f(\vec{u},\vec{v})|_{\vec{v}\perp\vec{u}}  \ge 0 && \text{(fermions)} \,, \\
& \sum_{a}\frac{7}{3840\pi|m_a|}\bigg[
|\vec{u}\cdot \vec{z}_a|^2|\vec{v}\cdot \vec{z}_a|^2
 -\frac{2}{7}|\vec{u}\cdot \vec{z}_a|^2
 -\frac{2}{7}|\vec{v}\cdot \vec{z}_a|^2\bigg]
+\gamma_s(\vec{u},\vec{v})|_{\vec{v}\perp\vec{u}} \ge 0 && \text{(scalars)} \,.
\end{align}
If $\gamma_f(\vec{u},\vec{v})|_{\vec{v}\perp\vec{u}}$ and $\gamma_s(\vec{u},\vec{v})|_{\vec{v}\perp\vec{u}}$ are below a critical value, these inequalities can only be satisfied if the first terms in the brackets are nonzero. Hence, for any choice of $\vec u$, $\vec v$ with $\vec u \cdot \vec v = 0$, there must then exist at least one particle satisfying both $\vec u \cdot \vec z_a \neq 0$ and $\vec v \cdot \vec z_a \neq 0$. This implies in particular that it is not consistent to have a theory in which $\gamma_{f}(\vec{u},\vec{v})|_{\vec{v}\perp\vec{u}}$ or $\gamma_{s}(\vec{u},\vec{v})|_{\vec{v}\perp\vec{u}}$ is small and the charge vectors of all particles are orthogonal to one another. We may rephrase this as the statement that we require the existence of bifundamentals for \emph{any} (orthogonal) basis choice for the $U(1)$ gauge fields.

\begin{figure}[t]
\centering
\vspace{-4em}
\includegraphics[scale=0.45]{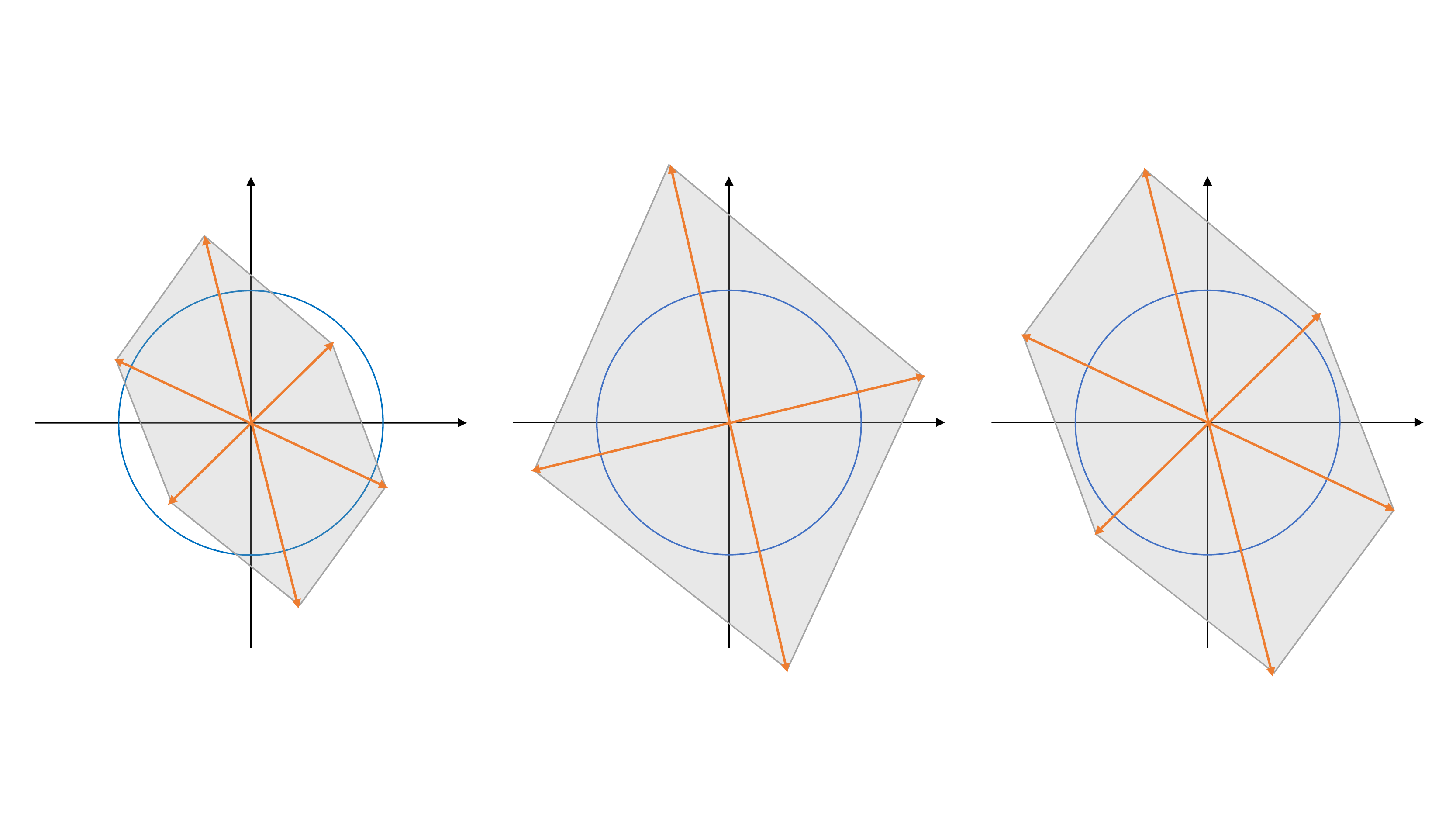}
\put(-305,108){$\scriptstyle z_1$}
\put(-163,108){$\scriptstyle z_1$}
\put(-20,108){$\scriptstyle z_1$}
\put(-355,182){$\scriptstyle z_2$}
\put(-213,182){$\scriptstyle z_2$}
\put(-70,182){$\scriptstyle z_2$}\\[-2em]
\caption{\label{fig:examples}\emph{Positivity constraints for 3D EFTs with two $U(1)$'s and particles with charge-to-mass vectors $\vec z_a$ (orange arrows). The first example does not satisfy the convex-hull condition (with the positivity bound indicated by the blue circle), and the second one does not have bifundamental particles for all basis choices of the $U(1)$ gauge fields. The third example is consistent with both positivity constraints.}}
\end{figure}
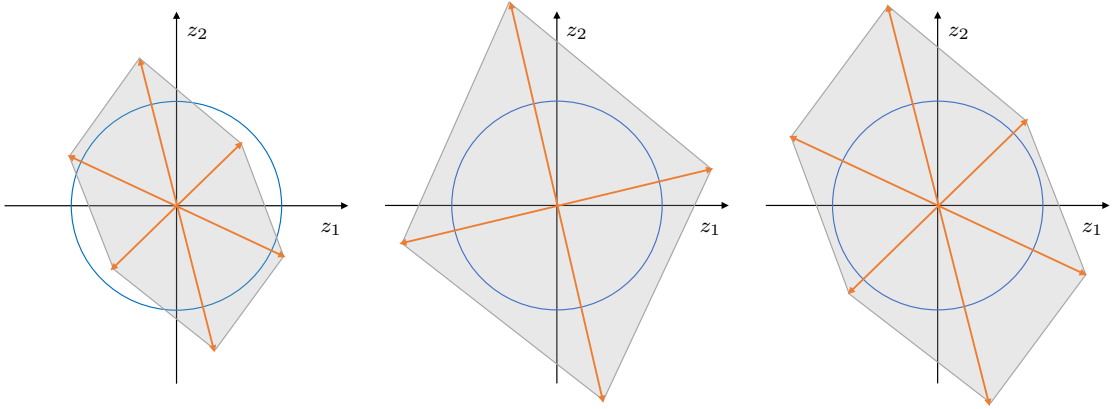

\medskip
As a simple example, consider a theory with two $U(1)$'s and take $u_i=\delta_{i1}$ and $v_i=\delta_{i2}$:
\begin{align}
& \sum_{a}\frac{1}{480\pi|m_a|}
\bigg[
z_{a1}^2z_{a2}^2
-\frac{3}{8}(z_{a1}^2+z_{a2}^2)
\bigg]
+\gamma_f(\delta_{i1},\delta_{i2})\geq0 && \text{(fermions)}\,,\\
&\sum_{a}\frac{7}{3840\pi|m_a|}
\bigg[
z_{a1}^2z_{a2}^2
-\frac{2}{7}(z_{a1}^2+z_{a2}^2)
\bigg]
+\gamma_s(\delta_{i1},\delta_{i2})\geq0 && \text{(scalars)}\,.
\end{align}
If $\gamma_f(\delta_{i1},\delta_{i2})$ and $\gamma_s(\delta_{i1},\delta_{i2})$ are sufficiently small, these inequalities can only be satisfied for nonzero $z_{a1}^2z_{a2}^2$, i.e., we require at least one bifundamental particle. This requirement together with the convex-hull condition is, for example, realized in a theory which has \mbox{(anti-)particles} with charge vectors $(\pm 1,\pm 1)$ and appropriately chosen masses.
However, the same argument can now also be repeated for any other $\vec u$, $\vec v$ satisfying $\vec u \cdot \vec v =0$, e.g., for the choice $u_i=\frac{1}{\sqrt{2}}(\delta_{i1}+\delta_{i2})\equiv\delta_{i+}$ and $v_i=\frac{1}{\sqrt{2}}(\delta_{i1}-\delta_{i2})\equiv\delta_{i-}$. This yields
\begin{align}
& \sum_{a}\frac{1}{480\pi|m_a|}
\bigg[
\frac{1}{4}(z_{a1}+z_{a2})^2(z_{a1}-z_{a2})^2
-\frac{3}{8}(z_{a1}^2+z_{a2}^2)
\bigg]
+\gamma_f(\delta_{i+},\delta_{i-})\geq0 && \text{(fermions)}\,,\\
&\sum_{a}\frac{7}{3840\pi|m_a|}
\bigg[
\frac{1}{4}(z_{a1}+z_{a2})^2(z_{a1}-z_{a2})^2
-\frac{2}{7}(z_{a1}^2+z_{a2}^2)
\bigg]
+\gamma_s(\delta_{i+},\delta_{i-})\geq0 && \text{(scalars)}\,.
\end{align}
In order for the first terms in the brackets to be nonzero, we also need at least one particle charged under both $A^\prime_{1\mu}=\frac{1}{\sqrt{2}}(A_{1\mu}+A_{2\mu})$ and $A^\prime_{2\mu}=\frac{1}{\sqrt{2}}(A_{1\mu}-A_{2\mu})$. A theory with only orthogonal charge vectors such as $(\pm 1,\pm 1)$ is therefore not consistent for sufficiently small $\gamma_f(\delta_{i+},\delta_{i-})$, $\gamma_s(\delta_{i+},\delta_{i-})$. The positivity constraints for EFTs with two $U(1)$'s are illustrated in Fig.~\ref{fig:examples}.

\medskip
To summarize, the positivity condition \eqref{bound_on_D_3d} for $\vec{u}=\vec{v}$ leads to bounds similar to the convex-hull type WGC bounds unless the UV-sensitive parameters $\gamma_f(\vec u,\vec u)$ and $\gamma_s(\vec u,\vec u)$ are large enough. Furthermore, the positivity condition for $\vec{u}\cdot\vec{v}=0$ requires the existence of bifundamental particles for all basis choices of the $U(1)$'s unless $\gamma_f(\vec u,\vec v)|_{\vec{v}\perp\vec{u}}$ and $\gamma_s(\vec u,\vec v)|_{\vec{v}\perp\vec{u}}$ are positive and large enough. The second type of condition turns out to be useful when we discuss the tower WGC later.

\section{Infrared Consistency in $D=4$}
\label{sec:4d}

Let us now move on to discuss causality and analyticity constraints in 4D EFTs. Since gravity is dynamical in 4D, such an analysis is somewhat more complicated than in the previously discussed 3D case. Nevertheless, we will be able to obtain positivity bounds similar to those obtained in the previous section, which we will again use to derive bounds on the charge-to-mass ratios of charged particles.

\subsection{Setup}
\label{setup4d}

The starting Wilsonian EFT (with cutoff $\Lambda$) is 
\begin{align}
\label{4deffaction}
& \Gamma = \int \d^4 x \sqrt{-g} 
\bigg[
\frac{M_4^2}{2} R 
- \frac{1}{4} \sum_i F_i^2 \bigg] 
+ \text{H.O.}
+ \left\{\begin{matrix*}[l]
  \Gamma_\text{scalar} \\
  \Gamma_\text{fermion} \end{matrix*}\right.
\,,
\end{align}
where $M_4$ is the 4D Planck mass and we consider scalars or fermions with actions
\begin{align}
\Gamma_\text{scalar} &= \int \d^4 x \sqrt{-g}\, \sum_a \left(-|D_{\mu} \phi_a|^2
- m_a^2 |\phi_a|^2\right), \\
\Gamma_\text{fermion} &= 
\int \d^4 x \sqrt{-g}\, \sum_a\bar\psi_a (-\slashed{D} - m_a) \psi _a\,.
\end{align}
We define the charge-to-mass ratio of a matter field as
\begin{align}
z_{ai}\equiv \frac{q_{ai}g_i M_4}{|m_a|}\,.
\end{align}

\medskip
Note that, unlike in 3D, there is no Chern-Simons term in \eqref{4deffaction}. However, as before, we allow higher-dimensional operators whose coefficients depend on the UV completion of the EFT. They may, for example, be generated by loops of heavy particles with masses above the cutoff scale and are therefore arbitrary from the low-energy point of view. In general, the operators are given by combinations of Riemann tensors and gauge field strengths (and derivatives thereof) but some of them can be eliminated by field redefinitions. The general form of the higher-dimensional operators is thus (see App.~\ref{simplify_action} for more details)
\begin{align}
\label{ho4d}
\text{H.O.}
= \sum_{i,j,k,l} \left[ c_{1ijkl}(F_i\cdot F_j)(F_k\cdot F_l) + c_{2ijkl}(F_i\cdot \tilde{F}_j)(F_k\cdot \tilde{F}_l) \right] +\sum_{i,j}c_{3ij}W^{\mu\nu\rho\sigma}F_{i\mu\nu}F_{j\rho\sigma}
\end{align}
up to terms with more than four derivatives. Here, $\tilde F_{i\mu\nu}=\frac{1}{2}\epsilon_{\mu\nu\rho\lambda}F_i^{\rho\lambda}$ is the dual gauge field strength, $W_{\mu\nu\rho\sigma} = R_{\mu\nu\rho\sigma} - \frac{1}{2}(g_{\mu[\rho}R_{\sigma]\nu}-g_{\nu[\rho}R_{\sigma]\mu}) + \frac{1}{6} R g_{\mu[\rho}g_{\sigma]\nu}$ is the Weyl tensor and $c_{1ijkl}$, $c_{2ijkl}$, $c_{3ij}$ are undetermined coefficients.

\medskip
As in 3D, we further integrate out charged matter in order to get the final EFT we are interested in. The final 1-loop 4-derivative effective action then reads (see App.~\ref{app:heatkernel} for the explicit computation)
\begin{align}
\Gamma_1 & =  \int \d^4 x \sqrt{-g} \left[ \frac{M_4^2}{2} R 
- \frac{1}{4} \sum_i F_i \cdot F_i +\sum_{i,j,k,l} C_{1ijkl}(F_i\cdot F_j)(F_k\cdot F_l) \right.
\nl
\qquad\qquad\quad\,
\left. +\sum_{i,j,k,l} C_{2ijkl}(F_i\cdot \tilde{F}_j)(F_k\cdot \tilde{F}_l)
+\sum_{i,j}C_{3ij}W^{\mu\nu\rho\sigma}F_{i\mu\nu}F_{j\rho\sigma} \right]
\,,
\end{align}
where we again used field redefinitions to eliminate some of the terms (see App.~\ref{simplify_action}).
The coefficients are given by
\begin{align}
C_{1ijkl} &= c_{1ijkl} + \frac{1}{2880\pi^2M_4^4} \sum_a \left( \frac{7}{8}z_{ai}z_{aj}z_{ak}z_{al} - z_{ai}z_{ak}\delta_{jl} + \frac{3}{4}\mathcal{I}_a \delta_{ik}\delta_{jl}\right)  \,, \label{4dcoeff1}\\
C_{2ijkl} &= c_{2ijkl} + \frac{1}{2880\pi^2M_4^4} \sum_a \left( \frac{1}{8}z_{ai}z_{aj}z_{ak}z_{al} - z_{aj}z_{ak}\delta_{il} + \frac{3}{4}\mathcal{I}_a \delta_{il}\delta_{jk}\right)  \,, \label{4dcoeff2}\\
C_{3ij} &= c_{3ij} - \sum_a \frac{z_{ai}z_{aj}}{2880\pi^2M_4^2} \label{4dcoeff3}
\end{align}
for the case of scalar matter and by
\begin{align}
C_{1ijkl} &= c_{1ijkl} + \frac{1}{2880\pi^2M_4^4} \sum_a \left( 2 z_{ai}z_{aj}z_{ak}z_{al} -\frac{11}{2} z_{ai}z_{ak}\delta_{jl} + \frac{9}{4}\mathcal{I}_a \delta_{ik}\delta_{jl}\right) \,, \label{4dcoeff4} \\
C_{2ijkl} &= c_{2ijkl} + \frac{1}{2880\pi^2M_4^4} \sum_a \left( \frac{7}{2}z_{ai}z_{aj}z_{ak}z_{al} -\frac{11}{2} z_{aj}z_{ak}\delta_{il} + \frac{9}{4}\mathcal{I}_a \delta_{il}\delta_{jk}\right)  \,, \label{4dcoeff5} \\
C_{3ij} &= c_{3ij} + \sum_a \frac{z_{ai}z_{aj}}{1440\pi^2M_4^2} \label{4dcoeff6}
\end{align}
for fermions, where $\mathcal{I}_a = 2\ln\frac{\Lambda}{|m_a|}-\gamma$ and $\gamma$ is the Euler-Mascheroni constant.
Similarly to our 3D analysis, we observe the structure $C_{1ijkl}, C_{2ijkl} \sim \mathcal{O}(z^4)+\mathcal{O}(z^2)+\mathcal{O}(z^0)$, $C_{3ij} \sim \mathcal{O}(z^2)+\mathcal{O}(z^0)$. Here, the scalar/fermion loops contribute to all three types of terms, while the $\mathcal{O}(z^0)$ terms receive a further contribution from the UV-sensitive operators \eqref{ho4d}. Furthermore, there are contributions from graviton and photon loops which are also of the order $\mathcal{O}(z^0)$. We refrain from computing these contributions explicitly and absorb them into the unknown coefficients $c_{1ijkl}$, $c_{2ijkl}$ and $c_{3ij}$ without loss of generality.

\medskip
Let us again point out the domain of validity of our results. Considering matter loops with photon and graviton legs, we find that our 1-loop effective action is valid in the perturbative regime
\begin{equation}
\alpha \equiv |qg| \ll 1
\,,
\qquad
\beta \equiv \frac{|m|}{M_4} \ll 1
\,,
\end{equation}
where we dropped the $(a,i)$ indices for simplicity.
Restricting to terms with at most four derivatives is valid in the weak-field regime
\begin{equation}
\frac{|qg F|}{m^2} \ll 1
\,,
\qquad
\frac{|R|}{m^2} \ll 1
\,.
\end{equation}
Since, in 4D, the charge-to-mass ratio $z$ satisfies $\displaystyle |z| = \frac{|qg| M_4}{|m|} = \frac{\alpha}{\beta}$, our EFT expansion is valid for a wide range of values
\begin{equation}
\alpha \ll |z| \ll \frac{1}{\beta}
\,
\end{equation}
including the regime $z \sim \mathcal{O}(1)$ around the WGC bound.

\subsection{Causality Constraints}

Let us now discuss subluminality constraints on the IR-effective Lagrangian
\begin{align}
\nonumber
\mathcal{L}&= \frac{M_4^2}{2} R
- \frac{1}{4} \sum_i F_i \cdot F_i +\sum_{i,j,k,l}\left[
C_{1ijkl}(F_i\cdot F_j)(F_k\cdot F_l)
+C_{2ijkl}(F_i\cdot \tilde{F}_j)(F_k\cdot \tilde{F}_l)
\right]
\\
\label{4DEFT}
&\quad
+\sum_{i,j}C_{3ij}W^{\mu\nu\rho\sigma}F_{i\mu\nu}F_{j\rho\sigma}\,.
\end{align}
Just as in the 3D case, we turn on background gauge fields and a background metric and then require subluminality of fluctuations on this background to constrain the EFT parameters.
However, in contrast to the 3D case, the graviton propagates in 4D and kinematically mixes with the photons in the presence of nontrivial electromagnetic backgrounds. To avoid technical complication due to such kinetic mixings, we follow Cheung and Remmen~\cite{Cheung:2014ega} and consider propagation in a thermal photon gas, where the electromagnetic fields have vanishing thermal average, $\overline{F_{i\mu\nu}}=0$, but nonzero, constant variance $\overline{F_{i\mu\nu}F_{j\rho\sigma}}\neq0$.

\medskip
In such a thermal photon gas, the part of the Lagrangian quadratic in the gauge field fluctuations takes the form
\begin{align}
\nonumber
\mathcal{L}&=-\frac{1}{4}\sum_{i}F_{i\alpha\beta}F_i^{\alpha\beta} +\sum_{i,j,k,l}\left(C_{1(ij)(kl)}\overline{F_{j\alpha\beta}F_{l\gamma\delta}}
+C_{2(ij)(kl)}\overline{\tilde{F}_{j\alpha\beta}\tilde{F}_{l\gamma\delta}}\right)F_i^{\alpha\beta}F_k^{\gamma\delta}
\\
&\quad
+\sum_{i,j}C_{3ij}\overline{W}_{\alpha\beta\gamma\delta}F_i^{\alpha\beta}F_j^{\gamma\delta}\,.
\end{align}
Here, we focus on the geometric-optics limit, where the photon wavelength is much shorter than the spacetime curvature scale. The indices $\alpha,\beta,\ldots$ are again for locally flat coordinates. We also performed a field redefinition to simplify the action. To make the argument more concrete, let us make the following ansatz for the photon background:\footnote{The energy density $\rho$ and pressure $p$ of a photon gas with temperature $T$ are given by $\rho=3p=\pi^2T^4/15$. As explained, e.g., in~\cite{Cheung:2014ega}, we therefore have $\overline{F_{\alpha\beta}F_{\gamma\delta}}=\overline{\tilde{F}_{\alpha\beta}\tilde{F}_{\gamma\delta}}
=\frac{\pi^2}{45}T^4(\delta_{\alpha\gamma}\delta_{\beta\delta}-\delta_{\alpha\delta}\delta_{\beta\gamma})
$ in the rest frame of the photon gas.}
\begin{align}
\overline{F_{i\alpha\beta}F_{j\gamma\delta}}=\overline{\tilde{F}_{i\alpha\beta}\tilde{F}_{j\gamma\delta}}=\frac{\pi^2}{45}T_{ij}^4(\delta_{\alpha\gamma}\delta_{\beta\delta}-\delta_{\alpha\delta}\delta_{\beta\gamma})\,, \label{photongas}
\end{align}
where the Kronecker delta $\delta_{\alpha\beta}$ breaks Lorentz invariance.
The symmetric matrix $T_{ij}^4$ specifies the properties of the photon gas. For example, when photons of all $N$ gauge fields are in thermal equilibrium such that they have the same temperature $T$, the matrix takes the form $T_{ij}^4=\delta_{ij}T^4$. Later, we will consider more general situations.

\medskip
Under the assumption \eqref{photongas},
the kinetic matrix for $2N$ helicity modes of $N$ photons simplifies in momentum space as
\begin{align}
\label{K4D}
&K_{ij}= -\delta_{ij}k^2+D_{ij}\delta_{\alpha\beta}k^\alpha k^\beta
\end{align}
with $D_{ij}$ defined by
\begin{align}
D_{ij}=\frac{4\pi^2}{45}\sum_{k,l}\left( C_{1(ik)(jl)}+C_{1(jl)(ik)}+C_{2(ik)(jl)}+C_{2(jl)(ik)} \right)T^4_{kl}\,,
\end{align}
where we dropped helicity indices because the dispersion is helicity-independent in our setup. We also used $\overline{W}_{\alpha\beta\gamma\delta}=0$ because the FRW spacetime sourced by the background photons is conformally flat. The kinetic matrix may be diagonalized to
\begin{align}
\widetilde{K}_{ij}={\rm diag}\Big(\big(k_0^2-|\vec{k}|^2\big)+D_1\big(k_0^2+|\vec{k}|^2\big),\ldots ,\big(k_0^2-|\vec{k}|^2\big)+D_N\big(k_0^2+|\vec{k}|^2\big)\Big)\,,
\end{align}
where the $D_i$'s are the eigenvalues of the matrix $D_{ij}$. Subluminality requires that all eigenvalues $D_i$ of $D_{ij}$ should be non-negative, which can be rephrased as
\begin{align}
\label{4D_causality1}
\sum_{i,j}D_{ij}u_iu_j\geq0
\end{align}
for an arbitrary real vector $u_i$.

\medskip
Finally, let us take a closer look at the constraint~\eqref{4D_causality1} for several photon gas setups. First, when all photons have the same temperature, $T_{ij}^4=\delta_{ij}T^4$, we obtain the condition
\begin{align}
\sum_{i,j,k}\left( C_{1(ik)(jk)}+C_{2(ik)(jk)} \right)u_iu_j\geq0
\quad
\forall u_i\,.
\end{align}
A stronger condition may be obtained by considering the case where photons of different gauge fields have different temperatures $T_i$, i.e., for $T_{ij}^4={\rm diag}(T_1^4,\ldots,T_N^4)$. By requiring subluminality for an arbitrary choice of the photon temperatures $T_i$, we arrive at the condition
\begin{align}
\sum_{i,j}\left( C_{1(ik)(jk)}+C_{2(ik)(jk)} \right)u_iu_j\geq0
\quad
\forall k, \,u_i\,.
\end{align}
In order to further generalize this bound, we observe that $T_{ij}^4$ is not $SO(N)$ invariant anymore when each photon has a different temperature. By rotating the photon basis or, equivalently, by considering the case when some linear combination of $N$ photons has a definite temperature, we obtain the condition
\begin{align}
\label{4Dcausality}
\sum_{i,j,k,l}\left(C_{1(ij)(kl)}+C_{2(ij)(kl)}\right)u_iu_kv_jv_l\geq0
\end{align}
for arbitrary real vectors $\vec{u}$ and $\vec{v}$, which we assume to be unit vectors without loss of generality. In this paper, we use the condition~\eqref{4Dcausality} to constrain the IR-effective Lagrangian. As we will see, the same condition arises from an analyticity argument under some assumptions.

\subsection{Analyticity Constraints}

We now discuss constraints from the analyticity of scattering amplitudes. In the setup~\eqref{4DEFT}, 4-point amplitudes with the forward-type helicity structure are given by
\begin{align}
\mathcal{M}(1_i^+,2_j^+,3_k^-,4_l^-)&=\mathcal{M}(1_i^-,2_j^-,3_k^+,4_l^+) \nll =\left( C_{1(ij)(kl)}+C_{1(kl)(ij)}+C_{2(ij)(kl)}+C_{2(kl)(ij)} \right)s^2
\nl +\text{(graviton exchange)}\,,
\\
\mathcal{M}(1_i^+,2_j^-,3_k^-,4_l^+)&=\mathcal{M}(1_i^-,2_j^+,3_k^+,4_l^-)\nll=\left( C_{1(ij)(kl)}+C_{1(kl)(ij)}+C_{2(ij)(kl)}+C_{2(kl)(ij)} \right)u^2\nl +\text{(graviton exchange)}\,,
\end{align}
where $C_{1ijkl}$ and $C_{2ijkl}$ are defined in Eqs.~\eqref{4dcoeff1}, \eqref{4dcoeff2}, \eqref{4dcoeff4} and \eqref{4dcoeff5}. The second term on the r.h.s.\ of each equation is from the single graviton exchange in the tree-level Einstein-Maxwell theory. As in the single photon case~\cite{Cheung:2014ega}, this contribution is singular $\sim s^2/t$ in the forward limit $t\to0$ and dominates over the 1-loop corrections from charged matter. Because of this singularity, it is not clear whether it is possible to derive a rigorous bound on higher-dimensional operators using analyticity arguments. However, in order to compare our multiple-photon setup with the single-photon case, let us follow~\cite{Cheung:2014ega} and compute the positivity bound on higher-dimensional operators by simply dropping the singular contribution due to graviton exchange.

\medskip
To apply the analyticity argument, it is convenient to introduce a linear combination,
\begin{align}
\nonumber
\mathcal{M}_{ijkl}&=
\mathcal{M}(1_i^+,2_j^+,3_k^-,4_l^-)
+\mathcal{M}(1_i^-,2_j^-,3_k^+,4_l^+)
+\mathcal{M}(1_i^+,2_j^-,3_k^-,4_l^+)
+\mathcal{M}(1_i^-,2_j^+,3_k^+,4_l^-)
\\
&=2\left( C_{1(ij)(kl)}+C_{1(kl)(ij)}+C_{2(ij)(kl)}+C_{2(kl)(ij)} \right)(s^2+u^2)\nl+\text{(graviton exchange)}
\,,
\end{align}
which is $s$-$u$ symmetric with respect to helicities. We further symmetrize the photon index as
\begin{align}
\mathcal{M}(\vec u,\vec v)=\sum_{i,j,k,l}u_iu_kv_jv_l\mathcal{M}_{ijkl}\,,
\end{align}
where $u_i$ and $v_i$ are real unit vectors. Just like in the 3D case, such a symmetric combination gives a positivity bound after using the optical theorem. Following the argument of Sec.~\ref{analyticity3d} and neglecting the contribution from graviton exchange, we arrive at the bound
\begin{align}
\label{bound_on_D}
\sum_{i,j,k,l}\left(C_{1(ij)(kl)}+C_{2(ij)(kl)}\right)u_iu_kv_jv_l\geq0
\quad
\forall
u_i
\,,
v_i
\,.
\end{align}
Although the argument here is not rigorous because of the singularity due to graviton exchange, we thus obtained the same bound as from the subluminality constraints.

\subsection{Bounds on Charge-to-Mass Ratios}

We now use the positivity conditions~\eqref{bound_on_D} on the EFT parameters to derive bounds on charge-to-mass ratios.
Substituting either \eqref{4dcoeff1}, \eqref{4dcoeff2} or \eqref{4dcoeff4}, \eqref{4dcoeff5}, we may compute the l.h.s.\ for our 4D setup as
\begin{align}
\nonumber
&M_4^4 \sum_{i,j,k,l}\left(C_{1(ij)(kl)}+C_{2(ij)(kl)}\right)u_iu_kv_jv_l
\\
\nonumber
&=\alpha_{f/s} \sum_{a}\bigg[
|\vec{u}\cdot \vec{z}_a|^2|\vec{v}\cdot \vec{z}_a|^2
- \frac{1}{2}\Big(|\vec{u}\cdot \vec{z}_a|^2+|\vec{v}\cdot \vec{z}_a|^2
+2(\vec{u}\cdot\vec{v})(\vec{u}\cdot \vec{z}_a)(\vec{v}\cdot \vec{z}_a)
\Big)
\bigg]
\\
&\quad
+\gamma_{f/s}(\vec{u},\vec{v})\,,
\end{align}
where
$\alpha_f=11/1440\pi^2$ for fermions and $\alpha_s=1/720\pi^2$ for scalars.
The functions $\gamma_{f/s}(\vec{u},\vec{v})$ are again defined such that they contain all $\mathcal{O}(z^0)$ contributions to the inequalities:
\begin{align}
\gamma_f(\vec{u},\vec{v}) &= \frac{9}{2880\pi^2} \sum_a \mathcal{I}_a \left( 1 + (\vec u \cdot \vec v)^2 \right) + M_4^4 \sum_{i,j,k,l}\left(c_{1(ij)(kl)}+c_{2(ij)(kl)}\right)u_iu_kv_jv_l\,,\\
\gamma_s(\vec{u},\vec{v}) &= \frac{3}{2880\pi^2} \sum_a \mathcal{I}_a \left( 1 + (\vec u \cdot \vec v)^2 \right)  + M_4^4 \sum_{i,j,k,l}\left(c_{1(ij)(kl)}+c_{2(ij)(kl)}\right)u_iu_kv_jv_l
\end{align}
with $\mathcal{I}_a = 2\ln\frac{\Lambda}{|m_a|}-\gamma$. 
For any choice of $\vec u,\vec v$, the positivity constraint~\eqref{bound_on_D} then implies nontrivial bounds on the charge-to-mass ratios when $\gamma_{f/s}(\vec{u},\vec{v})$ is in a certain range. Just as we did in 3D, let us now focus on the two illustrative cases $\vec{u}=\vec{v}$ and $\vec{u}\cdot\vec{v}=0$.

\medskip
First, we consider the case $\vec{u}=\vec{v}$ in which the positivity condition is reduced to
\begin{align}
\sum_{a}\alpha_{f/s}
|\vec{u}\cdot \vec{z}_a|^2\Big(|\vec{u}\cdot \vec{z}_a|^2-2\Big)+\gamma_{f/s}(\vec{u},\vec{u})\geq0\,.
\end{align}
This condition is trivially satisfied if $\gamma_{f/s}(\vec{u},\vec{u})$ is positive and large enough but provides nontrivial bounds whenever it is sufficiently small. Let us again take $\gamma_{f/s}(\vec{u},\vec{u})=0$ for illustration. The inequality then simplifies to
\begin{align}
\sum_{a}\alpha_{f/s}
|\vec{u}\cdot \vec{z}_a|^2\Big(|\vec{u}\cdot \vec{z}_a|^2-2\Big)\geq0\,,
\end{align}
which implies the existence of a super-extremal particle satisfying
\begin{align}
|\vec{u}\cdot\vec{z}_a|\geq \sqrt{2} \,.
\end{align}
Since we may take an arbitrary unit vector $\vec{u}$, we arrive at the convex-hull condition, which requires a super-extremal particle in any direction of the charge space.
Note that, for $\gamma_{f/s}(\vec{u},\vec{u})=0$, our bound on the charge-to-mass ratios is in fact numerically stronger than a super-extremality bound (which would only require $|\vec{u}\cdot\vec{z}_a|\geq \frac{1}{\sqrt{2}}$). The bound becomes stronger (weaker) if $\gamma_{f/s}(\vec{u},\vec{u})$ is negative (positive).

\medskip
Another illustrative example is the case $\vec{u}\cdot\vec{v}=0$. 
For concreteness, let us take $u_i=\delta_{i1}$ and $v_i=\delta_{i2}$, which yields
\begin{align}
\sum_{a}\alpha_{f/s}\bigg[
z_{a1}^2z_{a2}^2
-\frac{1}{2}z_{a1}^2-\frac{1}{2} z_{a2}^2
\bigg]+ \gamma_{f/s}(\delta_{i1},\delta_{i2}) \geq0\,.
\end{align}
For sufficiently small $\gamma_{f/s}(\delta_{i1},\delta_{i2})$, this inequality can only be satisfied if $z_{a1}^2z_{a2}^2$ is nonzero, i.e., it implies the existence of at least one bifundamental particle.
This is true unless $\gamma_{f/s}(\delta_{i1},\delta_{i2})$ satisfies
\begin{align}
\gamma_{f/s}(\delta_{i1},\delta_{i2})&\geq \frac{1}{2} \sum_{a}\alpha_{f/s}
(z_{a1}^2+z_{a2}^2)
\,.
\end{align}
We can repeat the above argument for any other choice of $\vec u, \vec v$ satisfying $\vec u \cdot \vec v=0$. Following the argument in Sec.~\ref{ex:3d}, we therefore conclude that, for sufficiently small $\gamma_{f/s}(\vec u,\vec v)$, bifundamental particles are required to exist for any orthogonal basis choice of the $U(1)$ gauge fields.

\section{Compactification and the Tower WGC}
\label{sec:comp}

In this section, we analyze causality and analyticity constraints of 4D EFTs compactified on a circle. It was shown in \cite{Heidenreich:2015nta} that a theory which satisfies the convex-hull condition proposed in \cite{Cheung:2014vva} does not necessarily satisfy it after compactification since then also charges under the KK $U(1)$ have to be considered. This was interpreted in \cite{Heidenreich:2015nta} as evidence for a stronger form of the WGC, i.e., the lattice WGC, which is robust under compactification. Here, we want to check whether we can conclude anything analogous from the study of infrared consistency conditions.
We will find that, in comparison to the results of the previous section, the causality and analyticity constraints indeed become stronger in the compactified theories. This suggests a particular version of the WGC in which the charge-to-mass ratios of an infinite tower of particles are bounded from below.

\subsection{Setup}
\label{sec:comp-setup}

Our starting point is a 4D EFT with metric $G_{MN}$ and either a scalar or a fermion charged under a $U(1)$ gauge field $A_M$. Here and in the following, we denote the 4D coordinates by $x^M=(x^\mu, x^3)$ with $\mu=0,1,2$ and the coordinate along the circle by $x^3$. We consider the effective action
\begin{equation}
\label{ea}
\Gamma = \int \d^4 x \sqrt{-G} \left( \frac{M_4^2}{2} R - \frac{1}{4}F^2 \right) + \text{H.O.} +  \left\{\begin{matrix*}[l]
  \Gamma_\text{scalar} \\
  \Gamma_\text{fermion} \end{matrix*}\right.\,,
\end{equation}
where ''H.O.`` denotes possible higher-derivative terms and
\begin{align}
\Gamma_\text{scalar} &= \int \d^4 x \sqrt{-G}\, \left(-\left|\partial_M \Phi + iqg_4 A_M \Phi\right|^2 - m^2 \left|\Phi\right|^2 \right)\,, \\
\Gamma_\text{fermion} &= 
\int \d^4 x \sqrt{-G}\, \bar\Psi (- \,\slash\!\!\!\! \nabla - i q g_4 \,\slash\!\!\!\! A - m) \Psi\,.
\label{eaf}
\end{align}

\medskip
We now compactify this theory on a circle with radius $r$.
To this end, we decompose the 4D metric $G_{MN}$ as
\begin{equation}
G_{\mu\nu} = \e^\lambda g_{\mu\nu} + r^2\e^{-\lambda}B_\mu B_\nu\,, \quad G_{\mu 3} = -r\e^{-\lambda}B_\mu\,, \quad G_{33} = \e^{-\lambda}\,.
\end{equation}
Here, $g_{\mu\nu}$ is the 3D metric, $\lambda$ is the radion, and $B_\mu$ is the graviphoton with field strength $H_{\mu\nu} = \partial_\mu B_\nu - \partial_\nu B_\mu$. The gauge field $A_M$ is decomposed into a 3D vector $A_\mu$ and an axion $A_3$. We then make the usual mode expansion
\begin{align}
g_{\mu\nu}(x^\mu,x^3) &= g^{(0)}_{\mu\nu}(x^\mu) + \sum_{n\neq 0} \frac{g^{(n)}_{\mu\nu}(x^\mu)}{\sqrt{\pi r}M_4} \e^{inx^3/r}\,,  \\ B_{\mu}(x^\mu,x^3) &= \sum_n \frac{B^{(n)}_{\mu}(x^\mu)}{\sqrt{\pi r }rM_4} \e^{inx^3/r}\,,  \\ \lambda(x^\mu,x^3) &= \sum_n \frac{\lambda^{(n)}(x^\mu)}{\sqrt{\pi r}M_4} \e^{inx^3/r}\,,  \\ A_\mu(x^\mu,x^3) &= \sum_n \frac{A_\mu^{(n)}(x^\mu)}{\sqrt{2\pi r}} \e^{inx^3/r}\,,  \\ A_3(x^\mu,x^3) &= \sum_n \frac{A_3^{(n)}(x^\mu)}{\sqrt{2\pi r}} \e^{inx^3/r}\,,
\end{align}
where the reality of the 4D fields imposes the conditions $g^{(n)*}_{\mu\nu}=g^{(-n)}_{\mu\nu}$, $B^{(n)*}_{\mu}=B^{(-n)}_{\mu}$, etc. and we have chosen the prefactors in the expansions such that the fields are canonically normalized for $\lambda=0$.
Analogously, we can expand the 4D scalar field,
\begin{equation}
\Phi(x^\mu,x^3) = \sum_n \frac{\phi^{(n)}(x^\mu)}{\sqrt{2\pi r}} \e^{inx^3/r}\,.
\end{equation}
The 4D spinor $\Psi$ is decomposed into two 3D spinors $\psi$ and $\chi$ with mode expansions
\begin{equation}
\label{spinor_4D_to_3D}
\psi(x^\mu,x^3) = \sum_n \frac{\psi^{(n)}(x^\mu)}{\sqrt{2\pi r}}\e^{inx^3/r}\,, \qquad \chi(x^\mu,x^3) = \sum_n \frac{\chi^{(n)}(x^\mu)}{\sqrt{2\pi r}}\e^{inx^3/r}\,.
\end{equation}

\medskip
Since $G_{MN}$ has two propagating degrees of freedom, we expect that the combined degrees of freedom of $g^{(n)}_{\mu\nu}$, $B^{(n)}_{\mu}$ and $\lambda^{(n)}$ should also equal two for each KK level $n$. For $n=0$, we have a massless spin-2 field, a massless vector and a massless scalar in 3D, which indeed adds up to two degrees of freedom. For each $n\neq 0$, $g^{(n)}_{\mu\nu}$ eats up the vector and the scalar via a St\"{u}ckelberg mechanism\footnote{We thank Gianluca Zoccarato for a useful discussion on this point.} (see, e.g., \cite{Hinterbichler:2011tt} for a review). A massive spin-2 field in 3D has two degrees of freedom and, hence, we again arrive at the expected number. Similarly, one can check that $A^{(n)}_3$ is eaten by $A^{(n)}_{\mu}$ for all $n\neq 0$. The degrees of freedom of $A^{(n)}_{\mu}$ and $A^{(n)}_3$ thus add up to two for each KK level, in agreement with the two degrees of freedom of $A_M$ in 4D.

\medskip
For simplicity, we will assume a suitable stabilization mechanism such that the zero modes of the radion $\lambda^{(0)}$ and the axion $A_3^{(0)}$ are stabilized at $\lambda^{(0)} = A_3^{(0)} = 0$. Their precise masses are irrelevant for our analysis since they are uncharged and the constraints we want to derive are only sensitive to loops of \emph{charged} particles.
The KK gravitons $g^{(n)}_{\mu\nu}$ and KK photons $A^{(n)}_{\mu}$, on the other hand, are charged under the KK $U(1)$ such that they generally contribute to the causality/analyticity constraints. We will see below, however, that there is a regime in which we can draw conclusions without having to know their precise contributions.
The different types of fields in the spectrum of the compactified theory are summarized in Table \ref{Tab:fields}.

\begin{table}[t]
\centering
\setlength{\tabcolsep}{12pt}
\renewcommand{\arraystretch}{1.2}
\begin{tabular}{cccc}
\toprule
field type & scalar case & fermion case & charge \\
\midrule
massless real & $B_\mu^{(0)}$, $A_\mu^{(0)}$ &  $B_\mu^{(0)}$, $A_\mu^{(0)}$ & $(0,0)$ \\
massive real & $\lambda^{(0)}$, $A_3^{(0)}$ &  $\lambda^{(0)}$, $A_3^{(0)}$ & $(0,0)$ \\
massive complex & $g_{\mu\nu}^{(n\neq 0)}$, $A_\mu^{(n\neq 0)}$ &  $g_{\mu\nu}^{(n\neq 0)}$, $A_\mu^{(n\neq 0)}$ & $(n,0)$ \\
 & $\phi^{(n)}$ &  $\psi^{(n)}$, $\chi^{(n)}$ & $(n,q)$ \\
\bottomrule
\end{tabular}
\vspace{5pt}
\caption{\label{Tab:fields} \emph{Spectrum of 3D fields and their charges under $B_\mu^{(0)}$, $A_\mu^{(0)}$.}}
\end{table}

\medskip
Our strategy in the next subsection will be to integrate out all massive fields in order to obtain a low-energy effective action which only depends on the massless gauge fields $A^{(0)}_\mu$ and $B^{(0)}_\mu$. Imposing causality/analyticity constraints as in the previous sections will then lead to inequalities for the charge-to-mass ratios of the massive fields with respect to the KK $U(1)$ and the original $U(1)$. As in the previous sections, we will perform the path integration in the one-loop approximation. It is therefore sufficient to restrict to terms in the action which are at most quadratic in any of the massive fields.

\medskip
Let us now rewrite the action \eqref{ea} in terms of the 3D fields, keeping in mind the above remarks. We define the 3D couplings
\begin{equation}
\label{couplings_3d}
M_3 = 2\pi r M_4^2 \,, \qquad g_3 = \frac{g_4}{\sqrt{2\pi r}}\,,\qquad g_\text{KK} = \frac{\sqrt{2}}{\sqrt{M_3}r}\,.
\end{equation}
The Einstein-Maxwell part of the action then reads\footnote{Here, we omit couplings to $\lambda^{(0)}$ and $A_3^{(0)}$ as well as couplings between the KK modes and derivatives of zero modes (such as $R^{(0)} g^{(n)*}g^{(n)}$) because they are not relevant for our analysis below.}
\begin{align}
\Gamma &\supset \int \d^3x \sqrt{-g^{(0)}} \left[ \frac{M_3}{2} R^{(0)} -\frac{1}{4} H^{(0)2} -\frac{1}{4} F^{(0)2} + \sum_{n\ge 1} \left(-\frac{1}{2}D_\lambda g_{\mu\nu}^{(n)*} D^\lambda g^{(n)\mu\nu} \right.\right. \nl \left.\left. + D_\lambda g_{\mu\nu}^{(n)*} D^\nu g^{(n)\mu\lambda} -\frac{1}{2} D_\mu g^{(n)*} D_\nu g^{(n)\mu\nu}- \frac{1}{2}D_\nu g^{(n)*\mu\nu}D_\mu g^{(n)}+\frac{1}{2}D_\mu g^{(n)*} D^\mu g^{(n)} \right.\right. \nl \left.\left. - \frac{n^2}{2r^2} (g_{\mu\nu}^{(n)*}g^{(n)\mu\nu}-g^{(n)*}g^{(n)}) - \frac{1}{2} |F^{(n)}|^2 - \frac{n^2}{r^2} A_{\mu}^{(n)*}A^{(n)\mu} \right) \right]\,, \label{em-kk}
\end{align}
where $F^{(n)}_{\mu\nu} = D_\mu A^{(n)}_\nu - D_\nu A^{(n)}_\mu$ and $D_\mu = \nabla_\mu + in g_\text{KK} B_\mu^{(0)}$.
We furthermore obtain the scalar action
\begin{equation}
\Gamma_\text{scalar} = \sum_n \int \d^3 x \sqrt{-g^{(0)}}\, \left(-\left|\partial_\mu \phi^{(n)} +  iqg_3 A^{(0)}_\mu \phi^{(n)} + in g_\text{KK}B^{(0)}_\mu \phi^{(n)}\right|^2 - m_n^2 \left|\phi^{(n)}\right|^2 \right)
\end{equation}
with masses
\begin{equation}
m_n = \sqrt{m^2+\frac{n^2}{r^2}} \label{m_n}
\end{equation}
and the fermion action
\begin{align}
\Gamma_\text{fermion} &= \sum_n \int \d^3 x \sqrt{-g^{(0)}} \bigg[ - \bar\psi^{(n)} (\,\slash\!\!\!\! \nabla+iqg_3\, \slash\!\!\!\! A^{(0)}+ing_\text{KK}\,\slash\!\!\!\! B^{(0)}) \psi^{(n)} \nl - \bar\chi^{(n)} (\,\slash\!\!\!\! \nabla+iqg_3 \,\slash\!\!\!\! A^{(0)}+ing_\text{KK}\,\slash\!\!\!\! B^{(0)}) \chi^{(n)} - m  (\bar\psi^{(n)} \chi^{(n)} + \bar\chi^{(n)} \psi^{(n)}) \nl - \bar\psi^{(n)} \left(\frac{n}{r} + \frac{i}{4\sqrt{2M_3}} \, \slash\!\!\!\! H^{(0)}\right) \psi^{(n)} - \bar\chi^{(n)} \left(-\frac{n}{r} - \frac{i}{4\sqrt{2M_3}} \, \slash\!\!\!\! H^{(0)}\right) \chi^{(n)} \bigg]\,, \label{3d-dirac}
\end{align}
where $\,\slash\!\!\!\! H^{(0)} = \gamma^\mu\gamma^\nu H^{(0)}_{\mu\nu}$. The mass eigenvalues of the fermions are $\pm m_n$ with $m_n$ again given by \eqref{m_n}. Also note that the scalars and fermions are charged under the two $U(1)$'s with $\vec{q_n} = (n,q)$.

\medskip
We finally comment on the regime of validity of our approach.
In order to observe compactification effects in our analysis, the KK scale $r^{-1}$ should lie below the cutoff of the 4D EFT. Furthermore, we expect to find the strongest constraints in the regime where the compactification radius is small in units of $m$ since this was also the case for the black hole arguments of \cite{Heidenreich:2015nta}. We therefore consider the following hierarchy of scales: 
\begin{equation}
\label{regime}
m \ll r^{-1} < \Lambda < M_4\,.
\end{equation}
Since we consider an EFT with cutoff $\Lambda$, we should keep all KK modes with masses $m_n \lesssim \Lambda$. For $m \ll \Lambda$, this implies that we should sum over all $n$ with $|n| \lesssim r\Lambda$. As discussed in Sec.~\ref{setup3d}, the one-loop approximation of the effective action is justified in the regime $|n| g_\text{KK}/\sqrt{m_n} \ll 1$. One can check that this is indeed the case for all $|n| \lesssim r\Lambda$ if we respect the hierarchy \eqref{regime}.

\subsection{Bounds on Charge-to-Mass Ratios}

Let us for the moment ignore the KK gravitons $g_{\mu\nu}^{(n\neq 0)}$ and KK photons $A_\mu^{(n\neq 0)}$ and assume that the only charged particles in the theory are the KK tower of scalars $\phi^{(n)}$ or fermions $\psi^{(n)}, \chi^{(n)}$. Integrating out this KK tower, we then obtain an effective action analogous to the one in Sec.~\ref{sec:3d}, where we now consider the special case of the gauge group $U(1)_\text{KK}\times U(1)$. As before, this yields inequalities of the form
\begin{align}
& \sum_n \frac{1}{|m_n|} \left(\lambda_1 z_{n1}^4 + \lambda_2 z_{n1}^2 
\right)+ \gamma_1 \ge 0 \,, \\
& \sum_n \frac{1}{|m_n|} \left(\lambda_3 z_{n2}^4 + \lambda_4 z_{n2}^2 
 \right) + \gamma_2 \ge 0 \,, \\
& \sum_n \frac{1}{|m_n|} \left(\lambda_5 z_{n1}^2 z_{n2}^2 + \lambda_6 z_{n1}^2 + \lambda_7 z_{n2}^2 
 \right) + \gamma_3 \ge 0 \,,
\end{align}
where the parameters
\begin{equation}
\gamma_1 \equiv 3840\pi\gamma_{f/s}(\delta_{i1},\delta_{i1})\,, \quad \gamma_2 \equiv 3840\pi\gamma_{f/s}(\delta_{i2},\delta_{i2})\,, \quad \gamma_3 \equiv 3840\pi\gamma_{f/s}(\delta_{i1},\delta_{i2})
\end{equation}
contain all $\calo(z^0)$ contributions as usual. We stress again that these contributions depend on the UV completion of the 4D EFT but also on the properties of the particles we integrated out. The latter implies in particular that the values of the $\gamma_i$ coefficients generally depend on the mass scales \eqref{regime}, i.e., $\gamma_i = \gamma_i (m,r,\Lambda)$. We will see below that, if $\gamma_3$ happens to drop below a critical value in a given EFT for some $m$, $r$, $\Lambda$, this implies a bound on the charge-to-mass ratios of a whole tower of 4D particles.

\medskip
The values of the $\lambda_i$'s depend on the spin and the couplings of the charged particles. 
As discussed in Sec.~\ref{sec:comp-setup}, we consider particles with charge vectors $\vec{q_n} = (n,q)$ and masses $m_n =\sqrt{m^2+\frac{n^2}{r^2}}$, where $|n| \lesssim r\Lambda$. The charge-to-mass ratios are therefore
\begin{equation}
\label{ctm}
z_{n1} = \frac{ng_\text{KK}\sqrt{M_3}}{\sqrt{m^2 + \frac{n^2}{r^2}}}\,, \qquad
z_{n2} = \frac{qg_3\sqrt{M_3}}{\sqrt{m^2 + \frac{n^2}{r^2}}}\,.
\end{equation}
The values of the $\lambda_i$'s are given in Table \ref{Tab:scalars_red} and computed in App.~\ref{app:reduction}.
Notice that, for scalars, the computation is straightforward: a scalar charged under a single $U(1)$ in 4D corresponds to a tower of scalars charged under $U(1)_\text{KK}\times U(1)$ in 3D. Therefore, the $\lambda_i$ coefficients are the same as in the $U(1)^2$ case without compactification, which was already discussed in Sec.~\ref{sec:3d}. However, for fermions, the $\lambda_i$ coefficients are different from those derived in Sec.~\ref{sec:3d} due to the appearance of extra interaction terms $\sim \slashed{H}^{(0)}$ in the 3D action \eqref{3d-dirac} which are not present in the standard Dirac Lagrangian (see App.~\ref{app:reduction} for details).

\medskip
We now include the effect of the KK gravitons and KK photons which we have neglected so far. According to \eqref{em-kk}, these particles have masses $\tilde m_n = \hat m_n = \frac{n}{r}$ and are charged under the KK $U(1)$ but not under the ordinary $U(1)$ such that
\begin{equation}
\tilde z_{n1}=\hat z_{n1}=\sqrt{2}\,, \qquad \tilde z_{n2}=\hat z_{n2}=0\,. \label{ctm-kk}
\end{equation}
Here, we dressed masses and charge-to-mass ratios with tildes for the KK gravitons and hats for the KK photons in order to distinguish them from the corresponding quantities for the scalars/fermions. Including the contributions due to loops of these particles, the inequalities become
\begin{align}
& \sum_n \frac{1}{|m_n|} \left(\lambda_1 z_{n1}^4 + \lambda_2 z_{n1}^2 
\right) + \sum_{n\neq 0} \frac{1}{|\tilde m_n|} \left(\tilde \lambda_1 \tilde z_{n1}^4 + \tilde \lambda_2 \tilde z_{n1}^2 
\right) + \sum_{n\neq 0} \frac{1}{|\hat m_n|} \left(\hat \lambda_1 \hat z_{n1}^4 + \hat \lambda_2 \hat z_{n1}^2
\right) \nl + \gamma_1 \ge 0 \,, \label{ineq1a} \\
& \sum_n \frac{1}{|m_n|} \left(\lambda_3 z_{n2}^4 + \lambda_4 z_{n2}^2
 \right) + \gamma_2 \ge 0 \,, \label{ineq1b} \\
& \sum_n \frac{1}{|m_n|} \left(\lambda_5 z_{n1}^2 z_{n2}^2 + \lambda_6 z_{n1}^2 + \lambda_7 z_{n2}^2 \right) + \sum_{n\neq 0} \frac{\tilde \lambda_6 \tilde z_{n1}^2 }{|\tilde m_n|} + \sum_{n\neq 0} \frac{\hat \lambda_6 \hat z_{n1}^2 }{|\hat m_n|} + \gamma_3 \ge 0 \,. \label{ineq1c}
\end{align}
Note that \eqref{ineq1b} is only sensitive to particles charged under $A_\mu^{(0)}$ and therefore unaffected by the KK gravitons and KK photons (apart from possible $\mathcal{O}(z^0)$ contributions to $\gamma_2$). On the other hand, \eqref{ineq1a} and \eqref{ineq1c} depend on particles charged under $B_\mu^{(0)}$ and thus receive corrections from the KK gravitons and KK photons, which are encoded in the coefficients $\tilde \lambda_i,\hat \lambda_i$. It is in principle possible to compute these coefficients, but we will see below that their exact values are not required for our argument.

\medskip
As in the previous sections, the inequalities \eqref{ineq1a}--\eqref{ineq1c} are trivially satisfied if the $\gamma_i$ coefficients are above a critical value but they yield nontrivial bounds on the charge-to-mass ratios for sufficiently small $\gamma_i$. For concreteness, let us make our usual assumption $\gamma_i=0$ in the following. For positive (negative) $\gamma_i$, the bounds on the charge-to-mass ratios become weaker (stronger).
Substituting the charge-to-mass ratios \eqref{ctm}, \eqref{ctm-kk} into the inequalities \eqref{ineq1a}--\eqref{ineq1c} and performing the sums, we find
\begin{align}
& \sum_n \frac{z_{n1}^4}{|m_n|} = \sum_{n\neq 0} \frac{\tilde z_{n1}^4}{|\tilde m_n|} = \sum_{n\neq 0} \frac{\hat z_{n1}^4}{|\hat m_n|} \simeq 8r \Psi^{(0)}\left(r\Lambda+1\right)+8r\gamma\,, \label{z-sums1} \\ & \sum_n \frac{z_{n1}^2}{|m_n|} = \sum_{n\neq 0} \frac{\tilde z_{n1}^2}{|\tilde m_n|} = \sum_{n\neq 0} \frac{\hat z_{n1}^2}{|\hat m_n|} \simeq 4r \Psi^{(0)}\left(r\Lambda+1\right)+4r\gamma\,,  \\ & \sum_n \frac{z_{n1}^2z_{n2}^2}{|m_n|} \simeq 2 m^2r^3 z_{02}^2 \left[2\zeta(3)+\Psi^{(2)}\left(r\Lambda+1\right)\right]\,,   \\ & \sum_n \frac{z_{n2}^4}{|m_n|} \simeq \frac{z_{02}^4}{m}\,,  \\ & \sum_n \frac{z_{n2}^2}{|m_n|} \simeq \frac{z_{02}^2}{m}\,, \label{z-sums2}
\end{align}
where $\Psi^{(n)}(x)=\frac{\d^n}{\d x^n}\frac{\Gamma^\prime(x)}{\Gamma(x)}$ is the $n$th polygamma function, $\gamma$ is the Euler-Mascheroni constant and we truncated the summation to modes lighter than the cutoff scale (i.e., $|n| \lesssim r\Lambda$)
and expanded in $mr$, which is consistent in the regime \eqref{regime}. Note that, if there is a large hierarchy $\Lambda \gg r^{-1}$, we have $\Psi^{(0)}\left(r\Lambda+1\right) \simeq \ln(r\Lambda)$. Our discussion below will be applicable both for moderately large $\Lambda \gtrsim r^{-1}$ and for $\Lambda \gg r^{-1}$, where in the latter case we will assume the regime $mr \ln(r\Lambda)\ll 1$. As a consequence, the leading order expressions for the inequalities are
\begin{align}
& 2\left( \lambda_1+\tilde\lambda_1+\hat\lambda_1 \right) + \lambda_2+\tilde\lambda_2+\hat\lambda_2 \ge 0 \,, \label{ineq2a} \\
& \lambda_3 z^4_{02}  + \lambda_4 z^2_{02} \ge 0 \,, \label{ineq2b} \\
& \lambda_7 z^2_{02} \ge 0 \,. \label{ineq2c}
\end{align}

\medskip
Whether the first inequality  \eqref{ineq2a} is satisfied or not depends crucially on the contributions of the KK gravitons and KK photons. Since we have not computed $\tilde \lambda_i, \hat \lambda_i$, we cannot draw any conclusions about the IR consistency of the EFT based on this inequality. The second inequality \eqref{ineq2b} is of the form we already found in the case without compactification. Consulting Table \ref{Tab:scalars_red}, we find that, for scalars, it is trivially satisfied (for our choice $\gamma_2=0$), while, for fermions, it leads to a WGC-like bound for the charge-to-mass ratio of the $n=0$ mode $z_{02}$,
\begin{equation}
z^2_{02} \left( z^2_{02} -\frac{1}{2} \right) \ge 0\,. \label{wgc-fermion}
\end{equation}

\begin{table}[t]
\centering
\setlength{\tabcolsep}{12pt}
\renewcommand{\arraystretch}{1.2}
\begin{tabular}{cccc}
\toprule
$\lambda_i$ & & scalar & fermion \\
\midrule
$\lambda_1$ &  &  $7$ & $- \frac{7}{8} $ \\
$\lambda_2$ &  &  $4$ & $ \frac{59}{2} $\\
$\lambda_3$ &  &  $7$ & $ 16 $ \\
$\lambda_4$ &  &  $4$ & $- 8 $ \\
$\lambda_5$ &  &  $7$ & $ \frac{17}{2}$ \\
$\lambda_6$ &  &  $ - 2$ & $- \frac{7}{2}$ \\
$\lambda_7$ &  &  $ - 2$ & $9$ \\
\bottomrule
\end{tabular}
\vspace{5pt}
\caption{\label{Tab:scalars_red} \emph{Values of the $\lambda_i$ coefficients for scalar and fermion case. We factorized out the common factor $\frac{1}{3840\pi}$.}}
\end{table}

\medskip
The most interesting inequality is \eqref{ineq2c}. Recall that it is due to IR consistency conditions mixing the ordinary $U(1)$ and the KK $U(1)$. We therefore expect this inequality to yield the strongest constraints among the three, analogously to the analysis of \cite{Heidenreich:2015nta} where black holes charged under both $U(1)$'s led to the strongest constraints. 
Interestingly, the dependence on the KK graviton and KK photon loops has completely vanished in \eqref{ineq2c} in the limit of small compactification radii. From Table \ref{Tab:scalars_red}, we can now read off the sign of $\lambda_7$:
\begin{equation}
\text{scalar:}\quad \lambda_7 < 0\,, \qquad \text{fermion:}\quad \lambda_7 > 0\,.
\end{equation}
Hence, in the scalar case, \eqref{ineq2c} is violated for \emph{all} values of $z_{02}$, i.e., the effective theory in the regime \eqref{regime}
is inconsistent in the IR. In the fermionic case, on the other hand, \eqref{ineq2c} is always satisfied such that the only nontrivial constraint on the charge-to-mass ratio is due to \eqref{wgc-fermion}.\footnote{Curiously, this is opposite to the results of \cite{Cottrell:2016bty}, where a black hole entropy calculation led to stronger constraints for fermions than for scalars.} Note that the constraints on fermions may be stronger in case that $\gamma_3 < 0 $. However, the corresponding inequalities can then still be satisfied by choosing a large enough $z_{02}$, while this is not possible in the scalar case. Let us also point out that further obstructions to satisfying the inequalities may exist both for scalars and fermions in regimes where the contributions from the KK gravitons and KK photons become relevant. We leave the difficult task of computing these contributions for future work and continue to discuss the scalar case in the following.

\subsection{The Tower WGC}

How can the IR inconsistency indicated by \eqref{ineq2c} be cured in the absence of fermions? A possible resolution of the problem would be to postulate a restriction on the mass scales such that considering the regime \eqref{regime} is not allowed in the first place. This is reminiscent of a minimal radius found in \cite{Heidenreich:2015nta} below which the convex-hull condition could not be satisfied anymore. We cannot exclude that such a restriction exists in some EFTs such that the problem discussed above is avoided in these theories.

\medskip
However, there is another way to satisfy the inequalities which seems to be more in line with existing examples in string theory (where a version of the WGC stronger than the convex-hull condition holds). As we will demonstrate in the following, this is possible if one replaces the single 4D scalar by a whole \emph{tower} of 4D particles whose masses and charges under the gauge field $A_M$ need to satisfy certain conditions. One can check that the one-loop corrections due to these extra particles then contribute to the effective action such that all three inequalities \eqref{ineq1a}--\eqref{ineq1c} can be satisfied simultaneously.
In particular, the term proportional to $\lambda_5$ in \eqref{ineq1c} is then not suppressed anymore compared to the other terms and thus helps to satisfy the inequality.

\medskip
To see this, consider a 4D EFT including $M$ scalars $\Phi_l$ with masses $m_l \in [m, m_\text{max}]$ and charges $q_l$ under the 4D gauge field $A_M$, where $l=0,\ldots,M-1$ is an index counting the 4D particles. After compactification, we then obtain a separate KK tower of 3D particles for each $l$.
The relevant inequality \eqref{ineq1c} thus becomes
\begin{align}
& \sum_{n,l} \frac{1}{|m_{nl}|} \left(\lambda_5 z_{nl1}^2 z_{nl2}^2 + \lambda_6 z_{nl1}^2 + \lambda_7 z_{nl2}^2 \right) + \sum_{n\neq 0} \frac{\tilde \lambda_6 \tilde z_{n1}^2 }{|\tilde m_n|} + \sum_{n\neq 0} \frac{\hat \lambda_6 \hat z_{n1}^2 }{|\hat m_n|} + \gamma_3 \ge 0 \, \label{ineq4c}
\end{align}
with $m_{nl}=\sqrt{m_l^2 + \frac{n^2}{r^2}}$ and
\begin{equation}
z_{nl1} = \frac{ng_\text{KK}\sqrt{M_3}}{\sqrt{m_l^2 + \frac{n^2}{r^2}}}\,, \qquad
z_{nl2} = \frac{q_lg_3\sqrt{M_3}}{\sqrt{m_l^2 + \frac{n^2}{r^2}}}\,.
\end{equation}
For scalars with $m_l$ much smaller than the KK scale, the sums over the KK states in \eqref{ineq4c} can be evaluated explicitly by expanding in $m_l r$ as in the single scalar case we considered before. On the other hand, for scalars with $m_l$ of the order of the KK scale $r^{-1}$ or larger, closed-form expressions for the sums are not available. However, we can still qualitatively understand their behavior up to $\mathcal{O}(1)$ factors by splitting the sums into three regimes: $n \sim m_l r$, $n \ll m_l r$ and $n \gg m_l r$, where in the last two regimes we can expand in $n/(m_lr)$ or $m_lr/n$, respectively. We thus find
\begin{align}
& z_{nl1}^2 = \frac{2n^2}{n^2+m_l^2r^2} \simeq 0\,, && z_{nl2}^2 = z_l^2 \frac{m_l^2r^2}{n^2+m_l^2r^2} \simeq z_l^2\,, && (m_lr \gg n) \notag \\
& z_{nl1}^2 = \frac{2n^2}{n^2+m_l^2r^2} \simeq \mathcal{O}(1)\,, && z_{nl2}^2 = z_l^2 \frac{m_l^2r^2}{n^2+m_l^2r^2} \simeq \mathcal{O}(1)z_l^2\,, && (m_lr \sim n) \notag \\
& z_{nl1}^2 = \frac{2n^2}{n^2+m_l^2r^2} \simeq 2\,, && z_{nl2}^2 = z_l^2 \frac{m_l^2r^2}{n^2+m_l^2r^2} \simeq 0\,, && (m_lr \ll n)
\end{align}
where $z_l = z_{0l2} = \frac{q_l g_3 \sqrt{M_3}}{|m_l|}=\frac{q_l g_4 M_4}{|m_l|}$ is the 4D charge-to-mass ratio. This yields
\begin{align}
& \sum_{n} \frac{1}{|m_{nl}|} \left(\lambda_5 z_{nl1}^2 z_{nl2}^2 + \lambda_6 z_{nl1}^2 + \lambda_7 z_{nl2}^2 \right) \notag \\
&\simeq
\left\{
  \begin{array}{ll}
    \lambda_7 \frac{z_{l}^2}{m_l} & \quad (m_l \ll r^{-1}, \Lambda \gtrsim r^{-1}) \vspace{1em} \\
    \lambda_7 \frac{z_{l}^2}{m_l} +4\lambda_6 r \ln(r\Lambda) & \quad (m_l \ll r^{-1}, \Lambda \gg r^{-1}) \vspace{1em} \\
    \mathcal{O}(1) \lambda_5 r z_{l}^2 + \mathcal{O}(1) \lambda_6r + \mathcal{O}(1) \lambda_7 r z_{l}^2 & \quad (m_l \sim r^{-1}, \Lambda \gtrsim r^{-1}) \vspace{1em} \\
     4 \lambda_6 r \ln(r\Lambda) & \quad (m_l \sim r^{-1}, \Lambda \gg r^{-1})\vspace{1em} \\
    \mathcal{O}(1) \lambda_5 r z_{l}^2 + \mathcal{O}(1) \lambda_6r + \mathcal{O}(1) \lambda_7 r z_{l}^2 & \quad (m_l \gg r^{-1}, \Lambda \gtrsim m_l)  \vspace{1em} \\
     4 \lambda_6 r \ln(\frac{\Lambda}{m_l}) & \quad (m_l \gg r^{-1}, \Lambda \gg m_l)\,.
  \end{array}
\right. \label{small-mass}
\end{align}
It is now straightforward to derive some basic properties the particle tower must have in order that the inequality \eqref{ineq4c} can be satisfied:
\begin{itemize}
\item Because of \eqref{small-mass} and Table \ref{Tab:scalars_red}, the KK sum for a single 4D particle $l$ contributes positively to the inequality \eqref{ineq4c} only when $(m_l, \Lambda)$ are in the third or fifth regimes of~\eqref{small-mass}. Moreover, the charge-to-mass ratio has to satisfy a bound $z_{l} \gtrsim \mathcal{O}(1)$ in order for the positive $\lambda_5$ term to dominate over the negative $\lambda_6$ and $\lambda_7$ terms.

\item
In the third and fifth regimes of \eqref{small-mass}, the scalar mass $m_l$ is near the cutoff scale, $m_l \lesssim \Lambda$.
In particular, the KK sums for scalars with $m_l \ll \Lambda$ always contribute negatively to the inequality \eqref{ineq4c}.

\item Suppose that the mass $m$ of the lightest scalar is much smaller than the KK and cutoff scales: $m\ll r^{-1},\Lambda$.
We can see from \eqref{small-mass} that the positive contribution from a scalar $l$ with a mass close to the cutoff $m_l\lesssim\Lambda$ is suppressed by a factor $mr \ll 1$ compared to the negative contribution of the lightest scalar with mass $m$, unless its charge-to-mass ratio $z_{l}$ is parametrically larger than that of the lightest scalar.
Assuming particles with finite charge-to-mass ratios, the number of particles in the tower thus needs to be at least of the order $(mr)^{-1}$. In the limit of small radii $mr \to 0$, this corresponds to a tower with an infinite number of particles.
\end{itemize}

\medskip
Note that the above constraints also imply a bound on the cutoff scale. We have seen that the tower needs to contain particles satisfying $z_l=\frac{q_l g_4 M_4}{m_l} \gtrsim \mathcal{O}(1)$, where $m_l$ is required to be of the order of the cutoff scale. It follows that $\Lambda \lesssim q_l g_4 M_4$, i.e., the cutoff needs to be much smaller than the 4D Planck scale in order for the EFT to be consistent. We thus also reproduce the magnetic WGC \cite{ArkaniHamed:2006dz} from our arguments.

\begin{figure}[t]
\centering
\includegraphics[scale=0.8]{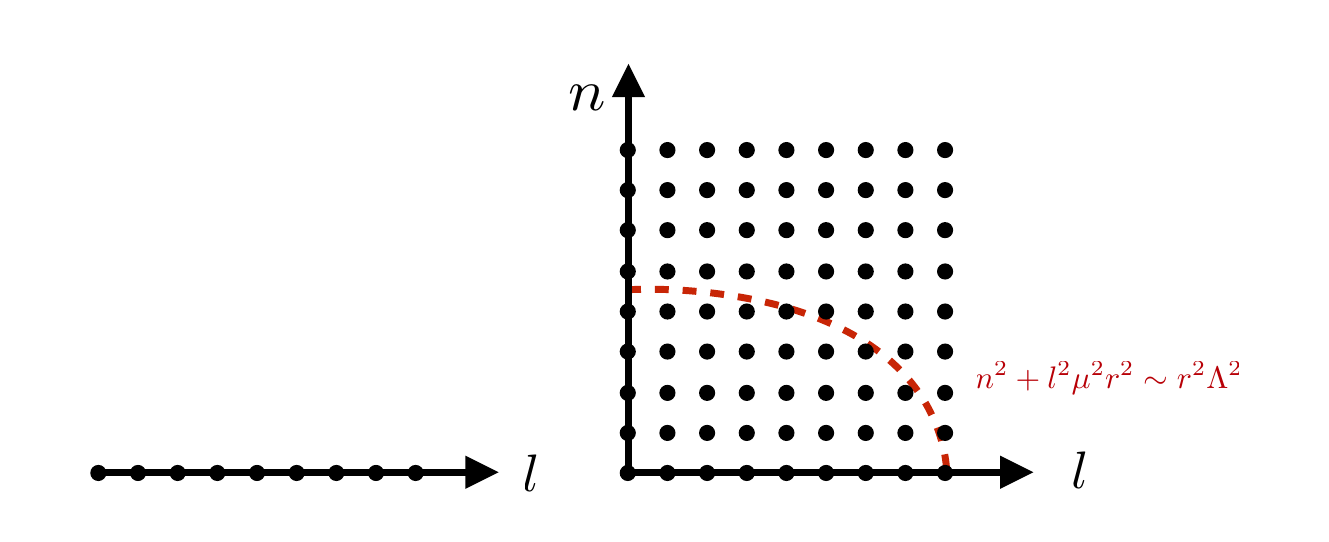}
\caption{\label{fig:lattice}\emph{Starting with a tower of 4D particles, after compactification the charges form a 2D lattice. Consistency requires to sum over all modes such that $m_{nl}^2=m^2+\frac{n^2}{r^2}+l^2\mu^2 \le \Lambda^2$. }}
\end{figure}

\medskip
As a simple example for the case $\Lambda \gg r^{-1}$, consider a 4D EFT including a tower of scalars $\Phi_l$ with masses $m_l=\sqrt{m^2+l^2\mu^2}$ and charges $q_l=(l+1)q$ under the 4D gauge field $A_M$. Here, $\mu$ denotes a mass scale by which the masses of the particles with different $l$ are separated. Analogously to our discussion in Sec.~\ref{sec:comp-setup}, the compactified theory then has a scalar particle spectrum labelled by indices $n,l$ with
\begin{equation}
\vec z_{nl} = \left( z_{nl1}, z_{nl2} \right) = \frac{\sqrt{M_3}}{m_{nl}}\left( n g_\text{KK}, (l+1)qg_3 \right)\,, \quad m_{nl} = \sqrt{m^2+\frac{n^2}{r^2}+l^2\mu^2}\,.
\end{equation}
The 3D particles thus fill a 2-dimensional charge lattice, see Fig.~\ref{fig:lattice}.

\medskip
Inserting these values into \eqref{ineq1a}--\eqref{ineq1c}, using \eqref{couplings_3d} and summing over all modes with $\frac{n^2}{r^2}+l^2\mu^2 \lesssim \Lambda^2$, we find
\begin{align}
& 3\lambda_1 + 2\lambda_2 \ge 0\,, \label{ineq3a} \\
& 3\lambda_3 q^2g_3^2 M_3 + 4\lambda_4 \mu^2 \ge 0 \,,  \\
& q^2g_3^2 M_3\left( \frac{1}{2}\lambda_5+\lambda_7\right)+ 2\lambda_6 \mu^2 \ge 0\, \label{ineq3b}
\end{align}
at leading order in an expansion in $(r\Lambda)^{-1}$, $\mu/\Lambda$.
We can now substitute the $\lambda_i$ values from Table \ref{Tab:scalars_red} and use again \eqref{couplings_3d} to express the inequalities in terms of 4D couplings. We thus find that \eqref{ineq3a}--\eqref{ineq3b} are satisfied if
\begin{equation}
\mu \le \sqrt{\frac{3}{8}} q g_4 M_4\,,
\end{equation}
i.e., the mass separation of the 4D particles is bounded from above.

\medskip
It is obvious from \eqref{ineq1a}--\eqref{ineq1c} that the above example is not the only possibility to satisfy the inequalities. Since they are not sensitive to each charge-to-mass ratio individually but only to their sum, more general distributions of particles will also suffice to satisfy them. For example, one could introduce variations between the mass separation of different particles in the tower or consider a tower with particles of different spins, etc. Another interesting possibility compatible with our constraints is to leave some of the charges unoccupied or filled by particles with negligible charge-to-mass ratios. This implies in particular that the tower of particles for which the charge-to-mass ratios are bounded from below need not necessarily occupy a charge lattice or even a charge sub-lattice. As long as the sum over the full tower behaves similarly to our example at leading order, our constraints can still be satisfied.

\medskip
The form of the WGC suggested by our analysis is thus stronger than the convex-hull condition but less restrictive than the (sub-)lattice WGC. It will be interesting to see whether there are examples in string theory which satisfy our version of the WGC but violate the stronger proposal of the (sub-)lattice WGC.

\section{Conclusions}
\label{concl}

In this work, we have argued for a specific version of the WGC in the presence of multiple $U(1)$ gauge fields by exploiting infrared consistency conditions of low-energy EFTs.
When the UV-sensitive EFT parameters $\gamma_{f/s}$ are in a certain range, our analysis leads to the following three constraints on the matter contents:
\begin{itemize}
\item The theories must contain particles consistent with the convex-hull type lower bounds on charge-to-mass ratios.
\item The theories must contain bifundamental particles in any basis choice for the $U(1)$ gauge fields.
\item The scalar theories must contain a tower of particles whose charge-to-mass ratios satisfy a lower bound.
\end{itemize}
This suggests that the convex-hull condition, which was originally motivated by black hole arguments, may not be strong enough. On the other hand, it is interesting that the constraints we find are flexible enough to not require the tower of particles to fill a full charge lattice. Instead, a sub-lattice or even a non-periodic occupation of charges are also consistent with our findings. Our version of the WGC is thus less restrictive than the previously proposed lattice WGC. It would be interesting to see whether there are string theory examples confirming such a behavior.

\medskip
As stated before, a crucial assumption of our work (and the earlier work \cite{Cheung:2014ega}) is that the $\gamma_{f/s}$ parameters encoding charge-independent corrections to the higher-derivative terms in the EFT are sufficiently small when evaluated at some cutoff scale $\Lambda$.
These parameters are sensitive to the UV-completion of the EFT and can therefore be interpreted as a quantum gravity input that the EFT analysis alone cannot fix. It is not surprising that such an extra ingredient is necessary to arrive at the above conclusions. After all, a key property of conditions separating the swampland from the landscape is precisely that they are \emph{not} visible from a pure low-energy perspective. Our results are therefore not a general ``proof'' of the WGC. Rather, we believe that our work, together with the earlier work \cite{Cheung:2014ega}, could provide a useful different perspective on the WGC by relating it to the values of the $\gamma_{f/s}$ parameters. An intriguing possibility is that the parameters are forced to be below the critical value in EFTs compatible with quantum gravity, or at least in certain classes thereof. In that case, we have shown that a specific version of the WGC automatically follows. We stress, however, that the converse is not true: it is in principle possible that the $\gamma_{f/s}$ parameters can take arbitrary values in quantum gravity theories, and that theories in which these parameters are large still satisfy the WGC for reasons unrelated to our constraints (see also a comment in \cite{Harlow:2015lma}).

\medskip
Our work is in the same spirit as a number of recent results showing that the original formulation of the WGC can be related to other claims which at first sight appear to be quite different.
Thus, depending on the considered setup, the WGC can be understood as a statement about black hole decay \cite{ArkaniHamed:2006dz}, field space variations \cite{Klaewer:2016kiy, Palti:2017elp}, CFT states \cite{Nakayama:2015hga, Harlow:2015lma, Benjamin:2016fhe, Montero:2016tif}, cosmic censorship \cite{Crisford:2017gsb, Cottrell:2016bty}, instabilities of AdS vacua \cite{Ooguri:2016pdq, Danielsson:2016mtx, Freivogel:2016qwc, Ooguri:2017njy}, or, as we argued here, the smallness of certain EFT parameters. In our point of view, it is far from obvious at the moment which of the many formulations of the WGC will ultimately turn out to be the most helpful, the easiest to prove or the most fundamental one.
 
\medskip
Our work suggests several opportunities for further research. Straightforward generalizations include an analysis of theories in dimensions greater than $4$, with a more general matter content, or with a nonzero cosmological constant. It would also be interesting to consider compactifications on manifolds other than a circle and check whether this yields further constraints in addition to those found in Sec.~\ref{sec:comp}.

\medskip
Another interesting route is to test our general arguments in concrete string compactifications. In particular, it would be nice to check explicitly whether there are obstructions to the assumptions that went into our analysis, for example, regarding the matter content, the hierarchy of scales, or the stabilization of the radion. One may also investigate how our analysis changes when the radion is left unstabilized and how this relates to the arguments of \cite{Palti:2017elp}, where a form of the WGC in the presence of massless scalar fields was conjectured.

\medskip
An important extension of our work would also be to derive the value of the $\gamma_{f/s}$ parameters in explicit string models. Unfortunately, higher-derivative corrections of the form necessary for our analysis are at present only partially known in string/M-theory compactifications (see, e.g., \cite{Grimm:2013gma, Grimm:2017okk} for the computation of some $R^2$ terms). The computation of these corrections is technically challenging but not impossible. This might allow us to prove that, at least for certain classes of compactifications, the WGC is indeed implied by analyticity and causality of the EFT.

\section*{Acknowledgements}

We would like to thank Billy Cottrell, Eran Palti, Pablo Soler and Gianluca Zoccarato for useful discussions. 
SA is supported in part by the Research Grants Council (RGC) of Hong Kong through grants HKUST4/CRF/13G, 604213, and 16304414.
DJ is supported in part by the DFG Transregional Collaborative Research Centre TRR 33 ``The Dark Universe''.
TN  is in part supported by Grant-in-Aid for Scientific Research (B) No. 17H02894 from the Japan Society for the Promotion of Science (JSPS).
GS is supported in part by the DOE grant DE-SC0017647 and the Kellett Award of the University of Wisconsin.

\newpage

\appendix

\section{Some Useful Formulae}
\label{app:formulae}

In $D=3$ dimensions, given 4 antisymmetric tensors $A$, $B$, $C$, $D$, one can show that 
\begin{equation}
\label{app:identity_3d}
A_{\mu\nu}B^{\nu\rho} C_{\rho\lambda}D^{\lambda\mu}
= \frac{1}{4} \left[ (A\cdot B) (C\cdot D) + (A\cdot D) (C\cdot B) \right]\,,
\end{equation}
where $A\cdot B = A_{\mu\nu} B^{\mu\nu}$. 
In $D=4$, we have instead
\begin{equation}
\label{app:identity_4d}
A_{\mu\nu}B^{\nu\rho} C_{\rho\lambda}D^{\lambda\mu} = \frac{1}{4} \left[(A\cdot B)(C\cdot D)+ (A\cdot D)(B\cdot C)+ (\tilde A\cdot C)(B\cdot \tilde D)\right]\,,
\end{equation}
where $\tilde A_{\mu\nu} = \frac{1}{2} \epsilon_{\mu\nu\rho\sigma}A^{\rho\sigma}$.

\section{Dualization}
\label{app:dualization}

We can dualize the effective action for multiple photons by introducing the same number of scalar fields $\phi_i$. This works as follows. Let us take the action 
\begin{equation}
\label{dual:action}
\Gamma = \sum_i \int \bigg( -\frac{1}{2} F_i \wedge \star F_i+\phi_i \d F_i + \alpha(F_i) \bigg)
\,,
\end{equation}
where the $F_i$'s are two-forms and the $\phi_i$'s some scalars for $i=1,...,N$. $\alpha(F_i)$ denotes collectively all higher-order operators involving the $F_i$'s. In our case, these will be the HO operators going like $F^4$.
We can integrate out these scalars in order to rewrite the previous action as the action for photons. Indeed, the equations of motion for the scalars, $\d F_i=0$, mean that we have (at least locally) $F_i=\d A_i$, where the $A_i$'s are one-forms. By substituting these equations of motion into the previous action we obtain the usual Maxwell action for multiple photons
\begin{equation}
\label{dual:action_photons}
\Gamma = \sum_i \int \bigg(  -\frac{1}{2} F_i \wedge\star F_i + \alpha(F_i) \bigg)
\,,
\end{equation}
where the $A_i$'s are to be interpreted as vector potentials. 

\medskip
On the other hand, we can integrate out the two-forms $F_i$ in order to obtain the dual action for scalars. First, we complete the square in \eqref{dual:action} to get
\begin{equation}
\label{dual:action2}
\Gamma = \sum_i \int \bigg( -\frac{1}{2}(F_i - \star \d\phi_i) \wedge\star (F_i - \star \d\phi_i) -\frac{1}{2} \d \phi_i \wedge\star \d \phi_i 
+ \alpha(F_i) \bigg)
\,.
\end{equation}
Then, we can substitute the equations of motion for $F_i$, $F_i = \star \d \phi_i + \mathcal{O}(\alpha)$, into the action \eqref{dual:action2}. Immediately, we see that the first term is $\mathcal{O}(\alpha^2)$, while the last term goes like $\alpha(\star \d\phi_i) + \mathcal{O}(\alpha^2)$. Therefore, \emph{up to} $\mathcal{O}(\alpha)$ operators, the dual effective action for scalars is
\begin{equation}
\Gamma = \sum_i \int \bigg( -\frac{1}{2}\d \phi_i \wedge\star \d \phi_i  + \alpha(\star \d\phi_i) \bigg)
\,.
\end{equation}

\section{Heat Kernel}
\label{app:heatkernel}

In this appendix, we show how to compute the one-loop effective action of gravity and $N$ gauge fields using the heat kernel expansion. Our discussion of the general method follows closely the review articles \cite{Vassilevich:2003xt, Avramidi:2000bm}.

\subsection{Scalars}

Let us begin with a brief review of the heat kernel approach. As a simple example, we first consider the partition function of a scalar field $\phi$,
\begin{equation}
Z[J] = \int \mathcal{D}\phi \,\e^{-S(\phi)} \simeq \int \mathcal{D}\varphi\, \e^{-S (\overline{\phi})-\int \d^D x \sqrt{g}\, \varphi(x)J(x)-\int \d^D x \sqrt{g}\, \varphi(x) \hat D \varphi(x)}\,,
\end{equation}
where $S(\phi)$ is the classical Euclidean action in $D$ dimensions,\footnote{We work in the Euclidean signature in the present and the next subsections when reviewing the heat kernel approach. Later, we translate the Euclidean results back to the Lorentzian signature for our application. Our convention for Wick rotation is the standard one: $t_E =i t_M$, $x_E=x_M$, $S_E = -i S_M$.}
$\hat D$ is a differential operator and $J$ is an (optional) source which is charged under the field. In the last step, we have split the field into a background and a fluctuation, i.e., $\phi\to \overline{\phi}+\varphi$, and expanded the action up to second order in $\varphi$. This approximation is valid as long as we restrict to computing the effective action at one-loop order.

\medskip
Because of the quadratic approximation, the functional integral is Gaussian. Setting the source term to zero and performing the functional integral, we arrive at
\begin{equation}
Z = \e^{-S(\overline{\phi})} (\det \hat D)^{-1/2}\,.
\end{equation}
The one-loop correction to the effective action is therefore
\begin{equation}
\delta\Gamma_1 = \frac{1}{2} \ln (\det \hat D)\,.
\end{equation}
It turns out that rewriting this in terms of the so-called heat kernel $K$ is very useful to explicitly perform the computation. The heat kernel is defined as
\begin{equation}
K(t,\hat D) = \e^{-\hat Dt}\,,
\end{equation}
which is to be understood as a matrix exponential. The name ``heat kernel'' is due to the fact that $K$ satisfies the heat conduction equation
\begin{equation}
\left(\partial_t + \hat D \right) K = 0\,. \label{heat-eq}
\end{equation}
Using the heat kernel, we can rewrite the one-loop correction to the effective action as
\begin{equation}
\label{HK:integral1}
\delta\Gamma_1 = \frac{1}{2} \ln (\det \hat D) = \frac{1}{2} \text{Tr} (\ln \hat D) = -\frac{1}{2} \text{Tr}\int_0^\infty \frac{\d t}{t} \e^{-\hat Dt} + \text{const.} = -\frac{1}{2} \text{Tr} \int_0^\infty \frac{\d t}{t} K + \text{const.}\,,
\end{equation}
where we defined $\text{Tr}\,K(x,y; t) \equiv \int \d^D x \sqrt{g} K(x,x; t)$.\footnote{Later, when we discuss fermions, we will include the trace over spinor indices into the definition of Tr, in addition to the coordinate integral. More explicitly, $\text{Tr}f(x,y)=\int \d^Dx \sqrt{g}\,\text{tr}\,f(x,x)$, where tr is for the trace over spinor indices, if any, of the function $f(x,y)$.} We also used that $\int_\epsilon^\infty \frac{\d t}{t} \e^{-\hat Dt}=-\gamma- \ln\epsilon -\ln \hat D$. In order to cancel the divergent constant term, we have to subtract the contribution from the trivial heat kernel $\int_\epsilon^\infty \frac{\d t}{t} \e^{-t}=-\gamma- \ln\epsilon$.

\medskip
Taking the differential operator to be $\hat D= -\nabla^2 + m^2$ in flat space $\mathbb{R}^D$, one checks that, in $D$ dimensions, \eqref{heat-eq} is solved for
\begin{equation}
K(x,y; t) = \frac{1}{(4\pi t)^{D/2}}\e^{-\frac{(x^\mu-y^\mu)^2}{4t}-tm^2} \label{heatkernel}
\end{equation}
with initial condition
\begin{equation}
K(x,y; 0) = \delta(x^\mu-y^\mu)\,.
\end{equation}
The last statement simply follows from the fact that \eqref{heatkernel} is a representation of the delta distribution in the limit $t\to 0$. We will see below how to obtain a solution for the heat kernel for more general differential operators on curved backgrounds.

\subsection{Fermions}

Instead of a scalar, let us now consider a fermion. The Euclidean action of a charged, massive fermion is
\begin{align}
& S(\bar\psi,\psi) = \int \d^D x \sqrt{g}\, \bar\psi (\,\slash\!\!\!\! D+m) \psi\,, \quad \,\slash\!\!\!\! D = \Gamma^\mu D_\mu = \Gamma^\mu\partial_\mu + i q g \Gamma^\mu A_\mu + \frac{1}{8}\Gamma^\mu[\Gamma_\alpha,\Gamma_\beta]\omega_\mu^{\alpha\beta} \,.
\end{align}
The behavior of the $\Gamma$ matrices under Wick rotation is nontrivial and described, for example, in \cite{vanNieuwenhuizen:1996tv}. Also note that $\bar\psi$ is an independent spinor in the Euclidean theory which is not related to the conjugate of $\psi$ \cite{vanNieuwenhuizen:1996tv}. For the path integral, this does not matter since there $\bar\psi$ and $\psi$ are treated as independent variables anyway.

\medskip
The partition function reads
\begin{equation}
Z = \int \mathcal{D}\bar \psi \mathcal{D}\psi \, \e^{-S(\bar\psi,\psi)} = \int \mathcal{D}\bar \psi \mathcal{D}\psi \, \e^{ \int \d^D x \sqrt{g}\, \bar\psi (-\,\slash\!\!\!\! D-m) \psi}\,.
\end{equation}
Here, $\bar \psi$ and $\psi$ should be treated as Grassmann variables because of their fermionic nature.
Evaluating the path integral, we thus find
\begin{align}
Z = \det (\,\slash\!\!\!\! D+m)\,.
\end{align}

\medskip
We now rewrite the Dirac-type operator $\,\slash\!\!\!\! D+m$ in terms of a Laplace-type operator $\hat D$, which we define such that $\hat D=(\,\slash\!\!\!\! D+m)(-\,\slash\!\!\!\! D+m)$.
This yields
\begin{align}
\label{HK:squaringDirac}
\hat D &= (\,\slash\!\!\!\! D+m)(-\,\slash\!\!\!\! D+m) 
= - g^{\mu\nu}D_\mu D_\nu + \frac{1}{4} R - \frac{iqg}{4}[\Gamma^\mu,\Gamma^\nu]F_{\mu\nu} + m^2\,,
\end{align}
where we used
\begin{align}
(\slash\!\!\!\! D)^2 &= \frac{1}{2}\Gamma^\mu\Gamma^\nu [D_\mu,D_\nu] + g^{\mu\nu} D_\mu D_\nu \,, \\
[D_\mu,D_\nu] &= \frac{1}{4} R_{\mu\nu \alpha\beta} \Gamma^{\alpha}\Gamma^{\beta} + i q g F_{\mu\nu} \,, \label{app:commutator}\\
R &= - \frac{1}{2} \Gamma^\mu\Gamma^\nu\Gamma^\rho\Gamma^\sigma R_{\mu\nu\rho\sigma}\,.
\end{align} 

\medskip
Let us now finally compute the effective action. We first observe that
\begin{equation}
\text{Tr} \ln (\pm \,\slash\!\!\!\! D + m) = \text{Tr} \left[ \ln(m) \pm \frac{\slash\!\!\!\! D}{m} - \frac{(\slash\!\!\!\! D)^2}{2m^2} \pm \ldots \right]\,.
\end{equation}
In even dimensions, the trace over an uneven number of $\Gamma$ matrices vanishes such that we have
\begin{equation}
\text{Tr} \ln (\,\slash\!\!\!\! D + m) = \text{Tr} \ln (-\,\slash\!\!\!\! D + m)\,.
\end{equation}
In odd dimensions, this is only true up to Chern-Simons terms, which have to be considered separately. Since our 3D argument focuses on the parity-invariant case, we neglect Chern-Simons terms in the following.
Hence,
\begin{align}
\delta\Gamma_1 &= -\ln Z = -\ln \det (\,\slash\!\!\!\! D+m) = -\text{Tr}\ln (\,\slash\!\!\!\! D+m) = -\frac{1}{2}\text{Tr}\ln (\,\slash\!\!\!\! D+m)-\frac{1}{2}\text{Tr}\ln (-\,\slash\!\!\!\! D+m) \nll = -\frac{1}{2}\text{Tr}\ln \left[-\,\slash\!\!\!\! D^2+m^2\right] = -\frac{1}{2}\text{Tr}\ln \hat D \,.
\end{align}
The one-loop correction to the effective action is therefore formally the same as in the scalar case, except for a different sign,
\begin{equation}
\label{HK:integral2}
\delta\Gamma_1 = -\frac{1}{2} \ln (\det \hat D) = -\frac{1}{2} \text{Tr} (\ln \hat D) = \frac{1}{2} \text{Tr} \int_0^\infty \frac{\d t}{t} K + \text{const}.
\end{equation}
Note that \text{Tr} contains an integral over $D$-dimensional space as well as a trace over the spinor indices here.

\subsection{The Effective Action}
\label{subsec:EFT3d}

Let us now compute the heat kernel $K$. As we saw above, we want to solve the heat conduction equation
\begin{equation}
\label{HK:conduction_equation}
(\de_t + \hat D) K(x,y;t) = 0
\end{equation}
for a general second-order differential operator of the form
\begin{equation}
\label{HK:differential_operator}
\hat D = 
- g^{\mu\nu} D_\mu D_\nu + U + m^2 = 
- (g^{\mu\nu} \de_\mu \de_\nu + a^\mu\de_\mu + b) + m^2
\,,
\end{equation}
where the covariant derivative may in general contain a Levi-Civita connection $\Gamma^\mu_{\rho\sigma}$, a spin connection and gauge connections. Let us denote the last two types of connections by $\omega_\mu$. We then have
\begin{equation}
\omega_\mu = \frac{1}{2} g_{\mu\nu} (a^\nu + g^{\rho\sigma} \Gamma^{\nu}_{\rho\sigma} \Id)
 \,,
\qquad
U = g^{\mu\nu} (\de_\mu\omega_\nu + \omega_\mu\omega_\nu - \omega_\sigma\Gamma^\sigma_{\mu\nu} ) - b  
\,,
\end{equation}
where $\Id$ is the identity operator in the vector bundle $V$.\footnote{From a geometric point of view, we are considering a smooth Riemannian manifold $M$ (with no boundaries $\de M = 0$) and a vector bundle $V$ over $M$. Here, we are studying a differential operator \eqref{HK:differential_operator} on the bundle $V$, which has connection $\omega_\mu$ \cite{Vassilevich:2003xt}.}

\medskip
In a curved space, a general ansatz for the heat kernel is provided by
\begin{equation}
\label{HK:Ansatz}
K(x,y;t) = \frac{1}{(4\pi t)^{D/2}} \sqrt{\Delta(x,y)}\, \e^{-\frac{\sigma(x,y)}{2t} - tm^2} \left(b_0(x,y)+ tb_2(x,y)+t^2b_4(x,y)+\ldots\right)
\,,
\end{equation}
where $b_i(x,y)$ are so-called heat kernel coefficients, $\sigma(x,y)$ is half of the square of the length of the geodesic connecting $x^\mu$ to $y^\mu$, and $\Delta(x,y)$ is the Van Vleck-Morette determinant
\begin{equation}
\Delta(x,y) = \frac{\det\left( - \frac{\de}{\de x^\mu} \frac{\de}{\de y^\nu} \sigma(x,y) \right)}{\sqrt{g(x)g(y)}}
\,.
\end{equation}
By substituting \eqref{HK:Ansatz} into \eqref{HK:conduction_equation}, one can solve the differential equation order by order in $t$.
In this way, one eventually obtains the coincidence limits of the heat kernel coefficients $b_i = b_i(x,x)$ \cite{Avramidi:2000bm}:
\begin{align}
\label{HK:b_0}
b_0 =&\  \Id  \,, \\
\label{HK:b_2}
b_2 =&\  P  \,, \\
\label{HK:b_4}
b_4 =&\ \frac{1}{2} U^2 - \frac{1}{6} RU  + \frac{1}{12} W_{\mu\nu} W^{\mu\nu} - \frac{1}{6} D^\mu D_\mu U + \Id \bigg( \frac{1}{180} R^2_{\mu\nu\rho\sigma} - \frac{1}{180}  R^2_{\mu\nu}  
+ \frac{1}{72} R^2 \nll + \frac{1}{30} D^\mu D_\mu R   \bigg)  \,, \\
\label{HK:b_6}
b_6 = &\
\frac{1}{6} P^3 
+ \frac{1}{12} P W_{\mu\nu} W^{\mu\nu} 
+ \frac{1}{12} P  D^\mu D_\mu P
- \frac{1}{90} (D_{\mu} W_{{\nu\rho}})( D^{\mu} W^{{\nu\rho}})
+ \frac{1}{180} (D_{\mu} W^{{\mu\nu}})( D^{\rho} W_{{\rho\nu}}) \nll
+\frac{1}{180} \left( -6 W_{\mu\nu} W^{\nu\rho} W_{\rho}^{\ \mu} 
+ 2  R^{\mu\nu} W_{\mu{\rho}} W^{{\rho}}_{\ \nu}   
+ 3R^{\mu\nu\rho\sigma} W_{\mu\nu} W_{\rho\sigma}	\right)		
\,,	
\end{align}
where
\begin{equation}
P \equiv \frac{R}{6} - U\,,
\qquad
W_{\mu\nu}  \equiv [D_\mu, D_\nu] 
\,.
\end{equation}
Notice that, for $b_6$, we restricted to terms containing up to 4 derivatives. We refer to \cite{Avramidi:2000bm} for the complete expression for $b_8$.

\medskip
Substituting \eqref{HK:Ansatz} in \eqref{HK:integral1}, \eqref{HK:integral2}, one finds that the expansion has divergent coefficients in front of $b_{i}$'s with $i \le D$, whereas they are finite for $i > D$. The finite part of the one-loop corrections to the effective action is then given by
\begin{equation}
\label{HK:effective_action}
\delta\Gamma_{1}^{\rm fin} = \pm \frac{1}{2(4\pi)^{D/2}} \sum_{i > D} m^{D-i} \Gamma\left(\frac{i}{2}-\frac{D}{2}\right)  B_{i}
\,,
\end{equation}
where the $+$ ($-$) is for scalars (fermions) and we defined
\begin{equation}
\label{HK:B_2r}
B_{i} = \left\{\begin{array}{lc}
\displaystyle-\int \d^D x \sqrt{g} \ \tr b_i(x,x) & \text{(Euclidean)}\,, 
\\[5mm]
\displaystyle\int \d^D x \sqrt{-g} \ \tr b_i(x,x) & \text{(Lorentzian)} \,.
\end{array}\right.
\end{equation} 
Notice in particular that the expression for $b_i(x,x)$ does not change after Wick rotation in our convention. The divergent part can also be computed in a similar way, once we specify a regularization scheme. We will calculate a part of it later when necessary.

\subsubsection{Fermion One-Loop Corrections}
We can now compute the effective action for the scalar and fermion case. Let us start with the latter. From here on, we work in the Lorentzian signature in the rest of this appendix.
As explained in the previous subsection, one has to first square the Dirac operator in order to get a Laplacian-like operator \eqref{HK:differential_operator}. In fact, from \eqref{HK:squaringDirac}, we can immediately recognize 
\begin{align}
\label{HK:nabla}
D_\mu &= \de_\mu + iqg A_\mu + \frac{1}{4} w_{\mu \alpha\beta} \Gamma^{\alpha}\Gamma^{\beta} \,,\\
\label{HK:U}
U &= \frac{R}{4} \Id - \frac{iqg}{2} F_{\mu\nu} \Gamma^\mu \Gamma^\nu 
\,.
\end{align}
Therefore, the commutator $W_{\mu\nu}$ is simply given by \eqref{app:commutator}. Here, $\Id$ is the $2^{[D/2]}$-dimensional unit matrix, where $[...]$ denotes the integer part. 

\medskip
Substituting these expressions into \eqref{HK:b_0}--\eqref{HK:b_6}, we can compute the heat kernel coefficients \eqref{HK:B_2r} up to 4 derivatives:\footnote{See~\cite{Ritz:1995nt} for the computation of $B_8$.}
\begin{align}
\label{HK:B_0}
B_0 & =	2^{[D/2]} \int  \d^D x \sqrt{-g} 			\,,	\\
\label{HK:B_2}
B_2 & =	- \frac{2^{[D/2]}}{12}	 \int  \d^D x \sqrt{-g}  R 			\,, \\
\label{HK:B_4}
B_4 & =	2^{[D/2]} \int  \d^D x \sqrt{-g} \left( \frac{R^2}{288} - \frac{R_{\mu\nu}^2}{180}  - \frac{7}{1440} R_{\mu\nu\rho\sigma}^2 + \frac{(qg)^2}{6} F^2 \right)				\,, \\
\label{HK:B_6}
B_6 & = - \frac{2^{[D/2]}}{360} \int \d^D x \sqrt{-g} \bigg( 5 (qg)^2  R F^2
+ 2 (qg)^2 R_{\mu\nu\rho\sigma} F^{\mu\nu}F^{\rho\sigma}  
- 26 (qg)^2 R_{\mu\nu} F^{\mu\rho} F^\nu_{\ \rho} 	\bigg)	\,,\\
\label{HK:B_8}
B_8 &= - \frac{2^{[D/2]}}{360} (qg)^4 \int \d^D x \sqrt{-g}  \left[ 14 F_{\mu\nu}F^{\nu\rho}F_{\rho\sigma}F^{\sigma\mu} - 5 (F^2)^2  \right]	
\,,
\end{align}
where we simplified the expressions by using integration by parts together with the equations of motion $\nabla_\rho F^{\rho\mu} = 0$ and the Bianchi identities $\nabla_{[\mu} F_{\rho\sigma]} = R_{[\mu\nu\rho]\sigma} = 0$. Here, $B_0$ and $B_2$ are corrections to the cosmological constant and the Planck mass, hence we may remove them by renormalization. Similarly, the $F^2$ term in $B_4$ may be removed by renormalization of the gauge coupling. On the other hand, $B_4$, $B_6$ and $B_8$ provide higher-derivative operators we are interested in (there are no four-derivative operators arising from $B_{i}$ with $i\geq10$). Upon an appropriate regularization when necessary, we may calculate the effective action up to four derivatives by using the above expressions.

\paragraph{3D effective action }
In $D=3$, $B_4$, $B_6$ and $B_8$ are classified into the finite part \eqref{HK:effective_action}, so that the 1-loop 4-derivative correction to the effective action is
\begin{align}
\delta\Gamma_1 = \ \int \d^3 x \sqrt{-g} 
\bigg[  &
\frac{(qg)^4}{1920 m^5 \pi} (F_{\mu\nu}F^{\mu\nu})^2
+ \frac{(qg)^2}{1152 m^3 \pi}   R F^2
\notag \\ &
- \frac{13 (qg)^2}{2880 m^3 \pi}  R_{\mu\nu} F^{\mu \rho} F^\nu_{\ \rho} 	 
+  \frac{(qg)^2}{ 2880 m^3 \pi} R_{\mu\nu\rho\sigma} F^{\mu\nu}F^{\rho\sigma}  
- \frac{1}{2304 m \pi} R^2
\notag \\ &
+ \frac{1}{1440 m \pi}  R_{\mu\nu}^2
+ \frac{7}{11520 m \pi} R_{\mu\nu\rho\sigma}^2 
\bigg]
\,, \label{HK:EFT_3d}
\end{align}
where we used the 3D identity~\eqref{app:identity_3d}.

\paragraph{4D effective action }
In $D=4$, $B_6$ and $B_8$ are in the finite part \eqref{HK:effective_action}, but $B_4$ is in the divergent part. In the cutoff regularization, the correction associated to $B_4$ can be computed as
\begin{align}
\delta\Gamma_1\ni -\frac{\cali}{2(4\pi)^2} B_4
\quad
{\rm with}
\quad
\cali \equiv \int_{1/\Lambda^2}^\infty \frac{\d t}{t} \e^{-tm^2}  =
\log\frac{\Lambda^2}{m^2} - \gamma + \calo(\Lambda^{-2}) \,.
\end{align}
Hence, the four-derivative corrections to the effective action are given by
\begin{align}
\delta\Gamma_1 = \ \int \d^4 x \sqrt{-g}
\bigg[ &
\frac{(qg)^4}{1440 \pi^2 m^4} (F_{\mu\nu}F^{\mu\nu})^2 
+ \frac{7(qg)^4}{5760 \pi^2 m^4} (F_{\mu\nu} \tilde F^{\mu\nu})^2 	
+ \frac{(qg)^2}{576 \pi^2 m^2}   R  F^2
\notag \\ &
- \frac{13 (qg)^2}{1440 \pi^2 m^2}  R_{\mu\nu} F^{\mu \rho} F^\nu_{\ \rho}  
+ \frac{(qg)^2}{1440 \pi^2 m^2}  R_{\mu\nu\rho\sigma} F^{\mu\nu}F^{\rho\sigma}  
\nonumber\\
&+\cali\left(
- \frac{R^2}{2304 \pi^2} 
+ \frac{R_{\mu\nu}^2}{1440 \pi^2}  
+ \frac{7}{11520 \pi^2}  R_{\mu\nu\rho\sigma}^2  
\right)
\bigg]
\,, \label{HK:finite_4d}
\end{align}
where we used the 4D identity~\eqref{app:identity_4d}.


\subsubsection{Scalar One-Loop Corrections}
We next discuss the scalar case. Let us integrate out the complex scalar in 
\begin{equation}
S = \int \d^D x \sqrt{-g} 
\bigg[
M_D^{D-2} \frac{R}{2}  
- \frac{1}{4} F^2
- |D_{\mu} \phi|^2
- m^2 |\phi|^2
\bigg]
\,, 
\qquad 
D_{\mu} = \de_\mu  + iqg A_{\mu}
\,.
\end{equation}
Partially integrating, we can rewrite this in terms of a Laplace-type operator with
$\hat D = -D^\mu D_\mu + m^2$ and $U=0$. As a consequence, the commutator is simply $W_{\mu\nu} = i qg F_{\mu\nu}$. By repeating the same steps as in the fermionic case, the heat kernel procedure gives the following 1-loop 4-derivative correction to the effective action for $D=3$:
\begin{align}
\delta\Gamma_1^{\rm fin} = \int \d^3 x \sqrt{-g}
\bigg[ &
\frac{7 (gq)^4}{15360 \pi m^5} (F^2)^2
  - \frac{(qg)^2}{1152 \pi m^3}  R F^2	
+ \frac{(qg)^2}{1440 \pi m^3}   R^{\mu\nu} F_{\mu\rho} F_{\ \nu}^{\rho} \nll
- \frac{(qg)^2}{2880 \pi m^3}  R^{\mu\nu\rho\sigma} F_{\mu\nu} F_{\rho\sigma}	
+ \frac{1}{576 \pi m}R^2  
- \frac{1}{1440 \pi m}  R^2_{\mu\nu} \nll
+ \frac{1}{1440 \pi m}  R^2_{\mu\nu\rho\sigma} 
\bigg]
\,.
\end{align}
Here, we used again the equations of motion for $F^{\mu\nu}$ and neglected terms corresponding to total derivatives.
For $D=4$, we find
\begin{align}
\delta\Gamma_1
 = 
 \int \d^4 x \sqrt{-g}
\bigg[ &
- \frac{q^2 g^2}{1152 \pi^2 m^2} R F^2					
- \frac{q^2 g^2}{1440 \pi^2 m^2} R^{\mu\nu} F_{\rho\mu} F^{\rho}_{\ \nu}   
- \frac{q^2 g^2}{2880 \pi^2 m^2} R^{\mu\nu\rho\sigma} F_{\mu\nu} F_{\rho\sigma}	
\nll
+ \frac{7 q^4 g^4}{23040 \pi^2 m^4} (F^2)^2
+ \frac{q^4 g^4}{23040 \pi^2 m^4} (F\cdot \tilde F)^2
\nonumber\\
&+\cali\left(
\frac{R^2}{1152 \pi^2}   
-  \frac{R^2_{\mu\nu}}{2880 \pi^2} 
+ \frac{R^2_{\mu\nu\rho\sigma}}{2880 \pi^2} 
\right)
\bigg]
\,,
\end{align}
where we again used the cutoff regularization.

\subsubsection{Matter Charged under Multiple $U(1)$'s}
\label{subsec:mult}

The generalization to the case with several fermions/scalars charged under multiple $U(1)$'s is then straightforward. Let us, for example, consider a
theory with fermions $\psi_a$ charged under $U(1)^N$:
\begin{align}
\label{startingEFT_multipleU1}
S &= \int \d^D x \sqrt{-g} 
\bigg[
\frac{M_D^{D-2}}{2} R  
- \frac{1}{4} \sum_{i} F_i \cdot F_i
- \sum_a \bar\psi_a (\slashed{D} + m_a) \psi_a 
\bigg]
\,, \\
D_{\mu} &=  \de_\mu 
+ \frac{1}{4} \omega_{\mu\alpha\beta} \Gamma^{\alpha} \Gamma^{\beta}
+ i \sum_i q_{a i} g_i A_{i\mu}
\,,
\end{align}
where the $a$-th fermion has a mass $m_a$ and a charge $q_{a i}$ under the $i$-th $U(1)$. Since the fermions do not mix, the total 1-loop correction to the effective action equals the sum of the terms obtained by integrating out each fermion individually. The latter are simply given by those obtained in the single $U(1)$ case, with the replacement $qg F \to \sum_i q_{a i} g_i F_i$. The same argument holds for scalars. Hence, the effective Lagrangian is
\begin{align}
\label{finalEFT}
\call &= 
  \frac{M_D^{D-2}}{2} R
- \frac{1}{4} \sum_{i} F_i \cdot F_i 
+ \sum_{i,j,k,l} \bigg( a_{1ijkl} (F_i \cdot F_j) (F_k \cdot F_l)
+ a_{2ijkl} (F_i \cdot \tilde F_j) (F_k \cdot \tilde F_l) \bigg)
\nl
+  \sum_{i,j} 
\bigg(  b_{1ij} (F_i \cdot F_j)  R
+  b_{2ij} F_{i\mu\rho} F_{j\nu}^{\ \ \rho}  R^{\mu\nu}
+  b_{3ij} F_{i\mu\nu} F_{j \rho\sigma}  R^{\mu\nu\rho\sigma}
\bigg) \nl
+ c_1 R^2 
+ c_2 R_{\mu\nu}^2  
+ c_3 R_{\mu\nu\rho\sigma}^2
\,,
\end{align}
with coefficients given in Tables \ref{Tab:scalars}--\ref{Tab:fermions}. There, we introduced
\begin{align}
\cali_a \equiv \int_{1/\Lambda^2}^\infty \frac{\d t}{t} \e^{-tm_a^2}  =
\log\frac{\Lambda^2}{m_a^2} - \gamma + \calo(\Lambda^{-2}) \,.
\end{align}

\begin{table}[t]
\centering
$
\begin{array}{ccc}
\toprule
\text{Scalars} & D=3 & D=4 \\
\midrule
a_{1ijkl} & \sum_a \frac{7 q_{ai}g_{i}q_{aj}g_{j}q_{ak}g_{k}q_{al}g_{l} }{15360 \pi m_a^5}  & \sum_a \frac{7 q_{ai}g_{i}q_{aj}g_{j}q_{ak}g_{k}q_{al}g_{l}}{23040 \pi^2 m_a^4} \\
a_{2ijkl} & - & \sum_a \frac{q_{ai}g_{i}q_{aj}g_{j}q_{ak}g_{k}q_{al}g_{l}}{23040 \pi^2 m_a^4}  \\
b_{1 ij} & - \sum_a  \frac{q_{ai}g_{i}q_{aj}g_{j}}{1152 \pi m_a^3}  & - \sum_a \frac{q_{ai}g_{i}q_{aj}g_{j}}{1152 \pi^2 m_a^2} \\
b_{2 ij} & - \sum_a  \frac{q_{ai}g_{i}q_{aj}g_{j}}{1440 \pi m_a^3}  & - \sum_a \frac{q_{ai}g_{i}q_{aj}g_{j}}{1440 \pi^2 m_a^2} \\
b_{3 ij} & - \sum_a  \frac{q_{ai}g_{i}q_{aj}g_{j}}{2880 \pi m_a^3}  & - \sum_a \frac{q_{ai}g_{i}q_{aj}g_{j}}{2880 \pi^2 m_a^2} \\
c_1 &  \sum_a  \frac{1}{576 \pi m_a} & \sum_a  \frac{\cali_a}{1152 \pi^2} \\
c_2 & - \sum_a  \frac{1}{1440 \pi m_a} & - \sum_a  \frac{\cali_a}{2880 \pi^2} \\
c_3 & \sum_a  \frac{1}{1440 \pi m_a} & \sum_a  \frac{\cali_a}{2880 \pi^2} \\
\bottomrule
\end{array}
$
\caption{\label{Tab:scalars} \emph{Scalar case.}}
\end{table}
\begin{table}[t]
\centering
$
\begin{array}{ccc}
\toprule
\text{Fermions} & D=3 & D=4 \\
\midrule
a_{1ijkl} & \sum_{a} \frac{q_{ai}g_{i}q_{aj}g_{j}q_{ak}g_{k}q_{al}g_{l}}{1920 \pi m_a^5}  &\sum_a \frac{q_{ai}g_{i}q_{aj}g_{j}q_{ak}g_{k}q_{al}g_{l}}{1440 \pi^2 m_a^4} \\
a_{2ijkl} & - & \sum_a \frac{7 q_{ai}g_{i}q_{aj}g_{j}q_{ak}g_{k}q_{al}g_{l}}{5760 \pi^2 m_a^4} \\
b_{1 ij} & \sum_a \frac{q_{ai}g_{i}q_{aj}g_{j}}{1152 \pi m_a^3}   & \sum_a \frac{q_{ai}g_{i}q_{aj}g_{j}}{576 \pi^2 m_a^2} \\
b_{2 ij} & - \sum_a \frac{13 q_{ai}g_{i}q_{aj}g_{j}}{2880\pi m_a^3}   & - \sum_a \frac{13 q_{ai}g_{i}q_{aj}g_{j}}{1440 \pi^2 m_a^2}  \\
b_{3 ij} & \sum_a \frac{q_{ai}g_{i}q_{aj}g_{j}}{2880\pi m_a^3}    & \sum_a \frac{q_{ai}g_{i}q_{aj}g_{j}}{1440 \pi^2 m_a^2} \\
c_1 & - \sum_a \frac{1}{2304 m_a \pi} & - \sum_a \frac{\cali_a}{2304 \pi^2}  \\
c_2 & \sum_a \frac{1}{1440 m_a \pi}   &  \sum_a \frac{\cali_a}{1440\pi^2} \\
c_3 & \sum_a \frac{7}{11520 m_a \pi}   & \sum_a \frac{7\cali_a}{11520\pi^2}   \\
\bottomrule
\end{array}
$
\caption{\label{Tab:fermions} \emph{Fermion case.}}
\end{table}

\subsection{Simplifying the Effective Action}
\label{simplify_action}

Since we restrict to terms in the effective action with not more than 4 derivatives, we can consistently use the leading-order Einstein equations in order to recast $R F^2$ and $R^2$ interactions in terms of $F^4$ terms (this is tantamount to applying a field redefinition) \cite{Cheung:2014vva}. This works slightly differently in 3D and 4D, as we show below. 

\medskip
In 3D, we can use the fact that the Weyl tensor vanishes, which yields the following 3D identity:
\begin{equation}
\begin{aligned}
R_{\mu\nu\rho\sigma} 
& =
 g_{\mu\rho}  R_{\sigma\nu} 
- g_{\mu\sigma}  R_{\rho\nu} 
- g_{\nu\rho}  R_{\sigma\mu} 
+  g_{\nu\sigma}  R_{\rho\mu} 
- \frac{1}{2} ( g_{\mu\rho} g_{\sigma\nu} - g_{\mu\sigma} g_{\rho\nu} ) R
\end{aligned}
\end{equation}
such that
\begin{equation}
R_{\mu\nu\rho\sigma}^2 = 4 R_{\mu\nu}^2 - R^2
\,.
\end{equation}
$R$ and $R_{\mu\nu}$ can be eliminated using the leading-order Einstein equations
\begin{equation}
R_{\mu\nu} - \frac{1}{2} g_{\mu\nu} R =  \frac{1}{M_3}  
\sum_{i} \left[
 F_{i\mu\rho}F_{i\nu}{}^\rho
-\frac{1}{4} g_{\mu\nu} F_i \cdot F_i
\right] 
\,.
\end{equation}
From this equation we find
\begin{equation}
R = 
- \frac{1}{2M_3} \sum_{i} F_i \cdot F_i
\,,
\qquad
R_{\mu\nu} =  
\frac{1}{M_3} 
\sum_{i} \bigg( 
 F_{i \mu\rho} F_{i \nu}^{\ \ \rho} 
- \frac{1}{2} g_{\mu\nu} F_i \cdot F_i 
\bigg)
\,.
\end{equation}
Using this together with the 3D identity \eqref{app:identity_3d}, the 1-loop effective Lagrangian can be recast as
\begin{equation}
\label{finalEFT_Fgauge}
\call = 
\frac{M_3}{2} R
- \frac{1}{4} \sum_{i} F_i \cdot F_i
+ \sum_{i,j,k,l} C_{ijkl} (F_i \cdot F_j) (F_k \cdot F_l)
\,,
\end{equation}
where
\begin{align}
C_{ijkl} &= 
a_{1ijkl} 
- \frac{b_{1ij}}{2M_3} \delta_{kl}
- \frac{b_{2ij}}{4M_3} \delta_{kl}
- \frac{b_{3ij}}{2M_3} \delta_{kl}
+ \frac{1}{4M_3^2} (c_1- c_3) \delta_{ij} \delta_{kl} 
\nl + \frac{b_{2ik}}{4M_3} \delta_{jl}
+ \frac{b_{3ik}}{M_3} \delta_{jl}
+ \frac{1}{4M_3^2} (c_2 + 4 c_3) \delta_{ik} \delta_{jl}
+c_{ijkl}
\,, \label{C's}
\end{align}
where $c_{ijkl}$ are the UV-dependent parameters introduced in the main text.

\medskip
In 4D, we can use the Gauss-Bonnet term (corresponding to a total derivative) to replace $R_{\mu\nu\rho\sigma}^2$ with
\begin{equation}
R_{\mu\nu\rho\sigma}R^{\mu\nu\rho\sigma} = \text{tot.\ der.} + 4 R^{\mu\nu} R_{\mu\nu} - R^2
\,.
\end{equation}
Moreover, we can use the Weyl tensor 
\begin{equation}
W_{\mu\nu\rho\sigma} 
= 
R_{\mu\nu\rho\sigma} 
- \frac{1}{2}
(g_{\mu\rho}R_{\sigma\nu} 
-g_{\mu\sigma}R_{\rho\nu}
-g_{\nu\rho}R_{\sigma\mu}
+g_{\nu\sigma}R_{\rho\mu})
+ \frac{1}{6} R (g_{\mu\rho}g_{\sigma\nu} - g_{\mu\sigma}g_{\rho\nu})
\end{equation}
to rewrite
\begin{align}
R^{\mu\nu\rho\sigma} F_{i\mu\nu} F_{j\rho\sigma} = 
W^{\mu\nu\rho\sigma} F_{i\mu\nu} F_{j\rho\sigma}
+ 2 R^{\sigma\nu} F_{i\mu\nu} F^{\mu}_{j\ \sigma}
- \frac{1}{3} R F_i \cdot F_j
\,.
\end{align}
The leading-order Einstein equations are
\begin{equation}
R_{\mu\nu} - \frac{1}{2} g_{\mu\nu} R = \frac{1}{M_4^2}  
\sum_{i} \bigg(
 F_{i\mu\rho}F_{i \nu}{}^\rho
- \frac{1}{4} g_{\mu\nu}  F_i \cdot F_i
\bigg)
\,,
\end{equation}
which implies
\begin{equation}
R  = 0 \,, 
\qquad
R_{\mu\nu} = \frac{1}{M_4^2}  
\sum_{i} \bigg(
 F_{i\mu\rho}F_{i \nu}{}^\rho
- \frac{1}{4} g_{\mu\nu} F_i \cdot F_i
\bigg)
\,.
\end{equation}
Using these expressions and the 4D identity \eqref{app:identity_4d}, the 1-loop effective action can be recast as 
\begin{align}
\call = 
& M_4^{2} \frac{R}{2} 
- \frac{1}{4} \sum_{i} F_i \cdot F_i 
+ \sum_{i,j,k,l} \bigg( C_{1 ijkl}  (F_i \cdot F_j) (F_k \cdot F_l)
+ C_{2ijkl} (F_i \cdot \tilde F_j) (F_k \cdot \tilde F_l) \bigg)
\nll
+ W^{\mu\nu\rho\sigma}  \sum_{i,j} (b_{3ij}+c_{3ij})  F_{i\mu\nu} F_{j\rho\sigma}
\,,
\end{align}
where
\begin{align}
\label{C's4d}
C_{1 ijkl} &= a_{1ijkl} + \frac{b_{2ik} + 2 b_{3ik}}{4 M_4^2}  \delta_{jl} 
+ \frac{c_2+4 c_3}{4 M_4^4} \delta_{ik} \delta_{jl}
+c_{1ijkl}\,, \\
C_{2 ijkl} &= a_{2ijkl} + \frac{b_{2ik} + 2 b_{3ik}}{4 M_4^2}  \delta_{jl}
+ \frac{c_2+4 c_3}{4 M_4^4} \delta_{ik} \delta_{jl}
+c_{2ijkl}
\,.
\end{align}
Again, $c_{1ijkl}$, $c_{2ijkl}$ and $c_{3ij}$ are the UV-dependent parameters introduced in the main~text.

\section{Details on Dimensional Reduction}
\label{app:reduction}

Here, we compute the one-loop corrections to the 3D effective action of Sec.~\ref{sec:comp} which arise from integrating out a KK tower of charged fermions/scalars.
The reduction ansatz is provided in the main text, see Sec.~\ref{sec:comp-setup}.

\subsection{Charged Fermions}

Let us begin with the fermion case. When we integrate out all KK modes associated to a single 4D fermion, it is convenient to keep the 4D spinor representation rather than to decompose it to the 3D one. More explicitly, we reformulate the fermion part of the action~\eqref{3d-dirac}  back to the form
\begin{align}
\label{app:reduced_action_simplified_-+++}
\Gamma & = \sum_{n} \int \d^3 x \sqrt{- g} 
\bar\Psi_{(n)} (- \cald - m) \Psi_{(n)}
\quad
{\rm with}
\quad
\Psi_{(n)}=\left(\psi^{(n)},\chi^{(n)}\right)\,,
\end{align}
where $\psi^{(n)}$ and $\chi^{(n)}$ are in 3D spinor representations as introduced in Eq.~\eqref{spinor_4D_to_3D}, and we define the operator $\cald$ in terms of the covariant derivative $D_\mu$ and its gravitational part $\nabla_\mu$ as
\begin{align}
\cald & \equiv 
\slashed{D} 
+ \frac{in}{r}  \Gamma^3
- \frac{rg_\text{KK}}{8} \slashed{H} \Gamma^3
\,, \\
 D_\mu \Psi_{(n)}
& = ( \nabla_\mu  + i qg_3A_\mu + ing_\text{KK}B_\mu)\Psi_{(n)} \nll =
(\de_\mu  + \frac{1}{4} {\omega}_{\mu \alpha\beta} \Gamma^\alpha \Gamma^\beta + i qg_3 A_\mu  + i ng_\text{KK} B_\mu)\Psi_{(n)}
\,.
\end{align}
Here and in the rest of this appendix, we omit the label $(0)$ for the zero modes of gauge fields and the metric. We also defined $\slashed{H} = H_{\mu\nu}\Gamma^\mu\Gamma^\nu$. Also note that $\Gamma^3$ is in the locally flat frame.

\medskip
By following the procedure described in App.~\ref{subsec:EFT3d}, we find that
\begin{equation}
\label{Dirac_squared}
\begin{aligned}
(\cald - m)(-\cald - m) & = -\cald^2 + m^2 
& = -  g^{\mu\nu} D^{H}_{\mu} D^{H}_{\nu} + U + m_n^2
\,,
\end{aligned}
\end{equation}
where $m_n^2 = m^2 + \frac{n^2}{r^2}$ are the KK masses, while
\begin{align}
\label{nabla}
D^H_\mu \Psi_{(n)} &= 
\left( D_\mu 
- \frac{rg_\text{KK}}{16} [\Gamma_\mu , \slashed{H}] \Gamma^3	
\right) \Psi_{(n)}	\,, \\
\label{U}
U &= 
\frac{ R}{4} 
- \frac{iqg_3}{2}\slashed{F} - \frac{ing_\text{KK}}{4}\slashed{H}
+ \frac{rg_\text{KK}}{16} (\nabla_\mu H_{\rho\sigma}) [\Gamma^\mu,\Gamma^\rho]\Gamma^\sigma \Gamma^3 		
- \frac{r^2g_\text{KK}^2}{32} H^2			
\,
\end{align}
up to terms vanishing by the leading-order equations of motion $\nabla^\mu H_{\mu\nu}=0$.

\medskip
The one-loop contribution to the final effective action will be given by $\sum_n \delta\Gamma_1^{(n)}$, where $\delta\Gamma_1^{(n)}$ is the one-loop contribution given by the $n$-th mode. This can be computed with the local heat kernel procedure for each $n$ separately since the modes do not mix. The result are the following operators $B_i$. Similarly to the case discussed in App.~\ref{subsec:EFT3d}, the operators arising from $B_0$ and $B_2$ may be removed by renormalization. $B_4$, $B_6$ and $B_8$ provide four-derivative operators in our interest as\footnote{We use the computer software \textbf{Cadabra} to work out the more complicated gamma matrix combinations \cite{Peeters:2007wn}.}
\begin{align}
B_4 = \int \d^3x \sqrt{- g} \ 
\bigg[ &
\frac{1}{72}  R^2  
- \frac{1}{45}  R^2_{\mu\nu} 
- \frac{7}{360}  R^2_{\mu\nu\rho\sigma} 
- \frac{1}{1536} r^4g_\text{KK}^4 (H^2)^2 
- \frac{1}{96} r^2g_\text{KK}^2  R H^2 \nll 
+ \frac{1}{24} r^2g_\text{KK}^2  H^{\rho\mu} H_{\rho}^{\ \nu}  R_{\mu\nu} 
\bigg] 
+ \sum_{i=\text{odd}} n^i (\ldots)
\,, \\
B_6 = \int \d^3x \sqrt{- g} 
\bigg[&
- \frac{1}{18} q^2g_3^2 F^2  R
+\frac{1}{144} n^2g_\text{KK}^2 H^2  R
-\frac{13}{384} n^2 r^2 g_\text{KK}^4 (H^2)^2
\nll 
-\frac{1}{45} q^2g_3^2 F^{\mu\nu} F^{\rho\lambda}  R_{\mu\nu\rho\lambda}
+\frac{7}{360} n^2g_\text{KK}^2 H^{\mu\nu} H^{\rho\lambda}  R_{\mu\nu\rho\lambda}\nll
-\frac{1}{16} q^2g_3^2 r^2g_\text{KK}^2 (F\cdot H)^2 
+\frac{13}{45} q^2g_3^2 F^{\mu\nu} F_{\mu}^{\ \rho}  R_{\nu\rho} + \frac{7}{180} n^2g_\text{KK}^2  H^{\mu\nu} H_{\mu}^{\ \rho}  R_{\nu\rho}
\bigg] \nll
+ \sum_{i=\text{odd}} n^i (\ldots)	
\,, \\
B_8 =
\int \d^3x \sqrt{- g} \  
\bigg[& 
- \frac{1}{45} q^4g_3^4 (F^2)^2 
+ \frac{7}{5760} n^4g_\text{KK}^4 (H^2)^2 
- \frac{17}{360} q^2g_3^2 n^2g_\text{KK}^2 (F \cdot H)^2 
\nll 
- \frac{2}{45}  q^2g_3^2 n^2g_\text{KK}^2 F^2 H^2 
\bigg]
+ \sum_{i=\text{odd}} n^i (\ldots)\,,
\end{align}
where we used the 3D identity \eqref{app:identity_3d}, the equations of motion $\nabla_\mu F^{\mu\nu} = \nabla_\mu H^{\mu\nu} = 0$ (equivalent to a field redefinition), and
\begin{align}
\label{3d_identity_nabla1}
& \nabla_\mu H_{\rho\sigma} \nabla^\mu H^{\rho\sigma} = 
H^{\mu\nu} H^{\rho\sigma}  R_{\mu\nu\rho\sigma} 
+ 2 H^{\mu\rho} H_{\rho}^{\ \nu}  R_{\mu\nu} 
\,, \\
\label{3d_identity_nabla2}
& \nabla_\mu H_{\rho\sigma} \nabla^\sigma H^{\mu\rho} = 
- \frac{1}{2}  H^{\mu\nu} H^{\rho\sigma}  R_{\mu\nu\rho\sigma} 
-  H^{\mu\rho} H_{\rho}^{\ \nu}  R_{\mu\nu} 
\,.
\end{align}
Here, we have omitted total derivatives on the right-hand sides and used again the leading-order equation of motion $\nabla_\mu H^{\mu\nu} = 0$. The same expressions hold for $F_{\mu\nu}$. We did not explicitly write down operators proportional to odd powers of $n$ since these terms will vanish after summing over $n \in [-n^*, n^*]$ with a cutoff $n^*$ given by $m_n\sim \Lambda$ (see also Sec.~\ref{sec:comp}).
We eventually find that the one-loop corrections in our interest are given by
\begin{align}
\label{red:EFT_final}
\delta\Gamma_1 = \int \d^3x \sqrt{- g}  
\bigg[ &
a_1 (F^2)^2
+ a_2 (H^2)^2 
+ a_3 F^2 H^2 
+ a_4 (F \cdot H)^2 
+ b_1  R F^2 
+ b_2 F^{\mu\nu} F_{\mu}^{\ \rho}  R_{\nu\rho}
\nll
+ b_3 F^{\mu\nu} F^{\rho\lambda}  R_{\mu\nu\rho\lambda}
+ b_4  R H^2 
+ b_5 H^{\rho\mu} H_{\rho}^{\ \nu}  R_{\mu\nu} 
+ b_6 H^{\mu\nu} H^{\rho\lambda}  R_{\mu\nu\rho\lambda} 
\nll + c_1  R^2  
+ c_2  R^2_{\mu\nu} 
+ c_3  R^2_{\mu\nu\rho\sigma} 
\bigg]
\end{align}
with coefficients
\begin{align}
\label{red:coef_ferm_4to3}
 a_1 &=  \frac{1}{16\pi} \sum_n \frac{z_{n2}^4}{60 |m_n| M_3^2}
\equiv a_{12222}
\,, \\
a_2 &=  \frac{1}{16\pi} \sum_n \bigg( 
\frac{1}{384 |m_n| M_3^2} 
+ \frac{13 z_{n1}^2}{384 |m_n| M_3^2} 
- \frac{7 z_{n1}^4}{7680 |m_n| M_3^2} \bigg)
\equiv a_{11111}
\,, \\
a_3 &= \frac{1}{16\pi} \sum_n \frac{z_{n2}^2 z_{n1}^2}{30 |m_n| M_3^2}
\,, \\
a_4 &= \frac{1}{16\pi} \sum_n \bigg(
\frac{z_{n2}^2}{16 |m_n| M_3^2}  
+ \frac{17 z_{n2}^2 z_{n1}^2}{480 |m_n| M_3^2}  \bigg)
\equiv a_{11212} + a_{11221} + a_{12112} + a_{12121}
\,, \\
b_1 &= \frac{1}{16\pi} \sum_n \frac{z_{n2}^2}{36 |m_n| M_3}
= \frac{5}{2} b_3 \equiv b_{122}
\,, \\
b_2 &=   - \frac{1}{16\pi} \sum_n \frac{13 z_{n2}^2}{90 |m_n| M_3} 
 = - 13 b_3 \equiv b_{222}
\,, \\
b_3 &=   \frac{1}{16\pi} \sum_n \frac{z_{n2}^2}{90 |m_n| M_3} 
\equiv b_{322}
\,, \\
b_4 &=   - \frac{1}{16\pi} \sum_n \bigg( - \frac{1}{48 |m_n| M_3} + \frac{z_{n1}^2}{288 |m_n| M_3}  \bigg)
\equiv b_{111}
\,, \\
b_5 &=   - \frac{1}{16\pi} \sum_n \bigg( \frac{1}{12 |m_n| M_3}  + \frac{7 z_{n1}^2}{360 |m_n| M_3}  \bigg)
\equiv b_{211}
\,, \\
b_6 &=    - \frac{1}{16\pi} \sum_n \frac{7 z_{n1}^2}{720 |m_n| M_3}
\equiv b_{311}
\,, \\
c_1 &= - \frac{1}{16\pi} \sum_n \frac{1}{72 |m_n|}
= - \frac{5}{7} c_3
\,, \\
c_2 &=  \frac{1}{16\pi} \sum_n \frac{1}{45 |m_n|}
= \frac{8}{7} c_3
\,, \\
c_3 &=  \frac{1}{16\pi} \sum_n \frac{7}{360 |m_n|}
\,.
\end{align}
Here, the charge-to-mass ratios are given by
\begin{equation}
z_{n1} = \frac{n g_\text{KK}\sqrt{M_3}}{|m_n|}\,, \qquad z_{n2} = \frac{qg_3\sqrt{M_3}}{|m_n|}\,.
\end{equation}

\subsection{Charged Scalars}

The computation for spin-0 matter fields is simpler since the reduction procedure does not yield any new interactions and leads to the same results in any dimension. Let us focus on the 4D $\to$ 3D case. We have a single complex scalar, charged under a $U(1)$:
\begin{equation}
\label{action_4d_scalar}
\Gamma = \int \d^4x \sqrt{-G} 
\left[
- |D_M\Phi|^2
- m^2 |\Phi|^2
\right]
\,.
\end{equation}
After compactification and reduction, the action is given by
\begin{equation}
\label{action_3d}
\Gamma = \sum_{n} \int \d^3x \sqrt{- g} 
\left[  
- |D_\mu\phi_n|^2 - m_n^2 |\phi_n|^2
\right]
\,,		
\end{equation}
where we absorbed a factor $\sqrt{2\pi r}$ into $\phi_n$, which now has the correct mass dimension in $D=3$. Furthermore, $m^2_n = m^2 + \frac{n^2}{r^2}$ are the KK scalar masses, and $D_\mu\phi_n = (\nabla_\mu + iq g_3 A_\mu + i n g_\text{KK} B_\mu) \phi_n$ is the covariant derivative.

\medskip
In order to integrate out the KK scalar tower, one has to isolate a Laplacian-like operator acting on $\phi_n$. This is achieved by integrating by parts (see Sec.~2 of \cite{Vassilevich:2003xt}). The equation of motion for each mode is therefore:
\begin{equation}
\label{laplacian}
D^\mu D_\mu \phi_n - m_n^2 \phi_n = 0\,.
\end{equation}
Since all KK modes are decoupled, we can compute the effective action for each one separately and take the sum over modes at the end. The computation follows the philosophy of App.~\eqref{subsec:mult}, where now each scalar mode is charged under the original $U(1)$ of the 4D theory (with charge $q$) and under the $U(1)_\text{KK}$ (with charge $n$).

\bibliographystyle{utphys}
\bibliography{groups}

\providecommand{\href}[2]{#2}\begingroup\raggedright\begin{thebibliography}{10}

\bibitem{ArkaniHamed:2006dz}
N.~Arkani-Hamed, L.~Motl, A.~Nicolis and C.~Vafa,  {\em {The String landscape,
  black holes and gravity as the weakest force}}, JHEP {\bf 0706} (2007) 060
[\href{http://www.arXiv.org/abs/hep-th/0601001}{{\tt hep-th/0601001}}].

\bibitem{Rudelius:2014wla}
T.~Rudelius,  {\em {On the Possibility of Large Axion Moduli Spaces}}, JCAP
  {\bf 1504} (2015), no.~04, 049
[\href{http://www.arXiv.org/abs/1409.5793}{{\tt 1409.5793}}].

\bibitem{delaFuente:2014aca}
A.~de~la Fuente, P.~Saraswat and R.~Sundrum,  {\em {Natural Inflation and
  Quantum Gravity}}, Phys. Rev. Lett. {\bf 114} (2015), no.~15, 151303
[\href{http://www.arXiv.org/abs/1412.3457}{{\tt 1412.3457}}].

\bibitem{Montero:2015ofa}
M.~Montero, A.~M. Uranga and I.~Valenzuela,  {\em {Transplanckian axions!?}},
  JHEP {\bf 08} (2015) 032
[\href{http://www.arXiv.org/abs/1503.03886}{{\tt 1503.03886}}].

\bibitem{Brown:2015iha}
J.~Brown, W.~Cottrell, G.~Shiu and P.~Soler,  {\em {Fencing in the Swampland:
  Quantum Gravity Constraints on Large Field Inflation}}, JHEP {\bf 10} (2015)
  023
[\href{http://www.arXiv.org/abs/1503.04783}{{\tt 1503.04783}}].

\bibitem{Brown:2015lia}
J.~Brown, W.~Cottrell, G.~Shiu and P.~Soler,  {\em {On Axionic Field Ranges,
  Loopholes and the Weak Gravity Conjecture}}, JHEP {\bf 04} (2016) 017
[\href{http://www.arXiv.org/abs/1504.00659}{{\tt 1504.00659}}].

\bibitem{Hebecker:2015rya}
A.~Hebecker, P.~Mangat, F.~Rompineve and L.~T. Witkowski,  {\em {Winding out of
  the Swamp: Evading the Weak Gravity Conjecture with F-term Winding
  Inflation?}}, Phys. Lett. {\bf B748} (2015) 455--462
[\href{http://www.arXiv.org/abs/1503.07912}{{\tt 1503.07912}}].

\bibitem{Bachlechner:2015qja}
T.~C. Bachlechner, C.~Long and L.~McAllister,  {\em {Planckian Axions and the
  Weak Gravity Conjecture}}, JHEP {\bf 01} (2016) 091
[\href{http://www.arXiv.org/abs/1503.07853}{{\tt 1503.07853}}].

\bibitem{Junghans:2015hba}
D.~Junghans,  {\em {Large-Field Inflation with Multiple Axions and the Weak
  Gravity Conjecture}}, JHEP {\bf 02} (2016) 128
[\href{http://www.arXiv.org/abs/1504.03566}{{\tt 1504.03566}}].

\bibitem{Rudelius:2015xta}
T.~Rudelius,  {\em {Constraints on Axion Inflation from the Weak Gravity
  Conjecture}}, JCAP {\bf 1509} (2015), no.~09, 020
[\href{http://www.arXiv.org/abs/1503.00795}{{\tt 1503.00795}}].

\bibitem{Heidenreich:2015wga}
B.~Heidenreich, M.~Reece and T.~Rudelius,  {\em {Weak Gravity Strongly
  Constrains Large-Field Axion Inflation}}, JHEP {\bf 12} (2015) 108
[\href{http://www.arXiv.org/abs/1506.03447}{{\tt 1506.03447}}].

\bibitem{Kooner:2015rza}
K.~Kooner, S.~Parameswaran and I.~Zavala,  {\em {Warping the Weak Gravity
  Conjecture}}, Phys. Lett. {\bf B759} (2016) 402--409
[\href{http://www.arXiv.org/abs/1509.07049}{{\tt 1509.07049}}].

\bibitem{Hebecker:2015zss}
A.~Hebecker, F.~Rompineve and A.~Westphal,  {\em {Axion Monodromy and the Weak
  Gravity Conjecture}}, JHEP {\bf 04} (2016) 157
[\href{http://www.arXiv.org/abs/1512.03768}{{\tt 1512.03768}}].

\bibitem{Palti:2015xra}
E.~Palti,  {\em {On Natural Inflation and Moduli Stabilisation in String
  Theory}}, JHEP {\bf 10} (2015) 188
[\href{http://www.arXiv.org/abs/1508.00009}{{\tt 1508.00009}}].

\bibitem{Baume:2016psm}
F.~Baume and E.~Palti,  {\em {Backreacted Axion Field Ranges in String
  Theory}}, JHEP {\bf 08} (2016) 043
[\href{http://www.arXiv.org/abs/1602.06517}{{\tt 1602.06517}}].

\bibitem{Hebecker:2016dsw}
A.~Hebecker, P.~Mangat, S.~Theisen and L.~T. Witkowski,  {\em {Can
  Gravitational Instantons Really Constrain Axion Inflation?}}, JHEP {\bf 02}
  (2017) 097
[\href{http://www.arXiv.org/abs/1607.06814}{{\tt 1607.06814}}].

\bibitem{Hebecker:2017wsu}
A.~Hebecker, P.~Henkenjohann and L.~T. Witkowski,  {\em {What is the Magnetic
  Weak Gravity Conjecture for Axions?}}, Fortsch. Phys. {\bf 65} (2017),
  no.~3-4, 1700011
[\href{http://www.arXiv.org/abs/1701.06553}{{\tt 1701.06553}}].

\bibitem{Hebecker:2017uix}
A.~Hebecker and P.~Soler,  {\em {The Weak Gravity Conjecture and the Axionic
  Black Hole Paradox}}, JHEP {\bf 09} (2017) 036
[\href{http://www.arXiv.org/abs/1702.06130}{{\tt 1702.06130}}].

\bibitem{Blumenhagen:2017cxt}
R.~Blumenhagen, I.~Valenzuela and F.~Wolf,  {\em {The Swampland Conjecture and
  F-term Axion Monodromy Inflation}}, JHEP {\bf 07} (2017) 145
[\href{http://www.arXiv.org/abs/1703.05776}{{\tt 1703.05776}}].

\bibitem{Ibanez:2015fcv}
L.~E. Ibanez, M.~Montero, A.~Uranga and I.~Valenzuela,  {\em {Relaxion
  Monodromy and the Weak Gravity Conjecture}}, JHEP {\bf 04} (2016) 020
[\href{http://www.arXiv.org/abs/1512.00025}{{\tt 1512.00025}}].

\bibitem{Brown:2016nqt}
J.~Brown, W.~Cottrell, G.~Shiu and P.~Soler,  {\em {Tunneling in Axion
  Monodromy}}, JHEP {\bf 10} (2016) 025
[\href{http://www.arXiv.org/abs/1607.00037}{{\tt 1607.00037}}].

\bibitem{Ooguri:2006in}
H.~Ooguri and C.~Vafa,  {\em {On the Geometry of the String Landscape and the
  Swampland}}, Nucl.Phys. {\bf B766} (2007) 21--33
[\href{http://www.arXiv.org/abs/hep-th/0605264}{{\tt hep-th/0605264}}].

\bibitem{Klaewer:2016kiy}
D.~Klaewer and E.~Palti,  {\em {Super-Planckian Spatial Field Variations and
  Quantum Gravity}}, JHEP {\bf 01} (2017) 088
[\href{http://www.arXiv.org/abs/1610.00010}{{\tt 1610.00010}}].

\bibitem{Palti:2017elp}
E.~Palti,  {\em {The Weak Gravity Conjecture and Scalar Fields}}, JHEP {\bf 08}
  (2017) 034
[\href{http://www.arXiv.org/abs/1705.04328}{{\tt 1705.04328}}].

\bibitem{Crisford:2017gsb}
T.~Crisford, G.~T. Horowitz and J.~E. Santos,  {\em {Testing the Weak Gravity -
  Cosmic Censorship Connection}},
\href{http://www.arXiv.org/abs/1709.07880}{{\tt 1709.07880}}.

\bibitem{Cottrell:2016bty}
W.~Cottrell, G.~Shiu and P.~Soler,  {\em {Weak Gravity Conjecture and Extremal
  Black Holes}},
\href{http://www.arXiv.org/abs/1611.06270}{{\tt 1611.06270}}.

\bibitem{Ooguri:2016pdq}
H.~Ooguri and C.~Vafa,  {\em {Non-supersymmetric AdS and the Swampland}},
\href{http://www.arXiv.org/abs/1610.01533}{{\tt 1610.01533}}.

\bibitem{Danielsson:2016mtx}
U.~Danielsson and G.~Dibitetto,  {\em {Fate of stringy AdS vacua and the weak
  gravity conjecture}}, Phys. Rev. {\bf D96} (2017), no.~2, 026020
[\href{http://www.arXiv.org/abs/1611.01395}{{\tt 1611.01395}}].

\bibitem{Freivogel:2016qwc}
B.~Freivogel and M.~Kleban,  {\em {Vacua Morghulis}},
\href{http://www.arXiv.org/abs/1610.04564}{{\tt 1610.04564}}.

\bibitem{Ooguri:2017njy}
H.~Ooguri and L.~Spodyneiko,  {\em {New Kaluza-Klein instantons and the decay
  of AdS vacua}}, Phys. Rev. {\bf D96} (2017), no.~2, 026016
[\href{http://www.arXiv.org/abs/1703.03105}{{\tt 1703.03105}}].

\bibitem{Danielsson:2017max}
U.~H. Danielsson, G.~Dibitetto and S.~C. Vargas,  {\em {A swamp of non-SUSY
  vacua}}, JHEP {\bf 11} (2017) 152
[\href{http://www.arXiv.org/abs/1708.03293}{{\tt 1708.03293}}].

\bibitem{Ibanez:2017kvh}
L.~E. Ibanez, V.~Martin-Lozano and I.~Valenzuela,  {\em {Constraining Neutrino
  Masses, the Cosmological Constant and BSM Physics from the Weak Gravity
  Conjecture}}, JHEP {\bf 11} (2017) 066
[\href{http://www.arXiv.org/abs/1706.05392}{{\tt 1706.05392}}].

\bibitem{Ibanez:2017oqr}
L.~E. Ibanez, V.~Martin-Lozano and I.~Valenzuela,  {\em {Constraining the EW
  Hierarchy from the Weak Gravity Conjecture}},
\href{http://www.arXiv.org/abs/1707.05811}{{\tt 1707.05811}}.

\bibitem{Hamada:2017yji}
Y.~Hamada and G.~Shiu,  {\em {Weak Gravity Conjecture, Multiple Point Principle
  and the Standard Model Landscape}}, JHEP {\bf 11} (2017) 043
[\href{http://www.arXiv.org/abs/1707.06326}{{\tt 1707.06326}}].

\bibitem{Montero:2017yja}
M.~Montero, A.~M. Uranga and I.~Valenzuela,  {\em {A Chern-Simons Pandemic}},
  JHEP {\bf 07} (2017) 123
[\href{http://www.arXiv.org/abs/1702.06147}{{\tt 1702.06147}}].

\bibitem{Montero:2017mdq}
M.~Montero,  {\em {Are tiny gauge couplings out of the Swampland?}}, JHEP {\bf
  10} (2017) 208
[\href{http://www.arXiv.org/abs/1708.02249}{{\tt 1708.02249}}].

\bibitem{Hebecker:2017lxm}
A.~Hebecker, P.~Henkenjohann and L.~T. Witkowski,  {\em {Flat Monodromies and a
  Moduli Space Size Conjecture}}, JHEP {\bf 12} (2017) 033
[\href{http://www.arXiv.org/abs/1708.06761}{{\tt 1708.06761}}].

\bibitem{Lust:2017wrl}
D.~Lust and E.~Palti,  {\em {Scalar Fields, Hierarchical UV/IR Mixing and The
  Weak Gravity Conjecture}},
\href{http://www.arXiv.org/abs/1709.01790}{{\tt 1709.01790}}.

\bibitem{Ibanez:2017vfl}
L.~E. Ibanez and M.~Montero,  {\em {A Note on the WGC, Effective Field Theory
  and Clockwork within String Theory}},
\href{http://www.arXiv.org/abs/1709.02392}{{\tt 1709.02392}}.

\bibitem{rudelius2017}
B.~Heidenreich, M.~Reece and T.~Rudelius,  {\em {The Weak Gravity Conjecture
  and Emergence from an Ultraviolet Cutoff}},
\href{http://www.arXiv.org/abs/1712.01868}{{\tt 1712.01868}}.

\bibitem{Brennan:2017rbf}
T.~D. Brennan, F.~Carta and C.~Vafa,  {\em {The String Landscape, the
  Swampland, and the Missing Corner}},
\href{http://www.arXiv.org/abs/1711.00864}{{\tt 1711.00864}}.

\bibitem{Cheung:2014vva}
C.~Cheung and G.~N. Remmen,  {\em {Naturalness and the Weak Gravity
  Conjecture}}, Phys. Rev. Lett. {\bf 113} (2014) 051601
[\href{http://www.arXiv.org/abs/1402.2287}{{\tt 1402.2287}}].

\bibitem{Heidenreich:2015nta}
B.~Heidenreich, M.~Reece and T.~Rudelius,  {\em {Sharpening the Weak Gravity
  Conjecture with Dimensional Reduction}}, JHEP {\bf 02} (2016) 140
[\href{http://www.arXiv.org/abs/1509.06374}{{\tt 1509.06374}}].

\bibitem{Montero:2016tif}
M.~Montero, G.~Shiu and P.~Soler,  {\em {The Weak Gravity Conjecture in three
  dimensions}}, JHEP {\bf 10} (2016) 159
[\href{http://www.arXiv.org/abs/1606.08438}{{\tt 1606.08438}}].

\bibitem{Heidenreich:2016aqi}
B.~Heidenreich, M.~Reece and T.~Rudelius,  {\em {Evidence for a sublattice weak
  gravity conjecture}}, JHEP {\bf 08} (2017) 025
[\href{http://www.arXiv.org/abs/1606.08437}{{\tt 1606.08437}}].

\bibitem{palti2018}
T.~Grimm, E.~Palti and I.~Valenzuela,  {\em {Infinite Distances in Field Spaces
  and Massless Towers of States}}, to appear.

\bibitem{Nakayama:2015hga}
Y.~Nakayama and Y.~Nomura,  {\em {Weak gravity conjecture in the AdS/CFT
  correspondence}}, Phys. Rev. {\bf D92} (2015), no.~12, 126006
[\href{http://www.arXiv.org/abs/1509.01647}{{\tt 1509.01647}}].

\bibitem{Harlow:2015lma}
D.~Harlow,  {\em {Wormholes, Emergent Gauge Fields, and the Weak Gravity
  Conjecture}}, JHEP {\bf 01} (2016) 122
[\href{http://www.arXiv.org/abs/1510.07911}{{\tt 1510.07911}}].

\bibitem{Benjamin:2016fhe}
N.~Benjamin, E.~Dyer, A.~L. Fitzpatrick and S.~Kachru,  {\em {Universal Bounds
  on Charged States in 2d CFT and 3d Gravity}}, JHEP {\bf 08} (2016) 041
[\href{http://www.arXiv.org/abs/1603.09745}{{\tt 1603.09745}}].

\bibitem{Fisher:2017dbc}
Z.~Fisher and C.~J. Mogni,  {\em {A Semiclassical, Entropic Proof of a Weak
  Gravity Conjecture}},
\href{http://www.arXiv.org/abs/1706.08257}{{\tt 1706.08257}}.

\bibitem{Cheung:2018cwt}
C.~Cheung, J.~Liu and G.~N. Remmen,  {\em {Proof of the Weak Gravity Conjecture
  from Black Hole Entropy}},
\href{http://www.arXiv.org/abs/1801.08546}{{\tt 1801.08546}}.

\bibitem{Sen:2011ba}
A.~Sen,  {\em {Logarithmic Corrections to N=2 Black Hole Entropy: An Infrared
  Window into the Microstates}}, Gen. Rel. Grav. {\bf 44} (2012), no.~5,
  1207--1266
[\href{http://www.arXiv.org/abs/1108.3842}{{\tt 1108.3842}}].

\bibitem{Sen:2014aja}
A.~Sen,  {\em {Microscopic and Macroscopic Entropy of Extremal Black Holes in
  String Theory}}, Gen. Rel. Grav. {\bf 46} (2014) 1711
[\href{http://www.arXiv.org/abs/1402.0109}{{\tt 1402.0109}}].

\bibitem{Adams:2006sv}
A.~Adams, N.~Arkani-Hamed, S.~Dubovsky, A.~Nicolis and R.~Rattazzi,  {\em
  {Causality, analyticity and an IR obstruction to UV completion}}, JHEP {\bf
  10} (2006) 014
[\href{http://www.arXiv.org/abs/hep-th/0602178}{{\tt hep-th/0602178}}].

\bibitem{Cheung:2014ega}
C.~Cheung and G.~N. Remmen,  {\em {Infrared Consistency and the Weak Gravity
  Conjecture}}, JHEP {\bf 12} (2014) 087
[\href{http://www.arXiv.org/abs/1407.7865}{{\tt 1407.7865}}].

\bibitem{Shiu:2015uva}
G.~Shiu, W.~Staessens and F.~Ye,  {\em {Widening the Axion Window via Kinetic
  and Stückelberg Mixings}}, Phys. Rev. Lett. {\bf 115} (2015) 181601
[\href{http://www.arXiv.org/abs/1503.01015}{{\tt 1503.01015}}].

\bibitem{Shiu:2015xda}
G.~Shiu, W.~Staessens and F.~Ye,  {\em {Large Field Inflation from Axion
  Mixing}}, JHEP {\bf 06} (2015) 026
[\href{http://www.arXiv.org/abs/1503.02965}{{\tt 1503.02965}}].

\bibitem{Saraswat:2016eaz}
P.~Saraswat,  {\em {Weak gravity conjecture and effective field theory}}, Phys.
  Rev. {\bf D95} (2017), no.~2, 025013
[\href{http://www.arXiv.org/abs/1608.06951}{{\tt 1608.06951}}].

\bibitem{Dunne:1998qy}
G.~V. Dunne,  {\em {Aspects of Chern-Simons theory}}, in {\em {Topological
  Aspects of Low-dimensional Systems: Proceedings, Les Houches Summer School of
  Theoretical Physics, Session 69: Les Houches, France, July 7-31 1998}}.
\newblock 1998.
\newblock
\href{http://www.arXiv.org/abs/hep-th/9902115}{{\tt hep-th/9902115}}.
\newblock

\bibitem{Deser:1983tn}
S.~Deser, R.~Jackiw and G.~'t~Hooft,  {\em {Three-Dimensional Einstein Gravity:
  Dynamics of Flat Space}}, Annals Phys. {\bf 152} (1984)
220.

\bibitem{Hinterbichler:2011tt}
K.~Hinterbichler,  {\em {Theoretical Aspects of Massive Gravity}}, Rev. Mod.
  Phys. {\bf 84} (2012) 671--710
[\href{http://www.arXiv.org/abs/1105.3735}{{\tt 1105.3735}}].

\bibitem{Grimm:2013gma}
T.~W. Grimm, R.~Savelli and M.~Weissenbacher,  {\em {On $\alpha'$ corrections
  in N=1 F-theory compactifications}}, Phys. Lett. {\bf B725} (2013) 431--436
[\href{http://www.arXiv.org/abs/1303.3317}{{\tt 1303.3317}}].

\bibitem{Grimm:2017okk}
T.~W. Grimm, K.~Mayer and M.~Weissenbacher,  {\em {Higher derivatives in Type
  II and M-theory on Calabi-Yau threefolds}},
\href{http://www.arXiv.org/abs/1702.08404}{{\tt 1702.08404}}.

\bibitem{Vassilevich:2003xt}
D.~V. Vassilevich,  {\em {Heat kernel expansion: User's manual}}, Phys. Rept.
  {\bf 388} (2003) 279--360
[\href{http://www.arXiv.org/abs/hep-th/0306138}{{\tt hep-th/0306138}}].

\bibitem{Avramidi:2000bm}
I.~G. Avramidi,  {\em {Heat kernel and quantum gravity}}, Lect. Notes Phys.
  {\bf M64} (2000)
1--149.

\bibitem{vanNieuwenhuizen:1996tv}
P.~van Nieuwenhuizen and A.~Waldron,  {\em {On Euclidean spinors and Wick
  rotations}}, Phys. Lett. {\bf B389} (1996) 29--36
[\href{http://www.arXiv.org/abs/hep-th/9608174}{{\tt hep-th/9608174}}].

\bibitem{Ritz:1995nt}
A.~Ritz and R.~Delbourgo,  {\em {The Low-energy effective Lagrangian for photon
  interactions in any dimension}}, Int. J. Mod. Phys. {\bf A11} (1996) 253--270
[\href{http://www.arXiv.org/abs/hep-th/9503160}{{\tt hep-th/9503160}}].

\bibitem{Peeters:2007wn}
K.~Peeters,  {\em {Introducing Cadabra: A Symbolic computer algebra system for
  field theory problems}},
\href{http://www.arXiv.org/abs/hep-th/0701238}{{\tt hep-th/0701238}}.

\end{thebibliography}\endgroup

\end{document}